\documentclass[useAMS,usenatbib]{mn2e}
\usepackage{epsfig}
\usepackage{natbib}
\usepackage{soul}
\usepackage{color}
\title[Molecular cloud distance determination from deep NIR survey extinction measurements]{Molecular cloud distance determination from deep NIR survey extinction measurements}
\author[]{J.J. Stead$^{1}$\thanks{E-mail:
phy2j2s@leeds.ac.uk; m.g.hoare@leeds.ac.uk}, M.G. Hoare$^{1}$\footnotemark[1]\\
$^{1}$School of Physics and Astronomy, University of Leeds, Leeds, LS2 9JT\\
}

\def\Bes{\rm Besan$\c{c}$\rm on}
\def\arcmin{^{\prime}}
\begin{document}

\date{}

\pagerange{\pageref{firstpage}--\pageref{lastpage}} \pubyear{2002}

\maketitle

\label{firstpage}

\begin{abstract}

Using near infrared UKIDSS Galactic Plane Survey data, we make extinction measurements to individual stars along the same line of sight as molecular clouds. Using an existing 3D extinction map of the inner Galaxy, that provides line of sight specific extinction-distance relationships, we convert the measured extinction of molecular clouds to a corresponding distance. These distances are derived independently from kinematic methods, typically used to derive distances to molecular clouds, and as such they have no near/far ambiguity. The near/far distance ambiguity has been resolved for 27 clouds and distances have been derived to 20 clouds. The results are found to be in good agreement with kinematic measurements to molecular clouds where the ambiguity has already been resolved, using HI self-absorption techniques.  

\end{abstract}
\begin{keywords}

(ISM:) Interstellar Medium (ISM): dust, extinction 

\end{keywords}
\section{Introduction}
\label{sec:intro}

$\indent$Molecular clouds are the birth places of massive stars. They form in an environment shielded from Galactic radiation and so their cold temperatures allow the formation of H$_2$. Star formation is triggered through gravitational instability inside the cloud, when a cloud fragment's gravitational energy exceeds its internal energy. As massive star forming regions are retentive of their natal cloud, until it is dispersed by the formation of HII regions, if the distance is known to the molecular cloud, then the distance is known to the massive star forming region.\\  
$\indent$The Red MSX Source (RMS) survey \citep{urquhart08} has carried out a series of multi-wavelength follow up observations, to identify genuine massive young stellar objects (MYSOs), and ultra compact (UC) HII regions. Initially, a large sample of MYSO candidates were colour-selected from the MSX point source catalogue \citep{egan03} by \citet{lumsden02}. To confirm the presence of a massive young star, we need to determine the luminosity of each source. Thus, we need an accurate measurement of each MYSO candidate's heliocentric distance. \\
$\indent$Every MYSO candidate in the RMS survey has had a kinematic distance determined, from $^{13}$CO observations \citep{urquhart08}, to the molecular cloud from which it was formed. Kinematic distances are derived by applying a Galactic rotational model to measured kinematic velocities. Excluding localised velocity perturbations, the kinematic velocity is the projection of the orbital velocity of a molecular cloud, about the Galactic centre, along the line of sight. Therefore for clouds within the Solar circle, there is not a unique solution to the derived distance and such clouds will possess a near/far kinematic ambiguity. Urquhart et al (in prep.) have used a recent catalogue of molecular clouds \citep{rathborne09} to identify the molecular clouds most likely associated with RMS sources. Using HI self-absorption and also absorption towards 21 cm continuum sources embedded in molecular clouds, \citet{roman09} resolved the near/far ambiguity to $\sim$90$\%$ of all molecular clouds identified by \citet{rathborne09}. Using data from \citet{roman09}, Urquhart et al (in prep.) have resolved the kinematic ambiguity to 186 RMS sources. \\
$\indent$Kinematic distance ambiguities, although shown to be solvable by \citet{roman09}, are not the only potential pitfall when attempting to derive distances to molecular clouds using kinematic techniques. The conversion of kinematic velocities into kinematic distances requires an accurate Galactic rotation model, and the assumption that the measured kinematic velocity is due entirely from rotation about the Galactic centre. Kinematic velocities infer distances of $\sim$4 kpc to stars found in certain portions of the Perseus spiral arm. However there is a discrepancy of a factor of two using luminosity distance estimates, $\sim$2.2 kpc \citep{humphreys78}, and by trigonometric parallax to 4 masers, 1.95$\pm$0.04 \citep{xu06}. Further disagreements have been found to 18 masers by \citet{reid09}, but parallax measurements are still sparse. Clearly there is much to gain from any distance determination method that is independent to Galactic rotation models. \\
$\indent$Molecular clouds are very optically thick. As such, stars behind molecular clouds will suffer much more extinction relative to stars in front of molecular clouds along the same line of sight. Several authors have used extinction measurements to perform a wide variety of tasks. Using cumulative star counts from 2MASS, \citet{froebrich05} have constructed relative extinction maps of the Galactic Plane. \citet{lombardi01} deredden 2MASS data to determine the near infrared colour excess of stars to produce extinction maps. They deredden all stars to a single, average intrinsic colour using the reddening vector of \citet{rieke85}. \citet{sale09} use IPHAS H$\alpha$ data to simultaneously determine extinction, intrinsic colour and distance estimates of early-A to K4 stars. They use these data to map extinction in three dimensions across the northern Galactic Plane. \citet{marshall06} also construct three dimensional extinction maps of the inner Galaxy using 2MASS data and the Stellar Population Synthesis Model of the Galaxy, developed in $\Bes$. They used the synthetic $\Bes$ data to provide the intrinsic colour and probable distances of colour selected giant stars, mapping the inner Galaxy in 15$\arcmin$x15$\arcmin$ tiles. \\
$\indent$Using the most reliable United Kingdom Infrared Deep Sky Survey (UKIDSS) Galactic Plane Survey (GPS) data, this work dereddens individual giant stars along a particular reddening track, to the point of intersection on an intrinsic giant locus. The dereddening process allows us to measure the extinction suffered by stars along the same line of sight as molecular clouds. We then determine distances, that are independent to Galactic rotation models, to molecular clouds using line of sight specific extinction-distance relationships, derived from an existing 3D extinction map of the inner Galaxy. We acquire the 3D extinction map from the previously mentioned \citet{marshall06} data. We utilise the UKIDSS GPS to essentially improve the resolution of the \citet{marshall06} data by analysing only stars that are situated directly along the same line of sight as the molecular cloud. \\
$\indent$In Section \ref{sec:data}, we describe the data used to derive the distance to each cloud, including the photometric data, and the data used to model the line of sight specific extinction-distance relationships. In Section \ref{sec:giants_select} we discuss how we colour-colour and colour-magnitude select a specific sample of stars, of similar spectral type, to determine the line of sight extinction. In Section \ref{sec:cloud_ident} we use the line of sight extinction measurements to identify the spatial extent of molecular clouds. In Section \ref{sec:method} we derive, in detail, the distance to two example molecular clouds. We derive distances to a further 19 molecular clouds and the results are presented in Section \ref{sec:results}. We discuss the results and make comparisons with previous authors in Section \ref{sec:conclusion}.

\section{Data}
\label{sec:data}

\subsection{The UKIDSS Galactic Plane Survey}
The UKIDSS Galactic Plane Survey (GPS, see \citet{lucas08}) covers, in the J(1.248$\mu$m), H(1.631$\mu$m) and K(2.201$\mu$m) filters, the region of the Galactic Plane accessible by the United Kingdom Infrared Telescope (UKIRT); $15^{o} < l < 107^{o}$ and 141$^{o} < l < 230^{o}$, $|$b$|$ $< 5^{o}$ and $-2^{o} <$ l $< 15^{o}$, $|$b$|$ $< 2^{o}$. \\
$\indent$The GPS data are obtained from the WFCAM Science Archive \citep{hambly08}. The median 5$\sigma$ depths are J=19.77, H=19.00 and K=18.05 (Vega system) in the second data release of the GPS \citep{warren07}. The survey depth is spatially variable however, typically decreasing longitudinally towards the Galactic centre and latitudinally towards the Galactic Plane. This occurs as the fields typically become more crowded towards these regions. \citet{lucas08} determine the modal depths in uncrowded fields to be J=19.4 to 19.65, H=18.5 to 18.75 and K=17.75 to 18.0. \\
$\indent$It is possible to use several different types of data quality cuts, provided by the WFCAM Science Archive, to select only the most reliable data. To do this we remove sources with saturated pixels, remove blended objects using ellipticity cuts, astrometry cuts remove sources mismatched between different filters, and finally photometric error cuts. The data cuts used in this paper are almost identical to those presented in \citet{lucas08}. The only alteration we make is a photometric error cut of 0.03 mag in each photometric filter, whereas \citet{lucas08} apply a 0.03 mag colour cut. The use of such data quality cuts means that the photometric depths of the data used in this paper are reduced to J$\sim$18.0, H$\sim$16.8 and K$\sim$16.1.

\subsection{The Stellar Population Synthesis Model of the Galaxy}
\label{sec:modelcdd-Aj}

We utilise the Stellar Population Synthesis Model of the Galaxy, developed in $\Bes$ \citep{robin03}, to model the distribution of giant stars, related to both the distance and extinction, along specific lines of sight. If an appropriate extinction-distance relationship is used to describe the extinction along the line of sight, then as will be shown, comparisons between synthetic and real data can be made to determine the distance to molecular clouds.\\
$\indent$A complete description of the $\Bes$ model inputs can be found in \citet{robin03}, however we summarise it here for completeness. The model contains four populations of stars, thin disc, thick disc, bulge and spheroid. Each of the four populations are described by a star formation rate history, an initial mass function, an age or age range, metallicity characteristics, kinematics, a set of evolutionary tracks, and includes a white dwarf population. The extinction is modelled, in terms of visual magnitudes per kiloparsec (mag kpc$^{-1}$), by a diffuse thin disc. It is also possible to insert discrete clouds with a specified A$_V$ and distance. 

\subsection{Distance-A$_V$ relationships}

\citet{marshall06} use 2MASS data and the $\Bes$ Galactic model to map the Galactic interstellar extinction distribution in three dimensions. This has been done to over 64,000 lines of sight, in the inner Galaxy, each separated by 15$^{\prime}$ (hereafter referred to as the M06 distributions). In this paper we use the M06 data to assess the extinction-distance relationships along specific lines of sight. Using the M06 data, the distribution of synthetic giant population is modelled by using a set extinction model to describe the thin disc (mag kpc$^{-1}$) and the insertion of discrete clouds at specific distances. \\
$\indent$The M06 data produce reliable 3D extinction maps and so provide good line of sight specific extinction-distance relationships. In some cases it is possible to detect, and assign a distance to, large molecular clouds using the M06 data alone. We use the increased sensitivity of the UKIDSS GPS, in comparison to 2MASS, to improve the spatial resolution of the M06 data, identifying only stars that are situated directly along the same line of sight as the target molecular cloud.

\subsection{The Galactic Ring Survey}
\label{sec:GRS}

It is essential to know the position and spatial size of each molecular cloud in order to identify the stars that are situated directly along the same line of sight. \citet{rathborne09} have identified 829 molecular clouds in the $^{13}$CO J=1-0 Galactic Ring Survey \citep[GRS][]{jackson06}, listing the Galactic position and the longitudinal and latitudinal FWHM of each cloud. The GRS covers a range of Galactic longitudes from 18$^o$ $<$ l $<$ 55.7$^o$ and $|b|<$1$^o$. The GRS is fully sampled with a pixel size of 22$\arcmin$$\arcmin$. The processed data cubes have a V$_{LSR}$ range from -5 to 135 km s$^{-1}$ and a spectral resolution of 0.13 km s$^{-1}$ for l $<$ 40$^o$. For larger Galactic longitudes the  V$_{LSR}$ range covers -5 to 85 km s$^{-1}$ and the spectral resolution is 0.21 km s$^{-1}$. \\
$\indent$The GRS data serve two purposes in this paper. For comparison with the extinction maps produced in section \ref{sec:Xmaps}, and to identify the Galactic position and size of several clouds in order to determine their distances. 

\section{Isolating Red Giants}
\label{sec:giants_select}

\begin{figure*}
\begin{center}
    \begin{tabular}{cc}
	\resizebox{70mm}{!}{\includegraphics[angle=0]{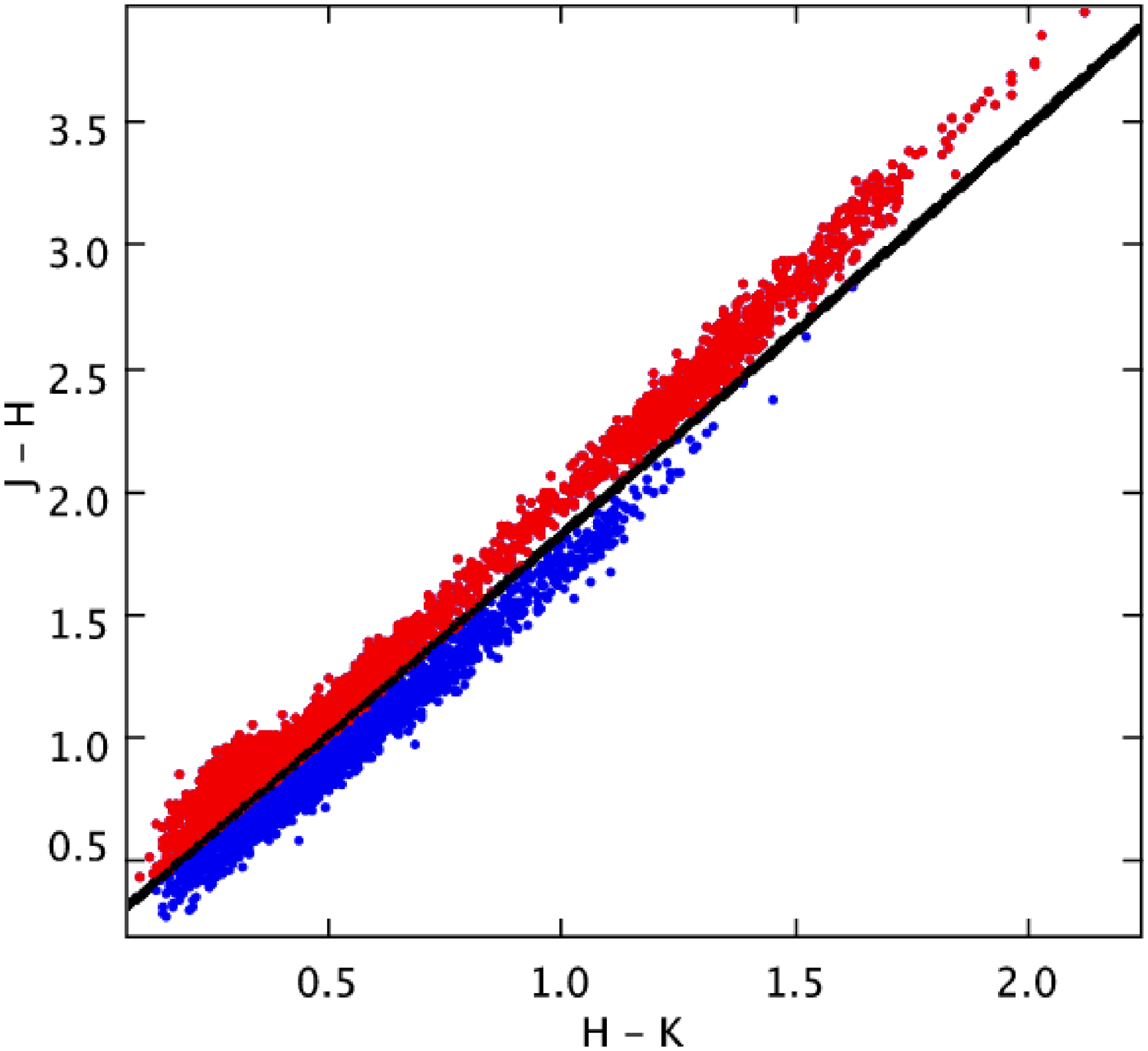}}&
     	 \resizebox{70mm}{!}{\includegraphics[angle=0]{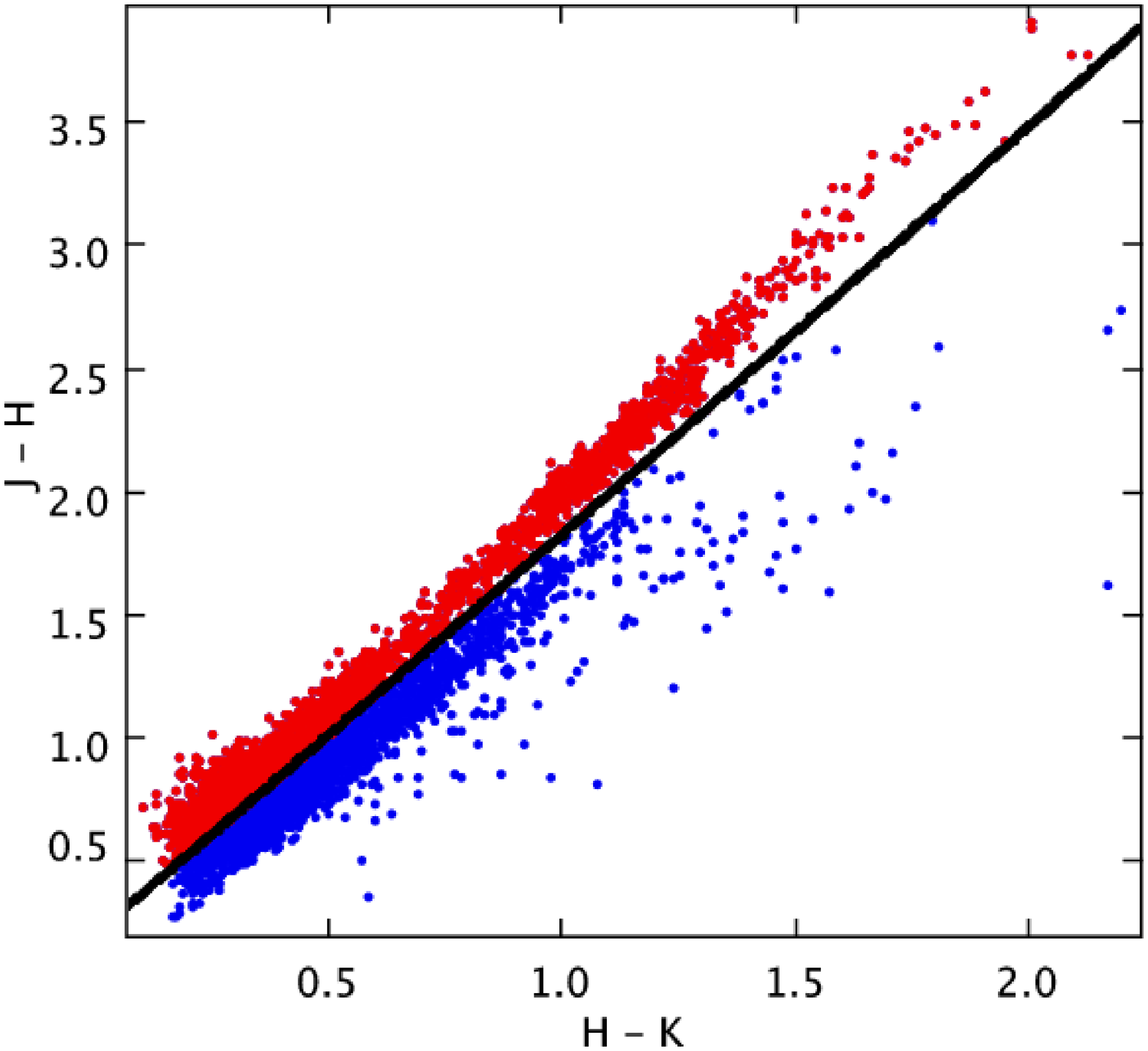}}\\
     	 \resizebox{70mm}{!}{\includegraphics[angle=0]{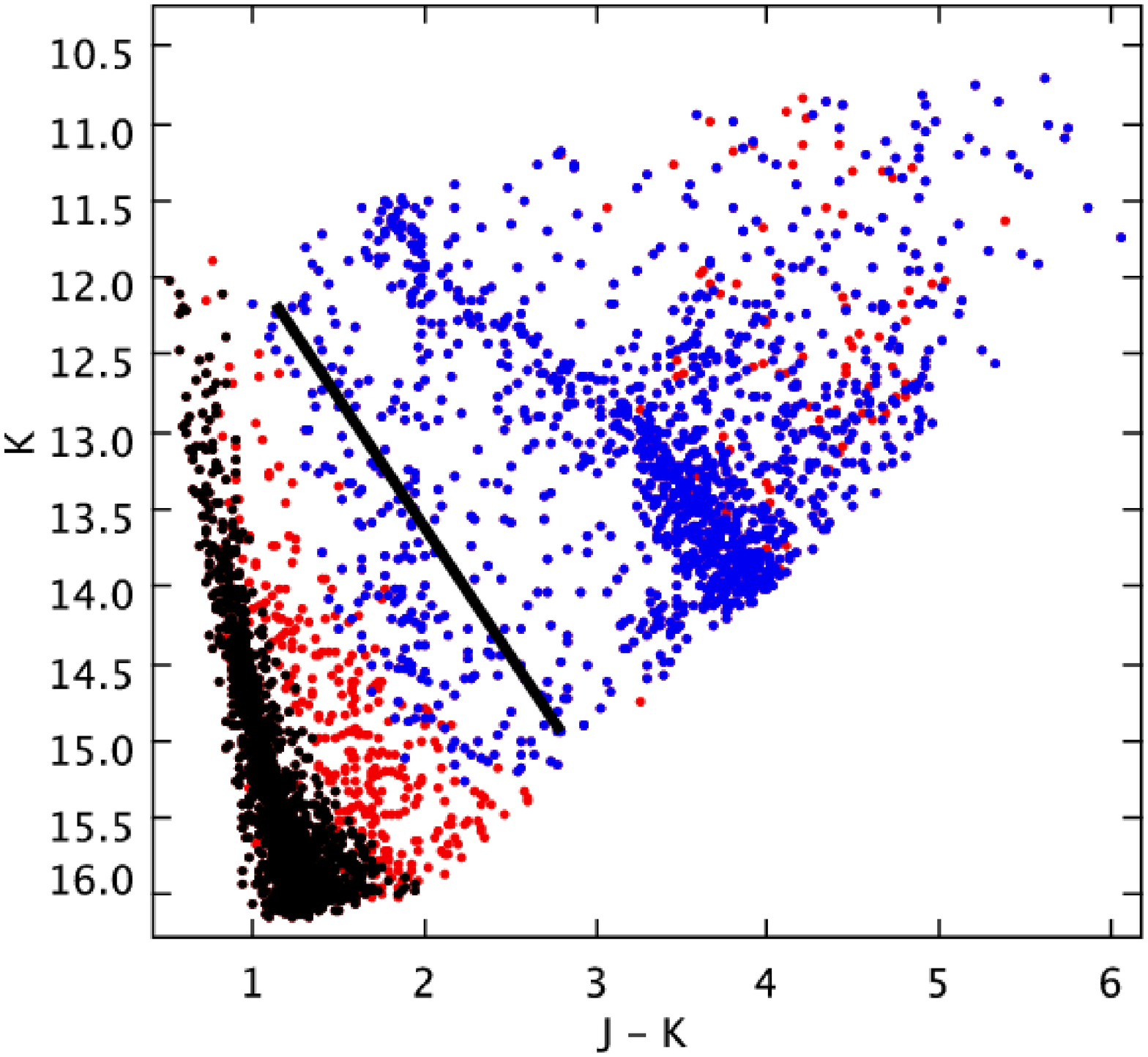}} &
	\resizebox{70mm}{!}{\includegraphics[angle=0]{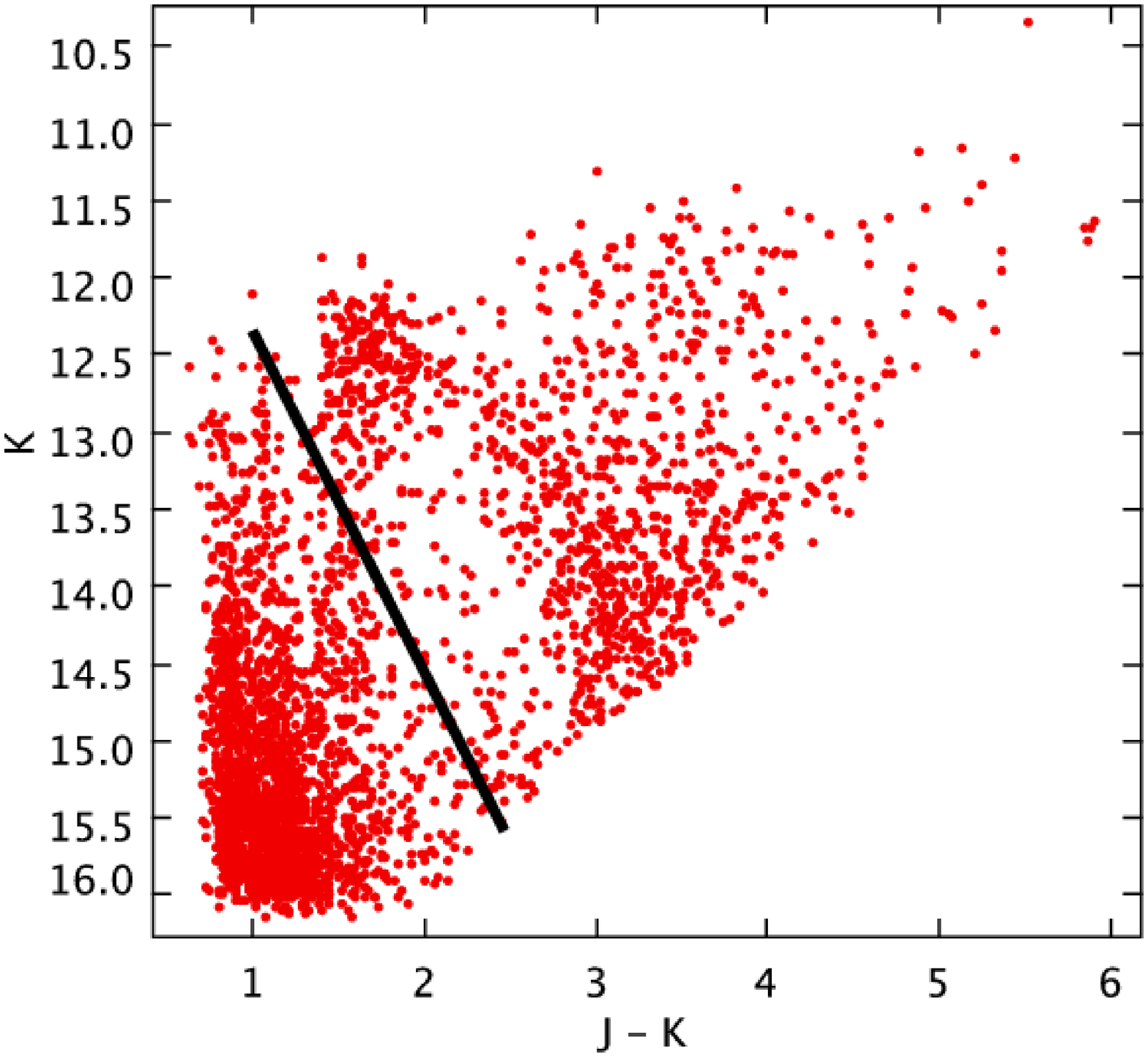}}
    \end{tabular} 
\caption[]{\small a (Top left): A CCD of $\Bes$ data modified as in \citet{stead09} to replicate the UKIDSS data centred on the RMS source G48.9897-00.2992 (see b). Stars that are redder than a G0III (red points) have been separated from the remaining data using a CK04 G0III reddening track (black line). b (Top right): Same as (a) except using UKIDSS data centred on the RMS source G48.9897-00.2992, covering an area of 60$^{\prime}$x6$^{\prime}$. c (Bottom left): A CMD of the remaining $\Bes$ data (red points in (a)). Dwarf and giant stars are displayed as black and red points respectively and form two clumps of data. A black line has been placed approximately equidistant from each clump separating the CMD into two halves. Data to the right of the black line have been selected as the final sample. The remaining red points are subgiants, bright giants and supergiants. d (Bottom right): A CMD of the remaining UKIDSS data (red points in (b)). Like in (c), a black line has split the CMD into two halves, data to the right of the black line have been selected as the final sample.}
\label{fig:CMDCCDXtract}
\end{center}
\end{figure*}

Unlike dwarfs stars, late type giant stars have lived long enough to have rotated around the Galaxy several times and have moved, in reference to the molecular clouds, from their original birth places in the spiral arms \citep{marshall06}. For this reason it can be considered that late type giant stars are ubiquitous throughout the entire Milky Way. As they are infrared bright, numerous, have known colours and are found everywhere, they have been used by many authors to create extinction maps of the Galaxy \citep{lombardi01,marshall06,lombardi09}. \\
$\indent$Using reliable photometry, it is possible to identify giant stars as they occupy known regions on colour-colour and colour-magnitude diagrams (CCDs/CMDs). Fig. \ref{fig:CMDCCDXtract} demonstrates this process with both $\Bes$ and UKIDSS data. The synthetic $\Bes$ data have been extracted from the same size and region of sky as the UKIDSS data, have been given a realistic extinction distribution \citep{marshall06} and have been reddened along UKIDSS reddening tracks \citep{stead09}, and given realistic photometric errors. (The reddening tracks used in this paper can be downloaded at www.ast.leeds.ac.uk/RMS/ReddeningTracks/). For these reasons the synthetic $\Bes$ data can be considered a realistic representation of the real UKIDSS data. \\
$\indent$On a CCD, the most reddened stars (H-K$>$0.8) appear to split into two separate branches of data. As the most reddened stars have to be the most intrinsically bright, these two branches will be primarily composed of late giants and early dwarfs. As late giants are intrinsically redder than early dwarfs, the upper, and therefore redder, branch of data will be composed of giants, the lower will be composed of dwarfs. \\
$\indent$A G0III stellar atmosphere model \citep{castelli04} is progressively reddened, using an extinction law of $\alpha$=2.14 \citep{stead09}, to create a reddening track, hereafter referred to as a CK04 G0III reddening track, that splits the CCD, plotted in Figs. \ref{fig:CMDCCDXtract} (a and b), into two halves. The upper half contains stars redder, and therefore of a later spectral type, than G0III. The CK04 G0III reddening track has been shifted down the y-axis by $\sim$0.04 to include stars that appear intrinsically bluer than a G0III but whose 1$\sigma$ errors would allow them to be considered as G0III or later candidates. \\
$\indent$As the synthetic $\Bes$ data can be considered an accurate representation of the real UKIDSS data, and the class and spectral type of each synthetic star is known, we can estimate the make-up of our final UKIDSS sample. Late giants have very similar colours to late dwarfs of the same spectral type and therefore the colour-colour cut data will be contaminated with a large number of late dwarf stars. As these stars are faint, they can not be observed through large amounts of reddening. In colour-colour space the late dwarfs blend with the late giants in the field. In colour-magnitude space however, there is a clear distinction between giant and dwarf stars. Fig. \ref{fig:CMDCCDXtract} (c) contains the previously selected sample of G0III and later candidates. Highlighted in black and red, in the synthetic data, are the dwarf and giant stars respectively, the remaining blue points consist of subgiants, bright giants and supergiants. The dwarf and giant stars form two clumps of data on the CMD. A black line has been placed, by eye, approximately equidistant from each clump separating the CMD into two halves. Although a small number of giants are positioned in the first half of the CMD, it is safe to assume that the second half of the CMD is free from dwarf stars. We select this second half as our final sample of giants, and repeat this halving of the CMD process with the real data in Fig. \ref{fig:CMDCCDXtract} (d). \\
$\indent$Should this method be applied to regions away from the Galactic Plane, where the line of sight extinction is lower and therefore the distinction between the giant and dwarf stars is less apparent, it may be necessary to statistically separate the two populations. However as there is a clear distinction between populations for all regions studied in this paper, such a process is not required. \\
\begin{table}
\centering
\caption{CMD/CCD III extraction}
\begin{tabular}{ c c c}
\hline
Data cut & $\Bes$ & UKIDSS \\
\hline
All & 6,591 & 6,638 \\
CCD & 3,792 & 3,329 \\
CMD & 1,339 & 1,135 \\
\hline
\end{tabular}
\label{tabz:CMDCDD}
\end{table}
$\indent$Table \ref{tabz:CMDCDD} contains the number of sources that remain after each cut. The $\Bes$ data do not model stars with an infrared excess. Such stars are abundant in star forming regions and are positioned to the right of the bulk of the data on a CCD, as such they are removed during the first data cut. In order to replicate the UKIDSS data, the $\Bes$ data are clipped at the minimum and maximum J, H and K UKIDSS magnitudes. Of the final 1,339 synthetic stars, 1,247, of a possible 1,626 from the total data set, are giants. Of the 1,247 giants in this final subset, 1,235 have a spectral type of G0 or later. Therefore the reddest giants, spectral type G0 or later, make up over 92$\%$ of the final UKIDSS sample.\\
$\indent$As shall be discussed in the following section, we deredden the final UKIDSS subset to an intrinsic giant locus on a CCD, created using CK04 stellar models, in order to derive distances to the molecular clouds they surround. As over half of the 1,247 giants in the final $\Bes$ subset have a spectral type between K0 and K1, we use a K0III reddening track to deredden the UKIDSS data. Using a K0III reddening track, created with the extinction law of \citet{stead09}, we calculate an approximate extinction-colour relationship, A$_V$$\sim$0.074(H-K). For each star a value of A$_J$ is determined directly and converted to a value of A$_V$ using the ratio A$_J$/A$_V$=0.2833, calculated using the \citet{cardelli89} extinction curve with R$_V$=3.1 (obtained from the Trilegal website http://stev.oapd.inaf.it/cgi-bin/trilegal). It is noted in the literature that the value R$_V$=3.1 may not be appropriate for molecular clouds, however for the purpose of this paper it provides only a scaling between A$_K$ and A$_V$ and does not effect the errors of any derived results.

\section{Molecular cloud identification}
\label{sec:cloud_ident}

To illustrate the dereddening process and how the presence of molecular clouds along lines of sight can be detected, we obtain two samples of stars extracted from two different, 15$\arcmin$x15$\arcmin$ areas of the sky. The first area has been centred on the RMS source G048.9897-0.2992 and therefore has a molecular cloud along the line of sight. The second area is centred on the Galactic coordinates G048.9897-0.7000. The area was visually selected from a region in the GRS that showed no obvious signs of $^{13}$CO, and therefore the line of sight extinction is likely to be free from dense molecular clouds.

\subsection{The dereddening method}

The stars along each line of sight have been colour-colour and colour-magnitude selected, as described in the previous section, to obtain a sample of stars primarily composed of red giants. We use a CK04 K0III reddening track to deredden each star in the sample created using the average extinction power law of \citet{stead09}. However to account for the error created by dereddening stars of differing spectral types along a K0III reddening track, instead of reddening tracks specific to the spectral type of each star, we increase the error in the average extinction power law to $\alpha$=2.14$^{+0.08}_{-0.08}$. Each star is dereddened from its position on a CCD, along the K0III reddening track, to the point of intersection on the giant locus. As the K0III reddening track has been created by convolving the progressively redder spectra of a CK04 K0III stellar model through the UKIDSS filters, the length of the reddening track traversed directly relates to the amount extinction each star suffers. The photometric errors relate to the error in the point of intersection with the giant locus, and therefore relate to the error in the determined extinction.
\begin{figure*}
  \begin{center}
    \begin{tabular}{cc}
      \resizebox{80mm}{!}{\includegraphics[angle=90]{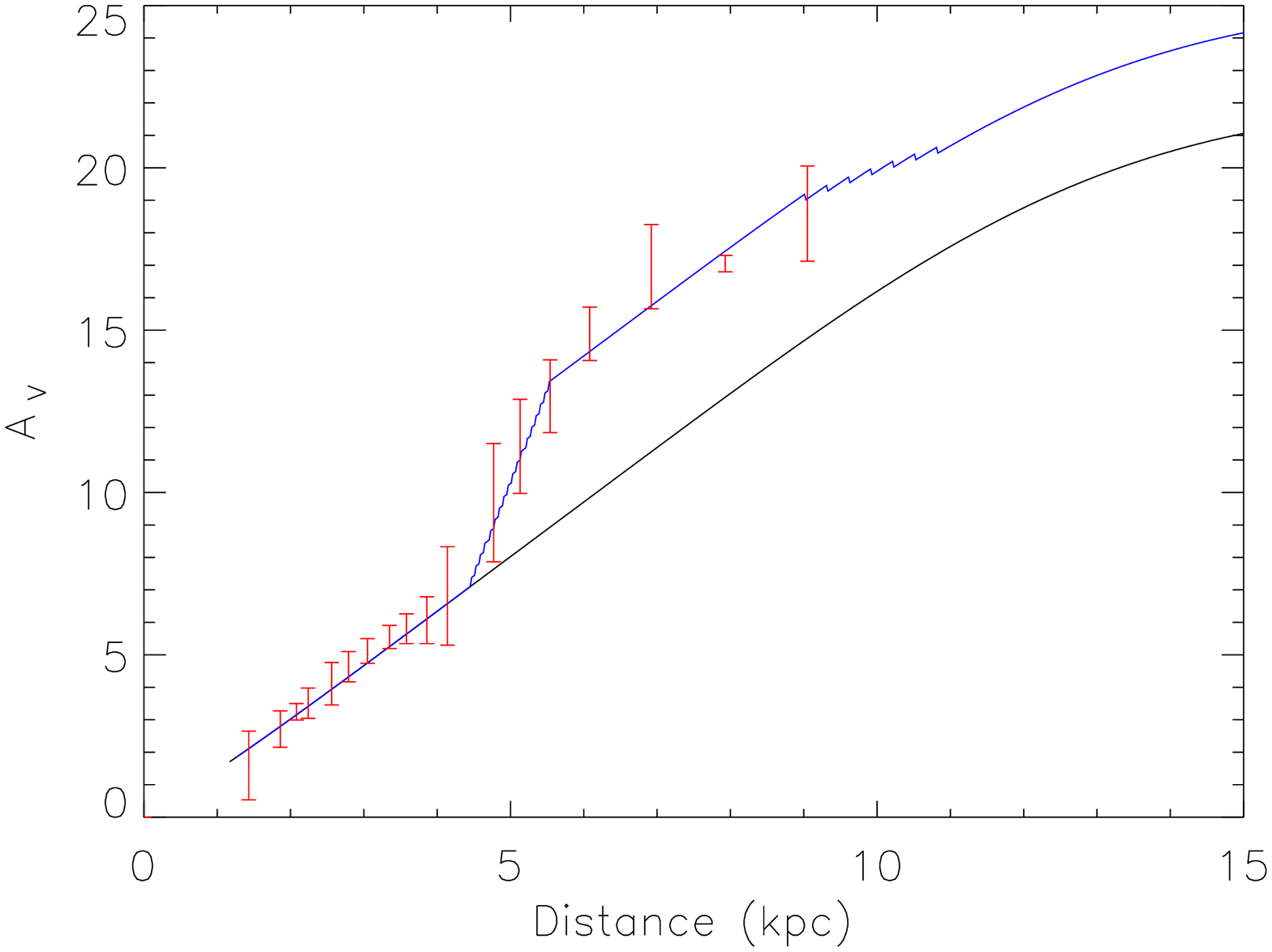}} &
     \resizebox{80mm}{!}{\includegraphics[angle=90]{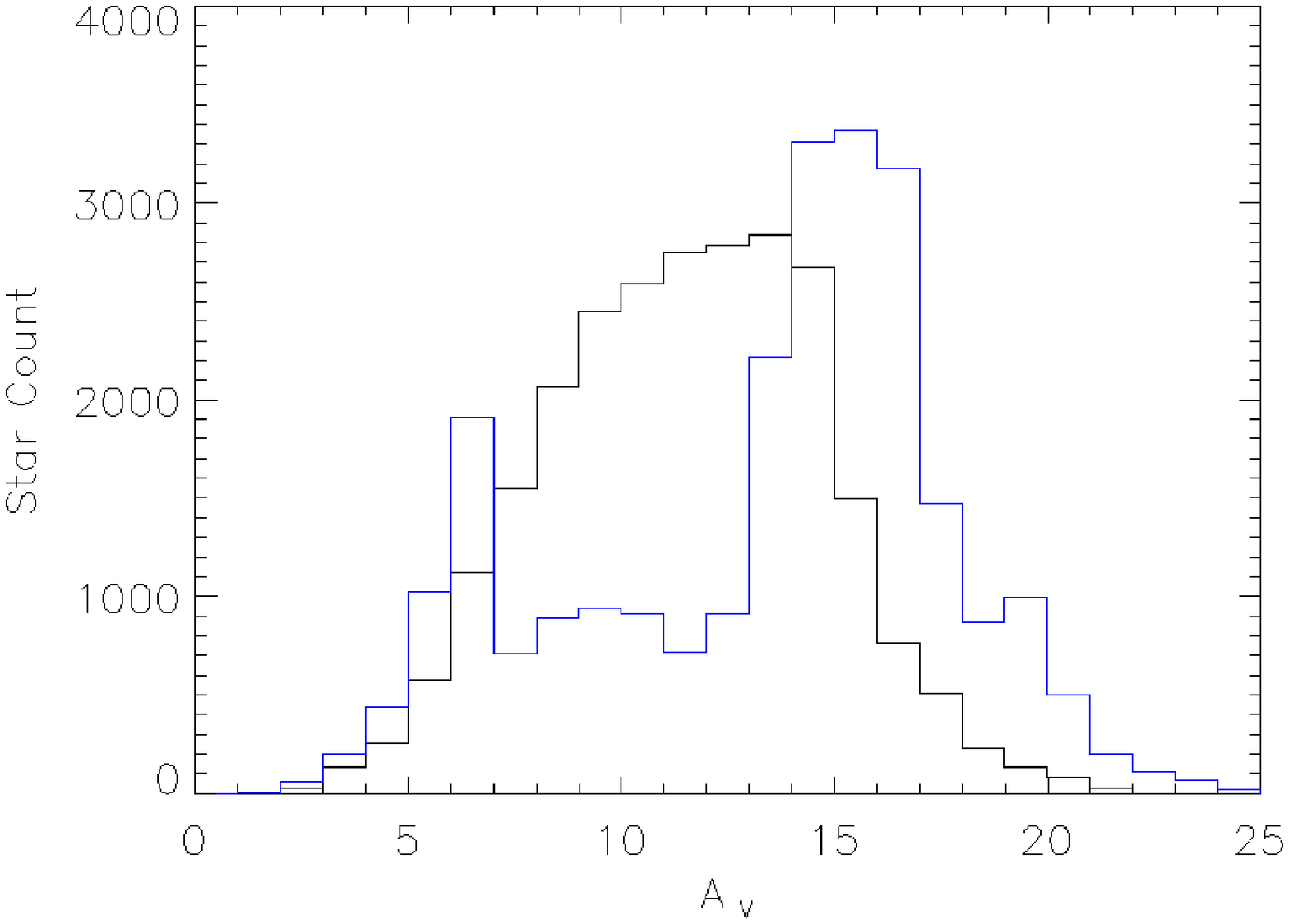}}
    \end{tabular}
    \caption{\small a (Left): A distance-A$_V$ plot containing synthetic data showing the extinction distribution model (blue line) created to reproduce the M06 data (red error bars) centred spatially closest to the RMS source G048.9897-0.2992. The black line represents the smooth extinction distribution model of 1.4mag kpc$^{-1}$. b (Right): A synthetic A$_V$ histogram of each extinction distribution presented in the distance-A$_V$ plot (Left panel). The blue data have a drop in star counts between A$_V$$\sim$7 and $\sim$13, corresponding to the sudden increase in extinction beginning at $\sim$4.5kpc in the distance-A$_V$ plot.}
    \label{fig:Av_dist}
  \end{center}
\end{figure*}

\subsection{Distance-A$_V$ plots and A$_V$ histograms}

A distance-A$_V$ plot for the region G48.9897-0.2992, using the $\sim$6,000 synthetic stars from Fig. \ref{fig:CMDCCDXtract} (a), is presented in Fig. \ref{fig:Av_dist} (a). It contains an extinction distribution model (EDM) created to follow the M06 data, extracted from the 15$\arcmin$x15$\arcmin$ tile that is centred spatially closest to the RMS source. At $\sim$5 kpc the EDM possesses a sharp rise in A$_V$. The position of this rise is consistent with the presence of a molecular cloud at the RMS kinematic distance of 5.2 kpc. \\
$\indent$Stars can also be binned into small groups representing small increments of extinction to construct A$_V$ histograms. When plotting an A$_V$ histogram of the synthetic data centred on the region G48.9897-0.2992, the sharp rise in extinction in Fig. \ref{fig:Av_dist} (a), at $\sim$5 kpc, represents a fall in the star count over this particular range of A$_V$. In comparison, the same synthetic data, reddened using a smooth extinction distribution model of 1.4 mag kpc$^{-1}$, produces a steady rise and fall in star counts of the over a similar range in the A$_V$ histogram (Fig. \ref{fig:Av_dist} (b)). If a distance-A$_V$ plot could be constructed that contained precise distance and A$_V$ measurements for each star in the field of view then, assuming no stars are situated inside the molecular cloud, the molecular cloud would be represented by a vertical rise in the extinction distribution at the exact distance to the molecular cloud. The corresponding A$_V$ histogram would contain a sharp fall and rise in star counts, with a gap equivalent to the total A$_V$ of the molecular cloud. \\
\begin{figure*}
  \begin{center}
    \begin{tabular}{cc}
      \resizebox{80mm}{!}{\includegraphics[angle=90]{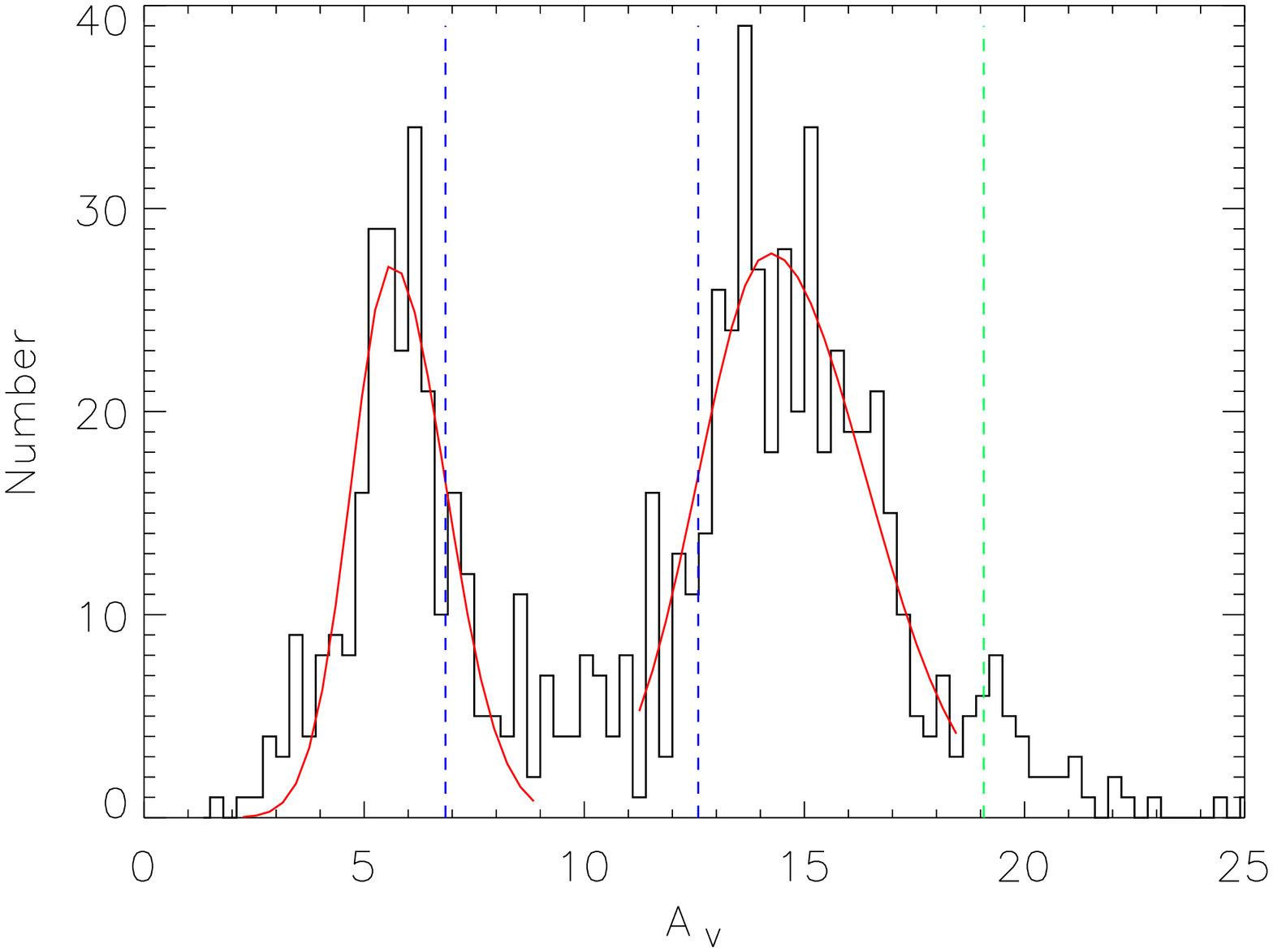}} &
     \resizebox{80mm}{!}{\includegraphics[angle=90]{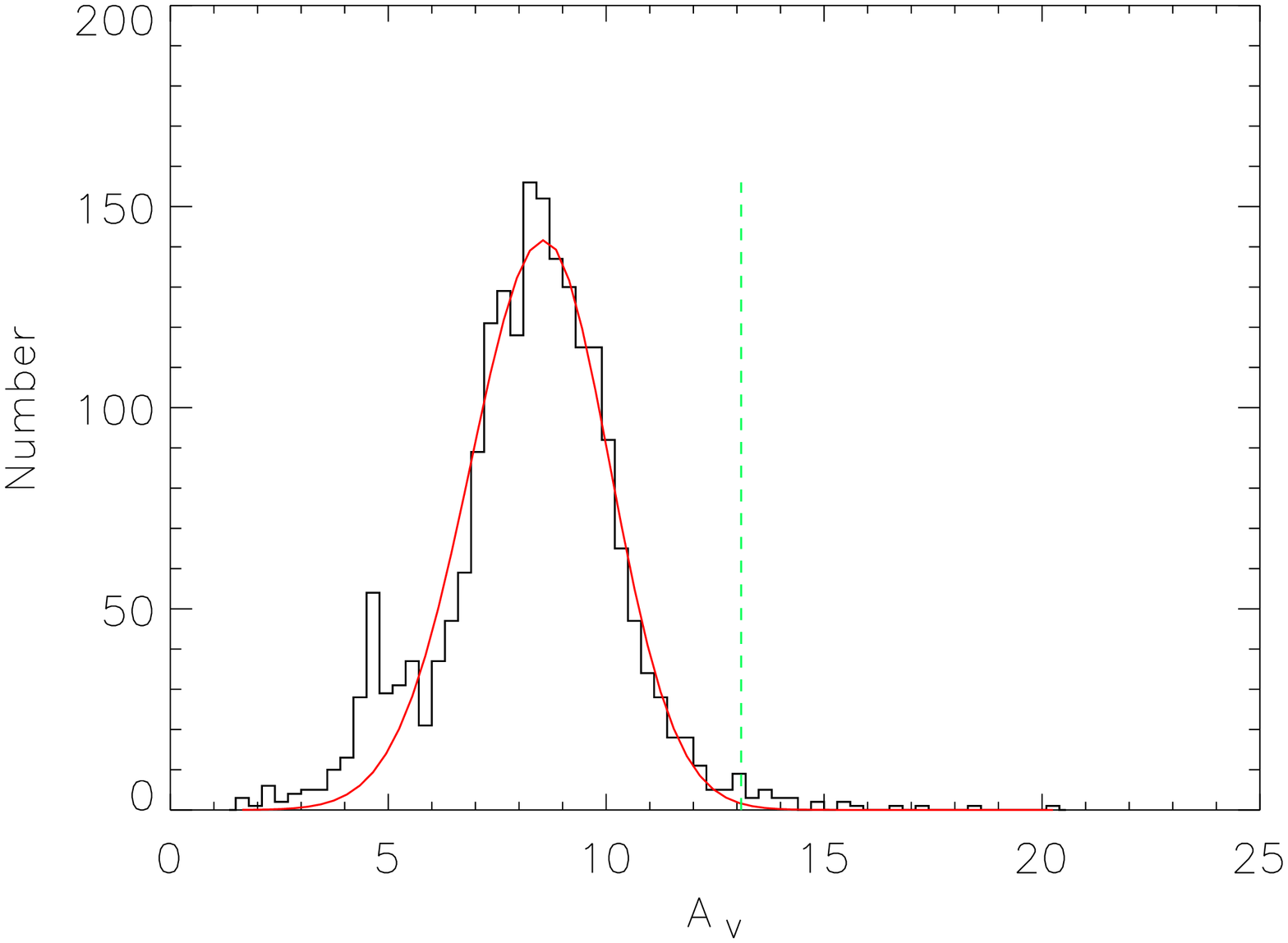}}
    \end{tabular}
    \caption{\small a (Left): An A$_V$ histogram of real stars extracted from a 15$\arcmin$x15$\arcmin$ region centred on the RMS source G048.9897-0.2992. We have defined the total line of sight extinction to the far side of the molecular cloud, as the 1$\sigma$ width (dashed blue line) on the left side of the skewed Gaussian (red line). (Right): An A$_V$ histogram of a 15$\arcmin$x15$\arcmin$ region centred on an unremarkable region of space (see text) at G048.9897-0.7000. The histogram shows no presence of a molecular cloud. We define the far line of sight extinction (see text) as the 3$\sigma$ deviation of all extinction measurements (dashed green line) of the fitted skewed Gaussian (red line).}
    \label{fig:two_fields}
  \end{center}
\end{figure*}
$\indent$Using only photometric data, it is possible to determine the amount of extinction a particular star suffers if accurate intrinsic colours are known. The amount of extinction suffered by a star is proportionate to how far the star moves, along a reddening track, on a colour-colour diagram (CCD), and can therefore be determined by dereddening individual stars along reddening tracks to their corresponding intrinsic colours. In this case the colour selected giants are dereddened to a giant locus. The giant locus has been generated using UKIDSS filter profiles and the CK04 stellar models \citep{stead09}. If we can accurately determine the amount of extinction suffered by every star along a particular line of sight, then by creating A$_V$ histograms the presence of a molecular cloud can be detected by the drop in star counts.

\subsection{Extinction maps}
\label{sec:Xmaps}

\begin{figure*}
  \begin{center}
    \begin{tabular}{c}
      \resizebox{145mm}{!}{\includegraphics[angle=90]{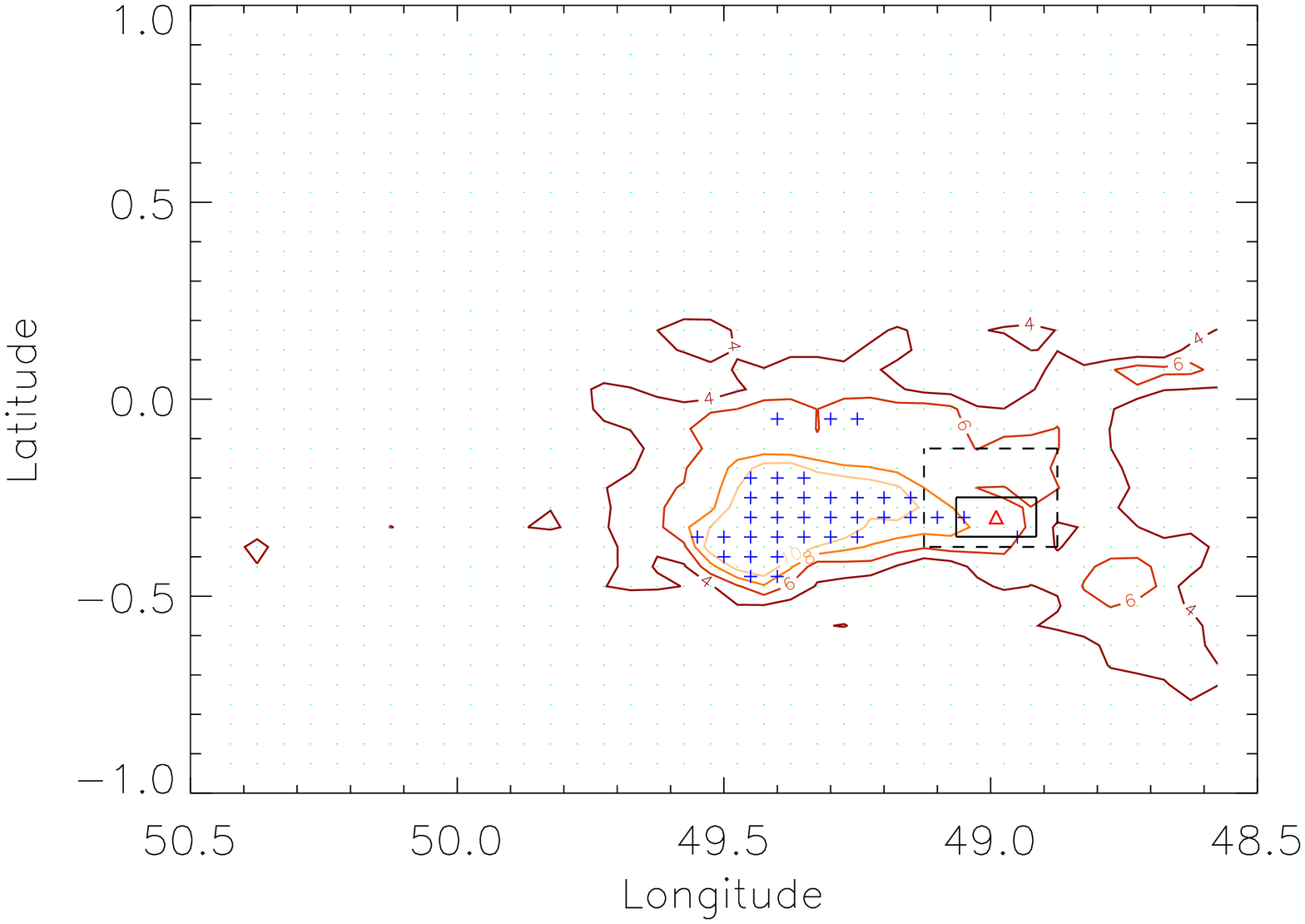}} \\
     \resizebox{145mm}{!}{\includegraphics[angle=90]{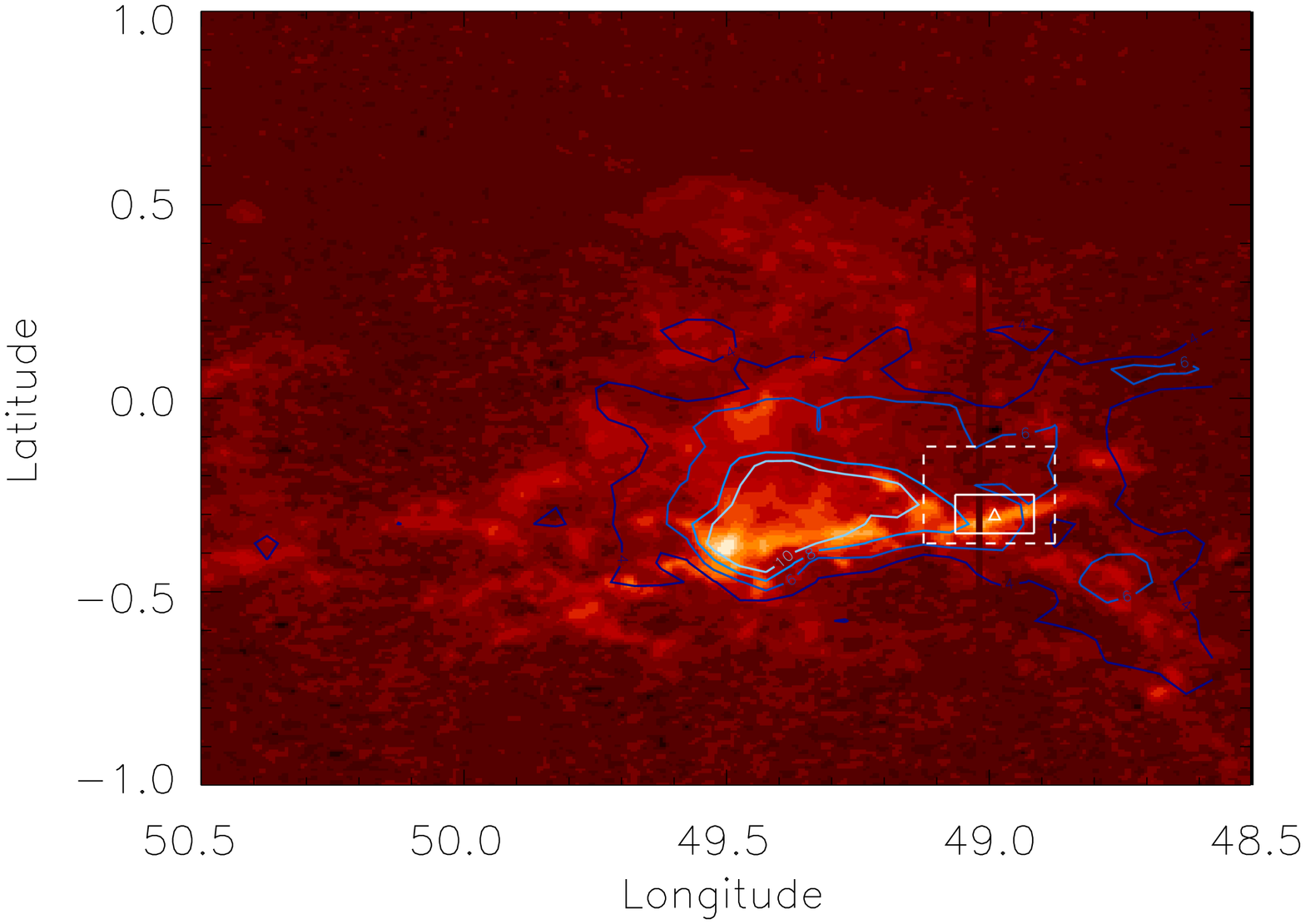}}
    \end{tabular}
    \caption{\small a (Top): A MCX map from the 2$^o$x2$^o$ region centred at G49.5+0.0. The 4, 6, 8 and 10 A$_V$ levels are plotted in various shades of red. The RMS source is plotted as a red triangle. Blue \textcolor{blue}{+} symbols represent regions where the extinction was too high to accurately determine the A$_V$ of detected clouds. The region of sky used to determine the M06 data is plotted as a dashed rectangle. A 9$\arcmin$x6$\arcmin$ region surrounding the RMS source is plotted as a solid rectangle.  b (Bottom): The same contour lines from (a) have been overlaid with a GRS integrated intensity image from the same region, created using a velocity range from 60 to 75 km s$^{-1}$.}
    \label{fig:Xmap_MCX}
  \end{center}
\end{figure*}

Fig. \ref{fig:two_fields} (a and b) contain A$_V$ histograms of real stars extracted from the two previously mentioned regions G048.9897-0.2992, centred on an RMS source, and G048.9897-0.7000, centred on a $^{13}$CO featureless region of space. As the A$_V$ histogram that is centred on an RMS source will contain a molecular cloud along the line of sight, like the blue histogram in Fig. \ref{fig:Av_dist} (b), there are two distinct peaks in the histogram, separated by a large dip in the source counts. The second region, that does not contain a molecular cloud, produces an A$_V$ histogram with a steady rise and fall in star counts, and has a similar morphology to the black histogram in Fig. \ref{fig:Av_dist} (b). As noted, the presence of a molecular cloud in the former histogram creates two distinct peaks in the histogram. The far edge of the first peak therefore represents the beginning of the molecular cloud. In the reverse manner, the near edge of the second peak therefore represents the end of the molecular cloud. To estimate the width of the gap, and therefore determine the A$_V$ of the molecular cloud, two skewed Gaussians are fit to each peak. The respective 1$\sigma$ deviations are used to define the edge of each peak, and therefore the difference between them is defined as the A$_V$ of the molecular cloud. In a similar fashion the far line of sight extinction is defined as the 3$\sigma$ deviation to the far side of the second peak. In Fig. \ref{fig:Av_dist} (b), as there is no molecular cloud and therefore no second peak, the 3$\sigma$ deviation is measured from a single skewed Gaussian. \\
$\indent$In this paper we have defined the far line of sight extinction as the 3$\sigma$ deviation of all extinction measurements. This should be distinguished from the total line of sight extinction, defined as the amount of extinction suffered along the line of sight towards the edge of the thin disk. Differences between the far and total line of sight extinction measurements can occur for two reasons. First, there may be a large amount of diffuse extinction behind the last detected cloud that would not fit well to a skewed Gaussian distribution. More importantly however, it could be that along the line of sight of the previously mentioned region G048.9897-0.7000, used to construct Fig. \ref{fig:two_fields} (b), there could be a molecular cloud with an optical depth large enough to prevent stars at the far side of the cloud being detected. Therefore the measured far line of sight extinction, that is dependent upon the depth of the photometry used, would be very different to the total line of sight extinction. However, the good agreement between real data and synthetic data (see section \ref{sec:avgap}, Fig. \ref{fig:Marshall_Avhist}) suggests that the far and total line of sight extinction are often the same. \\
$\indent$It is possible to repeat both of these processes in small increments for large fields to produce two different types of extinction map, far line of sight extinction (FLSX) maps and molecular cloud extinction (MCX) maps. The latter can be created by plotting A$_V$ histograms and identifying which regions show the presence of molecular clouds. The A$_V$ of each cloud can be measured and used to construct an extinction map, showing both the location and size of any clouds in the field of view. In comparison \citet{lombardi09}, by measuring the colour excess of background stars, evaluate the extinction at any location in the sky using a weighted average of extinction measurements to stars that are angularly close to a given location on the map.\\
$\indent$Using the above processes, both types of extinction maps have been created for a 2$^o$x2$^o$ region centred at G49.5+0.0, containing the RMS source G048.9897-0.2992 (The region was not centred closer to the RMS source due to incomplete UKIDSS data at l$<$48.5). Using the deep UKIDSS data, we are able to produce A$_V$ histograms for 6$\arcmin$x6$\arcmin$ tiles that are therefore over 6 times the resolution of the 15$\arcmin$x15$\arcmin$ M06 distributions. For each histogram we measure the far line of sight extinction, and depending whether or not the histogram contains the presence of a molecular cloud, the A$_V$ of the cloud is determined. Some histograms show the possible presence of more than one molecular cloud along the line of sight. In such cases we determine the far line of sight extinction to the far side of the most prominent molecular cloud. We analyse one 6$\arcmin$x6$\arcmin$ region of sky and then microstep by shifting the field of view by 3$\arcmin$, repeating the process for the entire 2$^o$x2$^o$ region. The large region is then again split into 6$\arcmin$x6$\arcmin$ tiles, however the extinction associated to each tile is taken from an average of the four surrounding microstepped 6$\arcmin$x6$\arcmin$ tiles. This process does not improve the resolution of the final extinction map, however it does smooth the extinction map, averaging out any anomalous A$_V$ measurements. \\
\begin{figure*}
  \begin{center}
    \begin{tabular}{c}
      \resizebox{145mm}{!}{\includegraphics[angle=90]{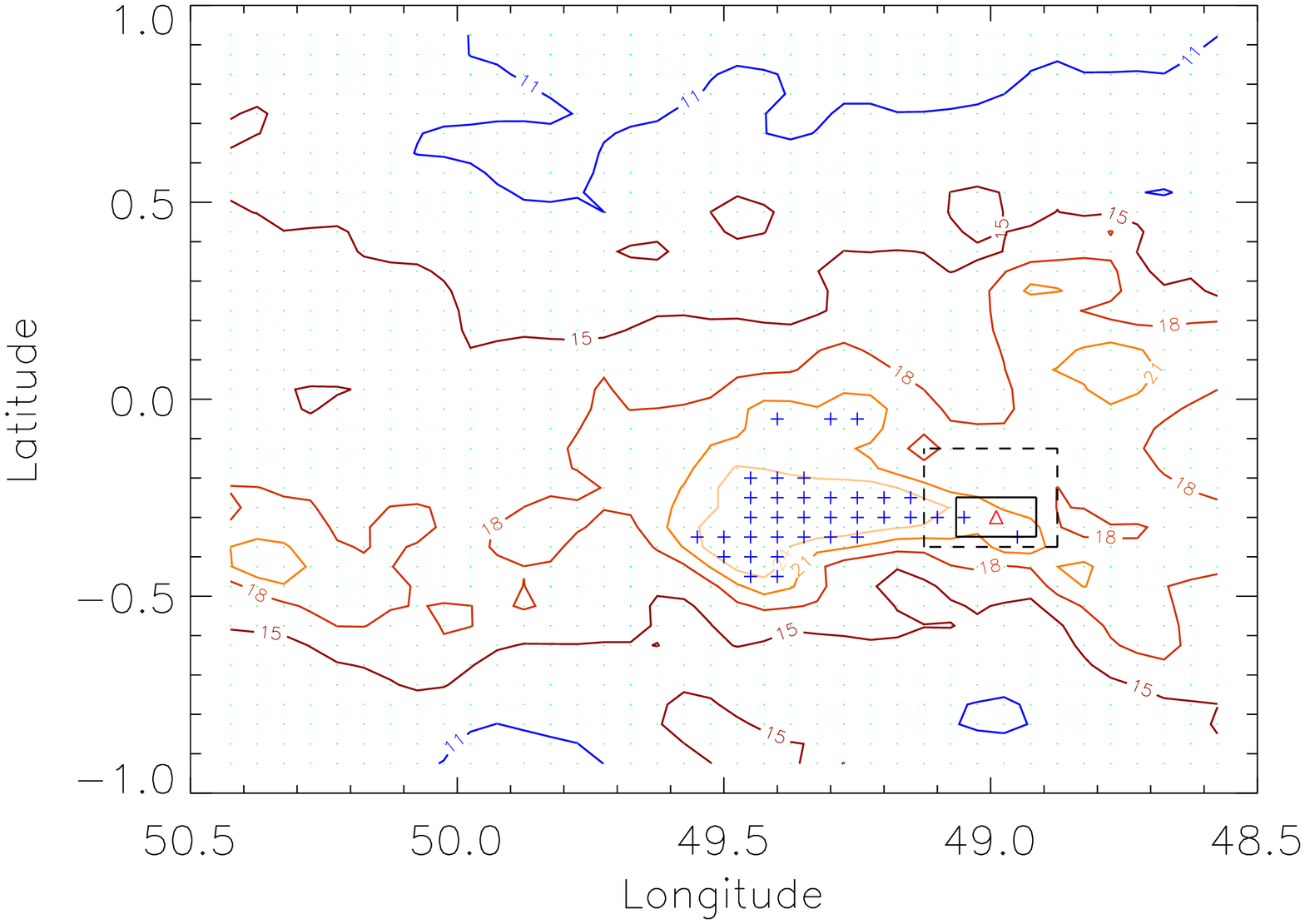}} \\
     \resizebox{145mm}{!}{\includegraphics[angle=90]{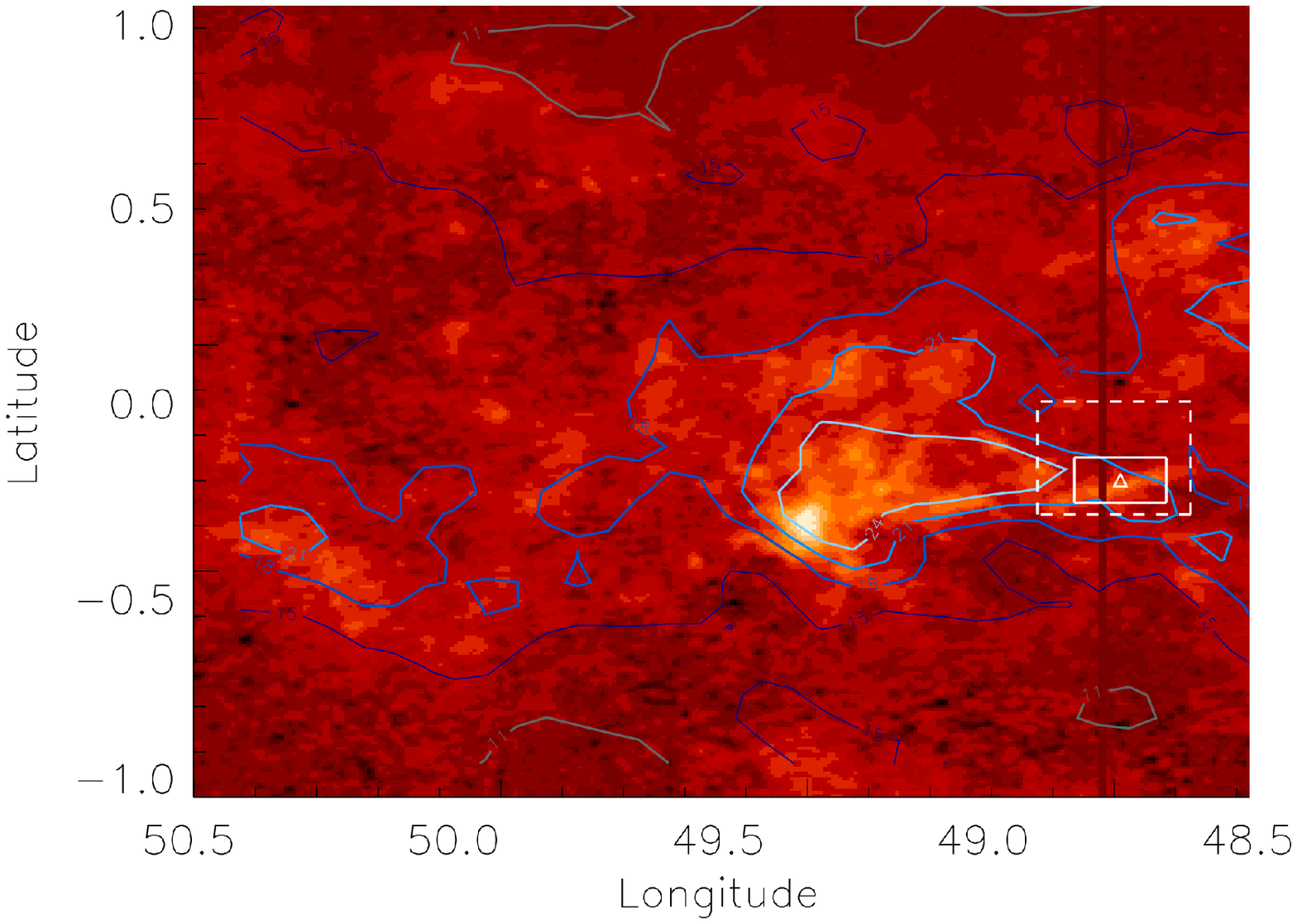}}
    \end{tabular}
    \caption{\small a (Top): A FLSX map from the 2$^o$x2$^o$ region centred at G49.5+0.0. The 15, 18, 21 and 24 A$_V$ levels are plotted in various shades of red. Blue contour lines represent the 11 A$_V$ level. The RMS source is plotted as a red triangle. Blue \textcolor{blue}{+} symbols represent regions where the extinction was too high to accurately determine the far line of sight extinction. The region of sky used to determine the M06 data is plotted as a dashed rectangle. A 9$\arcmin$x6$\arcmin$ region surrounding the RMS source is plotted as a solid rectangle.  b (Bottom): The same contour lines from (a) have been overlaid with a GRS integrated intensity image from the same region, created using the full velocity range available, from -5 to 85 km s$^{-1}$.}
    \label{fig:Xmap_FLSX}
  \end{center}
\end{figure*}
$\indent$Fig. \ref{fig:Xmap_FLSX} a contains a FLSX map from the 2$^o$x2$^o$ region centred at G49.5+0.0. Contour lines show the 11, 15, 18, 21 and 24 A$_V$ levels. The RMS source G048.9897-0.2992 (red triangle) is contained within the 21 A$_V$ level. Fig. \ref{fig:Xmap_FLSX} (b) contains the same extinction plot overlaid with a GRS integrated intensity image, from the same region, in order to draw a comparison between the positions of molecular clouds and the regions of high extinction traced out by the contours. The GRS image has been created using the full velocity range available from -5 to 85 km s$^{-1}$. The contour lines have mapped out the bulk of the extinction shown in the GRS data, and from this we can estimate the extent of the molecular cloud that contains the RMS source. There are two other regions surrounded by A$_V$=21 contour lines, at $\sim$G48.8+0.1 and $\sim$G50.4-0.4, in both cases there do appear to be other large groups of molecular clouds at these positions. \\
$\indent$Fig. \ref{fig:Xmap_MCX} a contains a MCX map from the 2$^o$x2$^o$ region centred at G49.5+0.0. Contour lines show the 4, 6, 8 and 10 A$_V$ levels of the measured depth of the molecular clouds. The RMS source G048.9897-0.7000 (red triangle) is contained within the 8 A$_V$ level. Fig. \ref{fig:Xmap_MCX} (b) contains the same extinction plot overlaid with the same GRS integrated intensity image, plotted in Fig. \ref{fig:Xmap_FLSX} (b), however only the velocity range 60 to 75 km s$^{-1}$ has been considered. This range corresponds to the approximate 3$\sigma$ width of the measured kinematic velocity of the molecular cloud associated with the RMS source \citep{urquhart08}. Like in Fig. \ref{fig:Xmap_FLSX} (b) the contour lines have mapped out the bulk of the extinction shown in the GRS data, and as before we can estimate the extent of the molecular cloud that contains the RMS source. Both types of extinction maps show very similar features, however in the MCX map, large areas are featureless due to the absence of molecular clouds. In all plots a 9$\arcmin$x6$\arcmin$ region surrounding the RMS source has been overlaid. Stars within this region have been selected to further analyse in the following section.

\section{The distance determination of molecular clouds}
\label{sec:method}

\subsection{The distance-A$_V$ relationship}

\begin{figure}
  \begin{center}
    \begin{tabular}{cc}
      \resizebox{80mm}{!}{\includegraphics[angle=90]{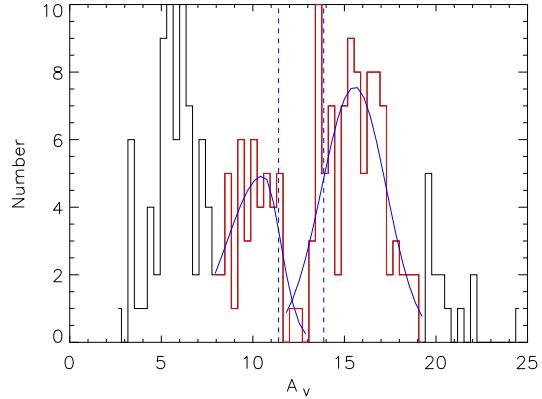}}
    \end{tabular}
    \caption{\small An A$_V$ histogram of real stars from a 9$\arcmin$x6$\arcmin$ region centred on the RMS source G048.9897-0.2992. The near and far sides of the molecular clouds are marked with dashed blue lines, derived from the 1$\sigma$ deviations of the Gaussians (solid blue lines) fitted to the red bins.}
    \label{fig:G48.9897_AvHist}
  \end{center}
\end{figure}

Fig. \ref{fig:G48.9897_AvHist} contains an A$_V$ histogram of real stars extracted from around the previously mentioned 9$\arcmin$x6$\arcmin$ region surrounding the RMS source G048.9897-0.2992. The morphology of the A$_V$ histogram is similar to the synthetic A$_V$ histogram, created using the EDM relating to the M06 data, presented in Fig. \ref{fig:Av_dist} (b - blue histogram). As both histograms are similar it indicates that the M06 data are an accurate representation of the extinction distribution along this particular line of sight. However there is an important feature worth noting, at A$_V$$\sim$12 there is a drop in source counts creating a gap in the histogram, occurring due to the presence of the molecular cloud along the line of sight. If we estimate the range of A$_V$ along the histogram that the cloud covers, and we have an accurate line of sight specific extinction-distance relationship, then it is possible to determine the distance to the cloud. The distance and the error in the distance to the cloud are derived by first measuring the A$_V$ values representing the beginning and end of the molecular cloud. These A$_V$ measurements are then converted to distances using the extinction-distance relationships provided by the M06 data. The distance to the cloud is determined as the average of these two measurements, and the error, the difference between them. The M06 error bars in A$_V$ are also considered and from this, we estimate the distance to the RMS source to be D =  5.5$\pm$0.4 kpc. This value is consistent with the RMS kinematic distance of 5.2$\pm$0.2 kpc. \\
$\indent$To calculate the certainty of detection of a molecular cloud in an A$_V$ histogram above the noise, a goodness of fit statistic is calculated as the square root of the summed squared residuals, presented in the following equation: 
\begin{equation}
  \beta^{2} = \frac{\sum_{i=1}^{n}(O_{i} - E_{i})^{2}}{n},
  \label{eq:no1}
\end{equation}
where O$_i$ is the actual star count and E$_i$ is the expected star count, and n is the number of bins in the histogram. The goodness of fit statistic is calculated for both Gaussians, fitted to each peak in the histogram. A single Gaussian is then fitted to the same, combined range of data across both previous peaks. A larger goodness of fit statistic is determined when fitting one Gaussian, thereby suggesting a poorer fit, than when using of two Gaussians.

\subsection{Measuring the A$_V$ gap}
\label{sec:avgap}

\begin{figure}
  \begin{center}
    \begin{tabular}{cc}
      \resizebox{80mm}{!}{\includegraphics[angle=90]{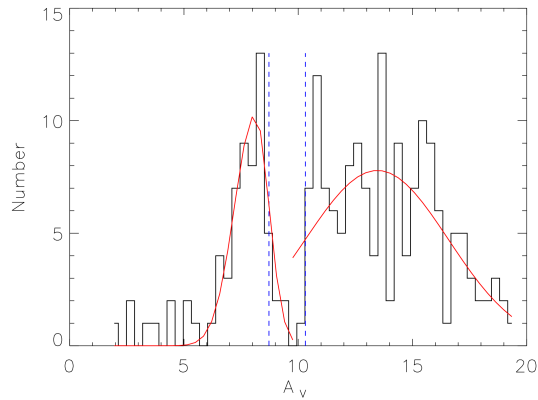}}
    \end{tabular}
    \caption{\small An A$_V$ histogram of synthetic stars has been used to test the reliability of using two Gaussians to measure the A$_V$ of a molecular cloud (see text). Each skewed Gaussian is shown as a red line and the corresponding 1$\sigma$ widths are shown as blue dashed lines. }
    \label{fig:synthetic_Avhist}
  \end{center}
\end{figure}

\begin{figure}
  \begin{center}
    \begin{tabular}{cc}
      \resizebox{80mm}{!}{\includegraphics[angle=90]{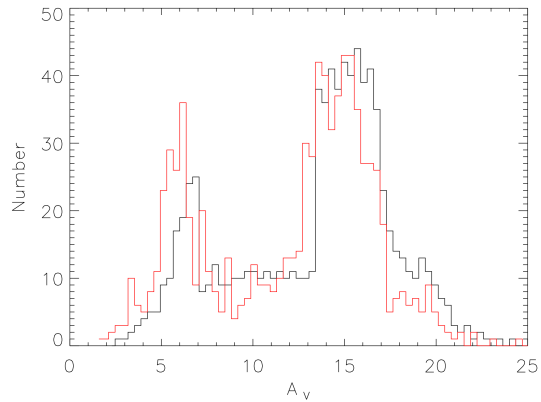}}
    \end{tabular}
    \caption{\small An A$_V$ histogram of real stars (red histogram) taken from the same region of sky that was used to create the Marshall data (see text). The synthetic histogram, containing the same number of sources as the real histogram, is presented in black.}
    \label{fig:Marshall_Avhist}
  \end{center}
\end{figure}

In order to test the reliability of using two Gaussians to measure the A$_V$ of a molecular cloud on an A$_V$ histogram, we have measured the dip in source counts in 2500 different synthetic A$_V$ histograms. Each synthetic histogram has been created using a smooth EDM (EDM1) of 1.5 mag kpc$^{-1}$ and contains a cloud with A$_V$=1.5 at 5.2 kpc. To model the effect of cloud patchiness, we also combine a smooth EDM (EDM2) of 1.5 mag kpc$^{-1}$ without a synthetic cloud along the line of sight. The ratio of EDM2 to EDM 1 is allowed to vary between 0\% and 40\%. Larger ratios than 40\% make it difficult to detect the synthetic cloud above the noise in the histogram. Realistic errors are given to the synthetic data, and from these 2500 measurements, we compute the gap to be A$_V$=1.6$\pm$0.5, suggesting that method used can reproduce the initial value of extinction used to create the synthetic cloud. An example synthetic histogram, produced using 20\% of data from the cloudless EDM (EDM2), is presented in Fig. \ref{fig:synthetic_Avhist}. \\
$\indent$Including this additional error in A$_V$ yields a distance to the RMS source G048.9897-0.2992 of D = 5.6$\pm$0.5 kpc. The additional error in the gap measurement of $\pm$0.5 mag has had little effect on the resultant distance and error. This is because the extinction-distance relationship towards this particular region (at $\sim$5 kpc), provided by the M06 data, is visually estimated from Fig. \ref{fig:Av_dist} (a) to be $\sim$5.5 mag kpc$^{-1}$. Therefore large increases in the error in A$_V$ will only translate into small errors in the derived distance. Towards regions where the extinction-distance relationship is shallower, the additional error in the gap measurement will translate into larger errors in the derived distance. All subsequent distances derived in this paper include the error in the extinction-distance relationship, and the $\pm$0.5 mag error in the gap measurement detailed above. \\
$\indent$Although Fig. \ref{fig:G48.9897_AvHist} does contain an expected sharp A$_V$ gap at A$_V$$\sim$12, the A$_V$ histogram begins to show a reduction in star counts at A$_V$$\sim$8, continuing until A$_V$$\sim$12. It is possible, due to the low number statistics, that this larger range is entirely caused by the molecular cloud and should be the measured dip. The distance assigned to this dip is D =  5.1$\pm$0.7 kpc. This value is still consistent with both the RMS kinematic distance and the distance previously derived. The error is smaller than might have been expected when measured from such a large range in A$_V$, this is due to the steep slope on the EDM reducing the range of possible distances. However, we do not believe that the molecular cloud is responsible for this broader A$_V$ gap. There are several molecular clouds, identified in the Galactic Ring Survey, within the vicinity of the molecular cloud that contains the RMS source. The majority of these clouds are at a similar kinematic distance to G048.9897-0.2992 and so may be responsible for the broadening of the A$_V$ gap. As the M06 data are not centred on the RMS source, the molecular cloud containing the RMS source is unlikely to be entirely responsible for the extinction measurements, along the line of sight, used to create the M06 data. \\
$\indent$The particular M06 data set used to model the extinction along the line of sight extinction towards the RMS source G048.9897-0.2992, is centred on the coordinates G049.00-00.25 and obtained from a 15$\arcmin$x15$\arcmin$ region. We have dereddened sources extracted from the same 15$\arcmin$x15$\arcmin$ region as the M06 data and compared, in Fig. \ref{fig:Marshall_Avhist}, a real A$_V$ histogram with a synthetic A$_V$ histogram. The synthetic histogram has been created by taking a random sample of $\Bes$ data equal to the number of real stars. Both histograms show a similar morphology, however the real histogram no longer has a second deep dip at A$_V$$\sim$12. It is only when we extract stars from a smaller region centred exactly on the molecular cloud that this drop appears. This suggests that the A$_V$ dip in Fig. \ref{fig:G48.9897_AvHist} is entirely due to the molecular cloud centred on the RMS source, and as such, the distance to the RMS source is, D = 5.6$\pm$0.5 kpc.\\
$\indent$The synthetic A$_V$ histogram in Fig. \ref{fig:Marshall_Avhist} was constructed with synthetic stars as distant as 15 kpc. As the real A$_V$ histogram also covers a similar range of A$_V$ as the synthetic A$_V$ histogram, it can be considered that for this particular line of sight, the far line of sight extinction and the total line of sight extinction, previously discussed in section \ref{sec:Xmaps}, are the same.

\subsection{RMS-GRS cloud associations}
\label{sec:RMSGRS}

As previously mentioned, Urquhart et al (in prep.) have identified the molecular clouds from \citet{rathborne09} that are associated with 292 RMS sources. As \citet{rathborne09} have listed the Galactic position and the longitudinal and latitudinal FWHM of each cloud, this information can be used in the same manner as discussed to derive distances, independent to kinematic methods, to RMS sources. The FWHM measurement of each cloud can be used to ensure we are observing stars directly in front and behind each cloud analysed. \\
$\indent$From the 292 RMS-GRS associations identified by Urqhart et al (in prep.), 173 have available J, H and K UKIDSS GPS data. To detect the presence of a molecular cloud in an A$_V$ histogram, enough foreground giant stars must be observed to notice the reduction in star counts per A$_V$ bin that signals the presence of a molecular cloud. Therefore there will be a lower limit to the derivable distances to molecular clouds. If 3 star counts per A$_V$ bin are needed to detect the presence of the foreground giants, in the synthetic histogram in Fig. \ref{fig:Marshall_Avhist} the first A$_V$ bin containing this number of sources occurs when A$_V$$\sim$3.5. Using the distance-A$_V$ plot in Fig. \ref{fig:Av_dist}, created with the M06 data, converts this A$_V$ value into a distance of $\sim$2 kpc. As Fig. \ref{fig:Av_dist} contains a very typical extinction distribution for the lines of sight considered in this paper, we therefore consider this a lower limit to the distances we can derive using this method and exclude clouds that have a near distance lower than 2 kpc. Furthermore, we exclude clouds within 27$^{\circ}$ of the Galactic Centre to avoid the long thin Galactic bar, since this is not included in the $\Bes$ model data. As the RMS and GRS kinematic distances have been derived using different Galactic rotation models, there are certain sources that do not possess an RMS kinematic ambiguity but are flagged as having so in the GRS catalogue. These clouds have also been excluded reducing our sample size to 74. It is not possible to assign an upper limit to the derivable distances, as the cut-off will be dependent upon the total extinction suffered and therefore photometric depth. Clouds at higher Galactic latitudes may suffer less extinction than those at lower latitudes and it will therefore be possible to 'see further' along these lines of sight. We have assigned an initial upper limit of 9 kpc, and due to this upper limit, we have 34 clouds in our final sample that have a near distance larger than 2 kpc and a far distance smaller than 9 kpc. \\
$\indent$When using the GRS FWHM measurement to identify stars along the same line of sight as a molecular cloud, the area observed is very important. A large star count is needed to detect the presence of the cloud in an A$_V$ histogram, however, assuming each cloud to possess a spherical shape, stars along the line of sight of the edge of the cloud will suffer less extinction than those along the line of sight of the centre. Furthermore, the larger the area the greater the chance of contamination from overlapping clouds. To avoid excessive noise in the resultant A$_V$ histogram we need to limit the size of the area observed, yet keep it large enough to detect enough stars. The size of each region observed is done on a cloud-by-cloud basis and we typically observe between a 0.6 and 1.0 $\sigma$ width of the cloud. Finally, an EDM is created using the M06 data set that is spatially closest to the centre of the molecular cloud.

\subsection{Solving the kinematic distance ambiguity}
\label{sec:soveambig}

\begin{figure*}
  \begin{center}
    \begin{tabular}{cc}
      \resizebox{80mm}{!}{\includegraphics[angle=90]{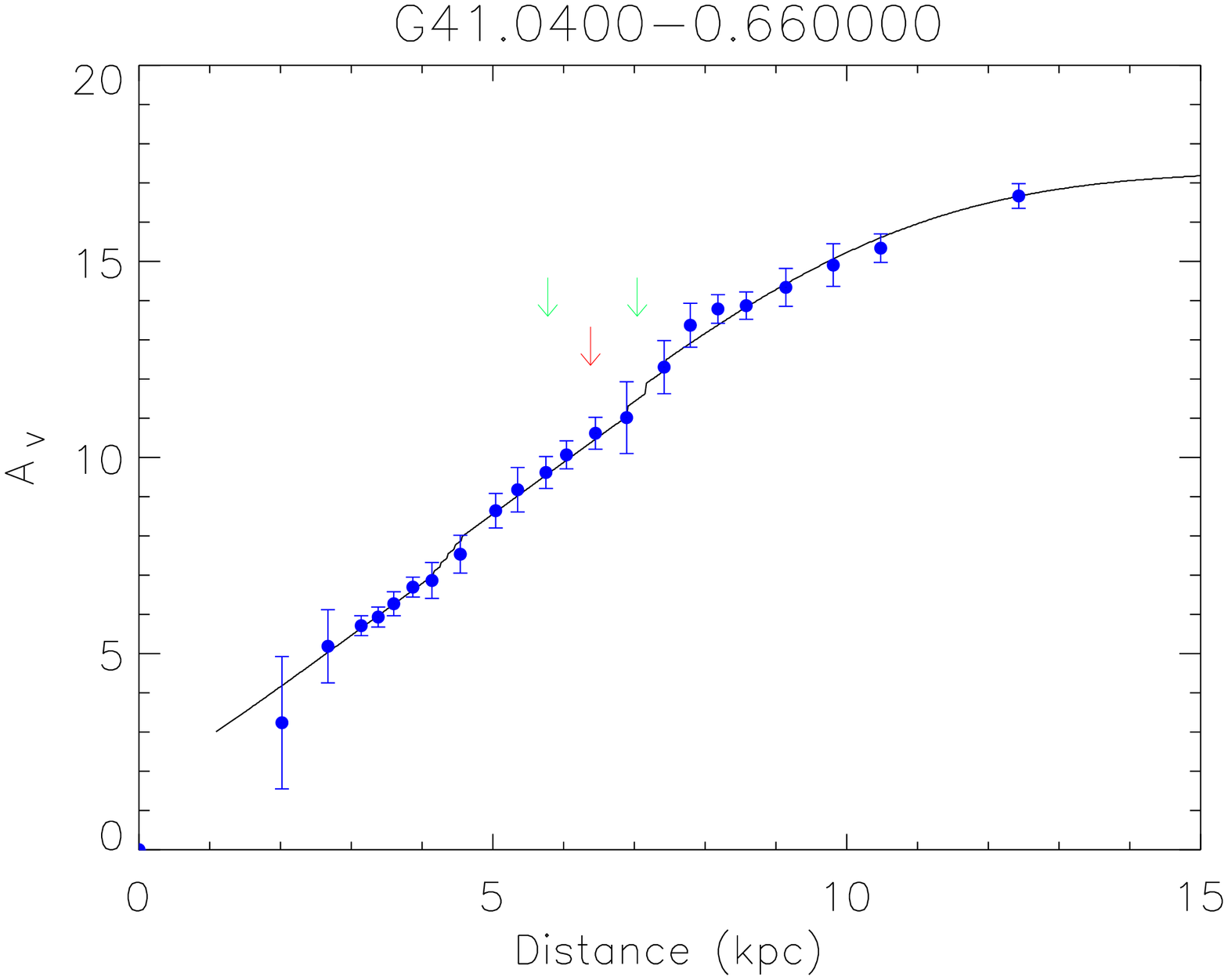}} &
      \resizebox{80mm}{!}{\includegraphics[angle=90]{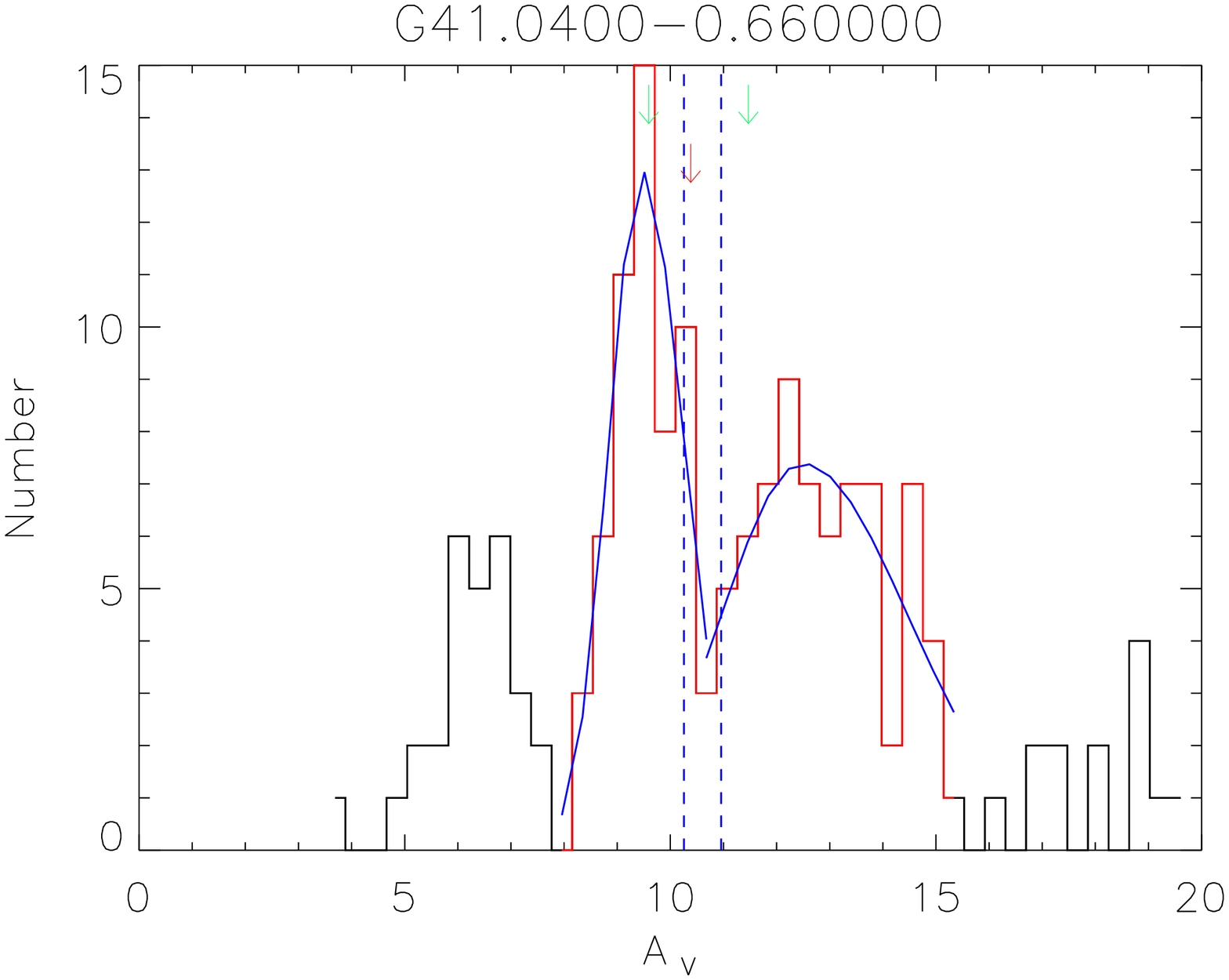}}
    \end{tabular}
    \caption{\small a (Left): A distance-A$_V$ plot containing the extinction distribution model (black line) created to reproduce the M06 data (blue error bars) along the line of sight of the GRS molecular cloud G041.04-0.66. The GRS and two RMS kinematic distances are plotted as red and green arrows respectively. b (Right): An A$_V$ histogram of real stars from a small region centred on the GRS molecular cloud G041.04-0.66 contains a dip in star counts at A$_V$$\sim$11. Two Gaussians have been fitted to each peak separated by the dip. The 1$\sigma$ widths have been used to determine the distance to the molecular cloud.}
    \label{fig:G41.04_Av_dist}
  \end{center}
\end{figure*}

The previous region surrounding the RMS source G048.9897-0.2992 is particularly simple to analyse for two reasons. First, from inspection of the M06 data, there appears to be only one, very sharp rise in extinction. For this reason the distance to the cloud, using Fig. \ref{fig:Av_dist} (a), can be visually estimated to be 5.0$\pm$0.5 kpc. Second, the RMS source has a kinematic velocity located at a tangent point in the Galaxy and therefore has no near/far distance ambiguity. \\
$\indent$As a second demonstration we have extracted data along the line of sight of the molecular cloud GRSMC G041.04-0.66. The cloud has been associated with the RMS source G041.0780-00.6365 Urqhart et al (in prep.) and has two RMS kinematic distances of 5.8$\pm$1.1 and 7.0$\pm$1.1 kpc. Fig. \ref{fig:G41.04_Av_dist} (a) contains a distance vs A$_V$ plot of the extinction distribution model used to reproduce the M06 data assigned to the cloud. Unlike Fig. \ref{fig:Av_dist} (a), Fig. \ref{fig:G41.04_Av_dist} (a) does not contain a sudden jump in extinction at the position of any of the kinematic distances. Therefore the resolution of the M06 data is too low to distinguish between the near and far RMS kinematic distances. However, the M06 data still provide a reliable description of how the extinction relates to distance along the line of sight.\\
$\indent$Fig. \ref{fig:G41.04_Av_dist} (b) contains an A$_V$ histogram of the dereddened giants extracted from a 1$\sigma$ region of the molecular cloud. The histogram has two clear dips in star counts at A$_V$$\sim$8 and A$_V$$\sim$11 that each separate two distinct peaks. We use the extinction distribution to plot the kinematic distances on the A$_V$ histogram. The RMS and GRS kinematic distances are plotted as green and red arrows respectively. We first measure the dip that appears to be associated with the kinematic distances, and determine the distance to be D = 6.9$\pm$0.5. Goodness of fit statistical tests have been applied to each Gaussian which have suggested that the dip is a real molecular cloud detection. This distance is consistent with the GRS kinematic distance D = 6.38 kpc and the far RMS distance, D = 7.0 kpc. However as the near RMS distance is D = 5.8 kpc, the GRS distance is close to the average of the two RMS distances. Differences between the RMS and GRS kinematic distances occur due to the choice of the Galactic rotation model used.  \\
$\indent$As previously mentioned, there is a second, equally well defined dip in the star count in Fig. \ref{fig:G41.04_Av_dist} (b), signalling the presence of a second molecular cloud. This second dip appears at A$_V$$\sim$8 and is situated close to the near RMS kinematic distance. The distance to this second cloud has been determined to be D = 4.8$\pm$0.7 kpc. Although this value disagrees with the near RMS kinematic distance at the 1$\sigma$ level, there still remains some level of ambiguity to the derived distance to this molecular cloud. However, upon spatial examination of the entire GRS catalogue, it appears the region of sky previously studied also overlaps with a second molecular cloud G041.04-0.26, with a GRS distance D = 4.7 kpc. As the RMS source is spatially nearer the GRS cloud G041.04-0.66, the assigned distance to the RMS source is D = 6.9$\pm$0.5 kpc. \\
$\indent$We have repeated the method on the remaining 33 molecular clouds. All UKIDSS data towards each molecular cloud have been colour-colour and colour-magnitude cut to select a majority population of late giants. Goodness of fit tests have been performed on each and the results are detailed in the following section.

\section{Results and Discussion}
\label{sec:results}

\begin{figure*}
\begin{center}
    \begin{tabular}{cc}
\resizebox{80mm}{!}{\includegraphics[angle=90]{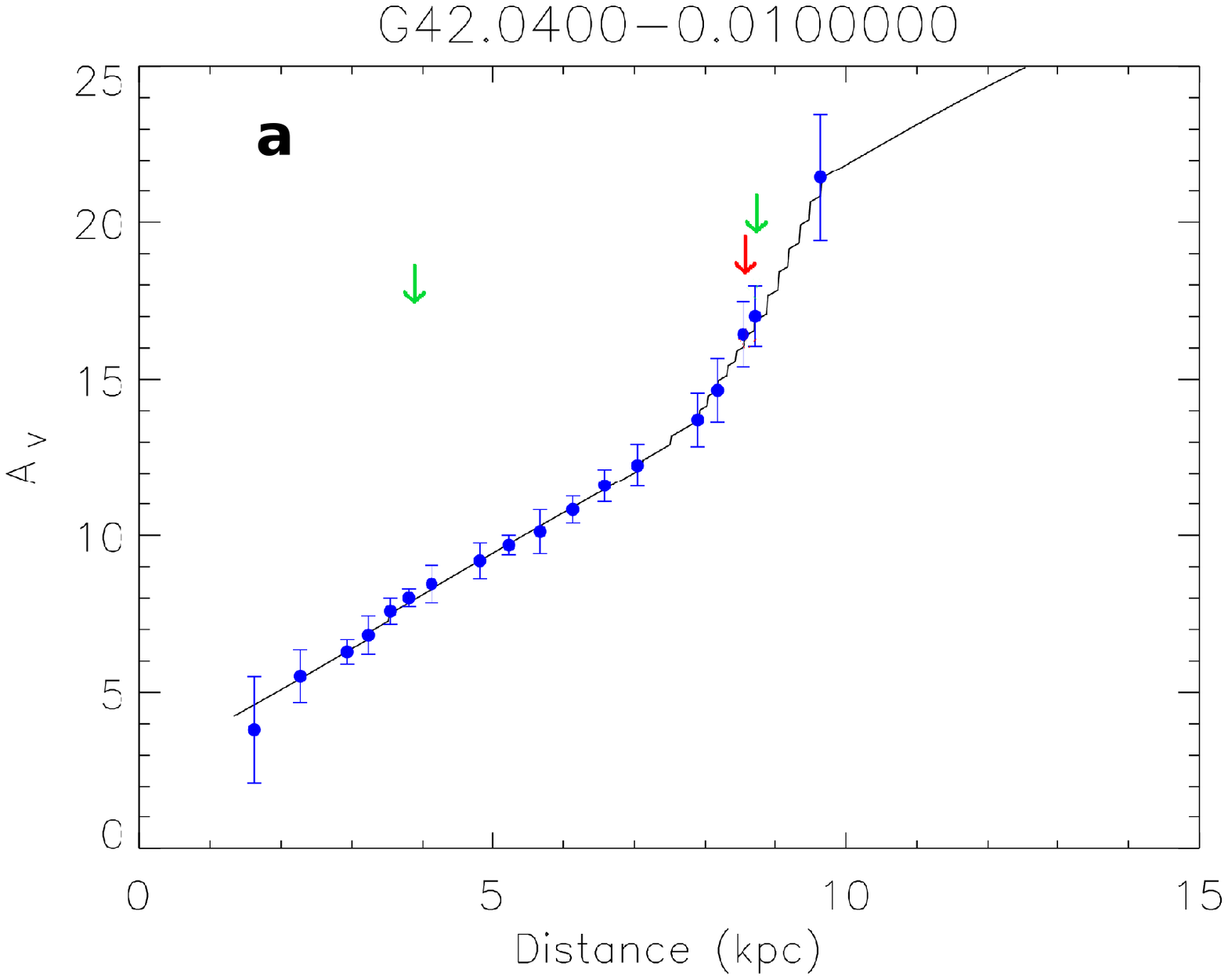}}&
\resizebox{80mm}{!}{\includegraphics[angle=90]{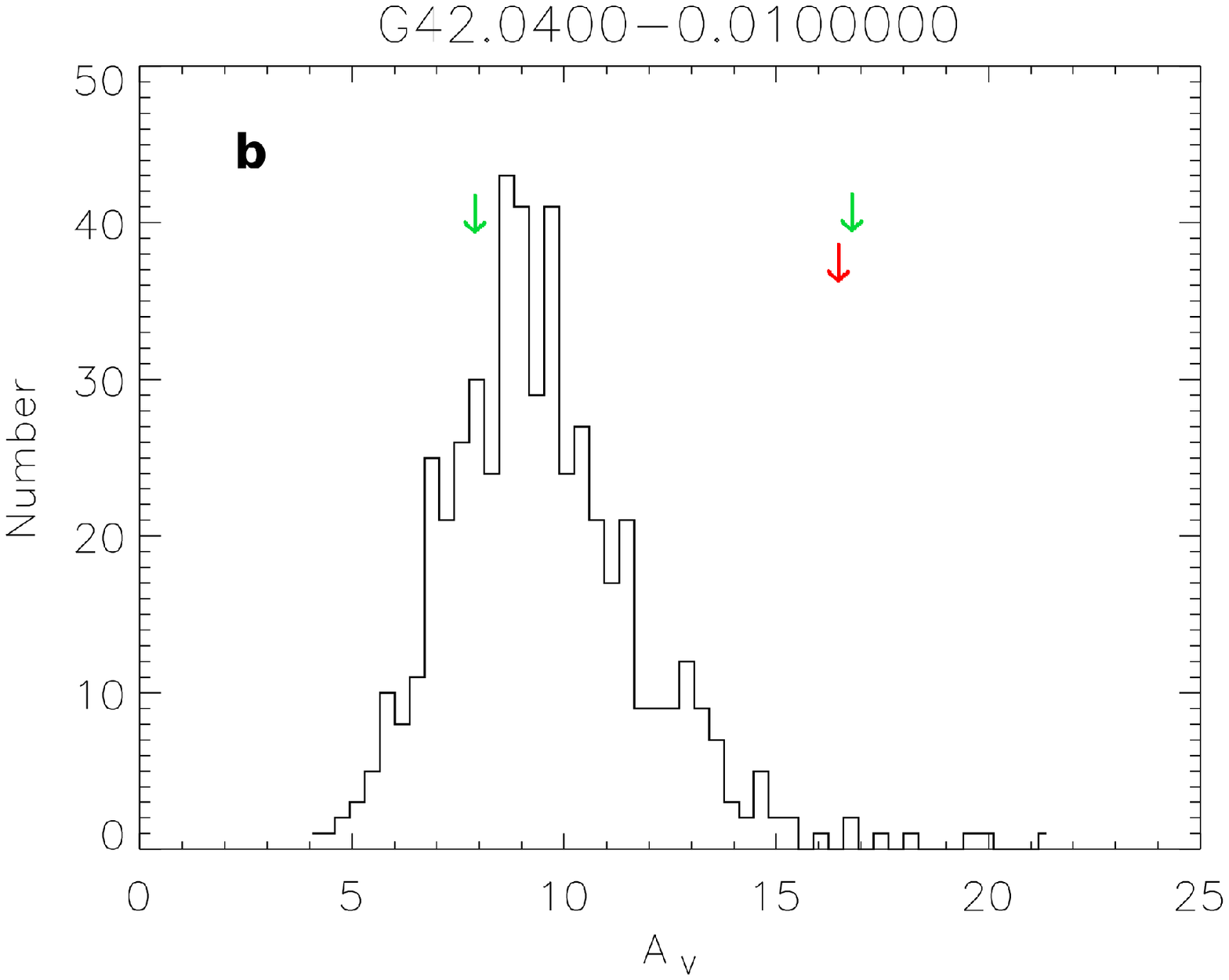}} \\
\resizebox{80mm}{!}{\includegraphics[angle=90]{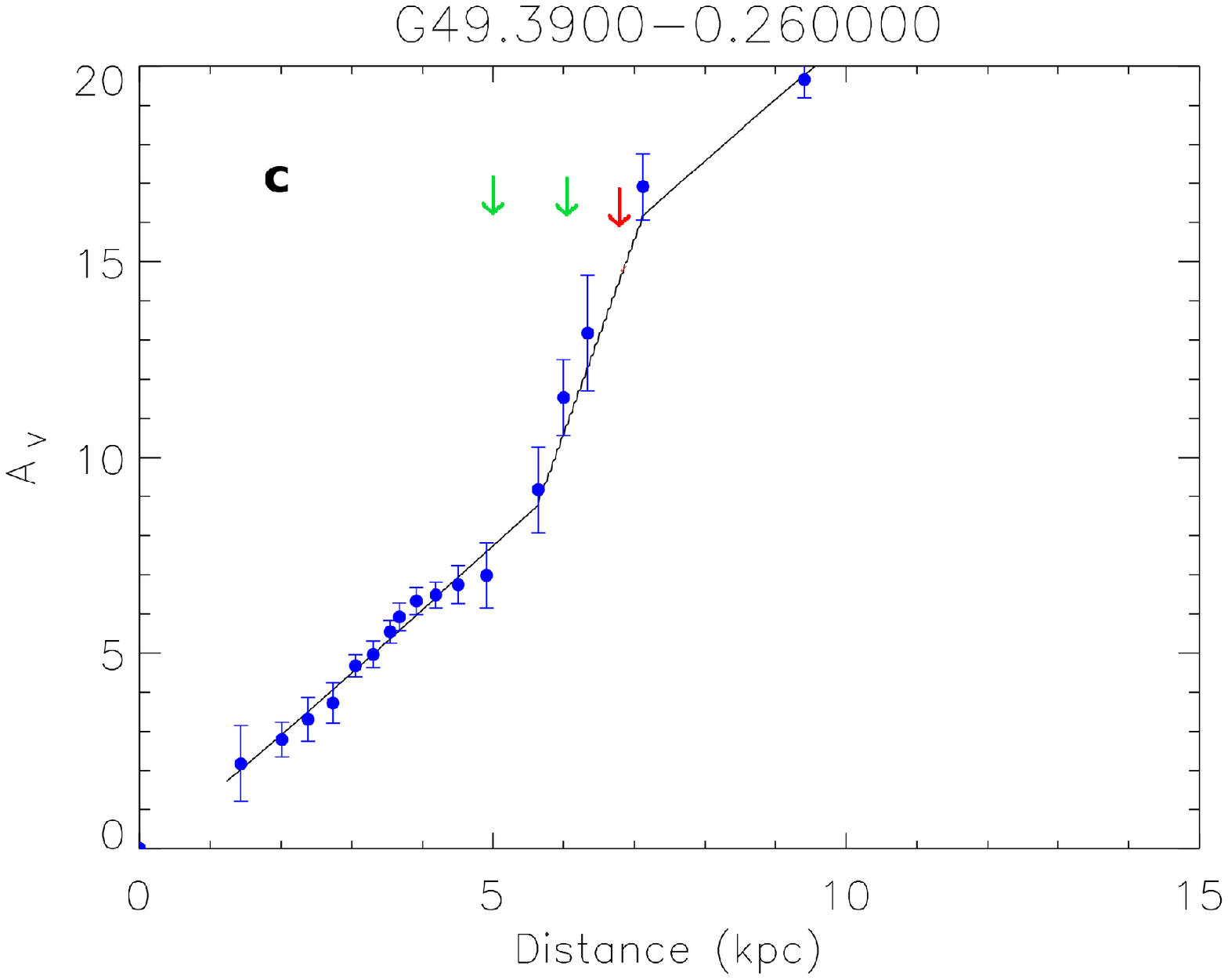}}&
\resizebox{80mm}{!}{\includegraphics[angle=90]{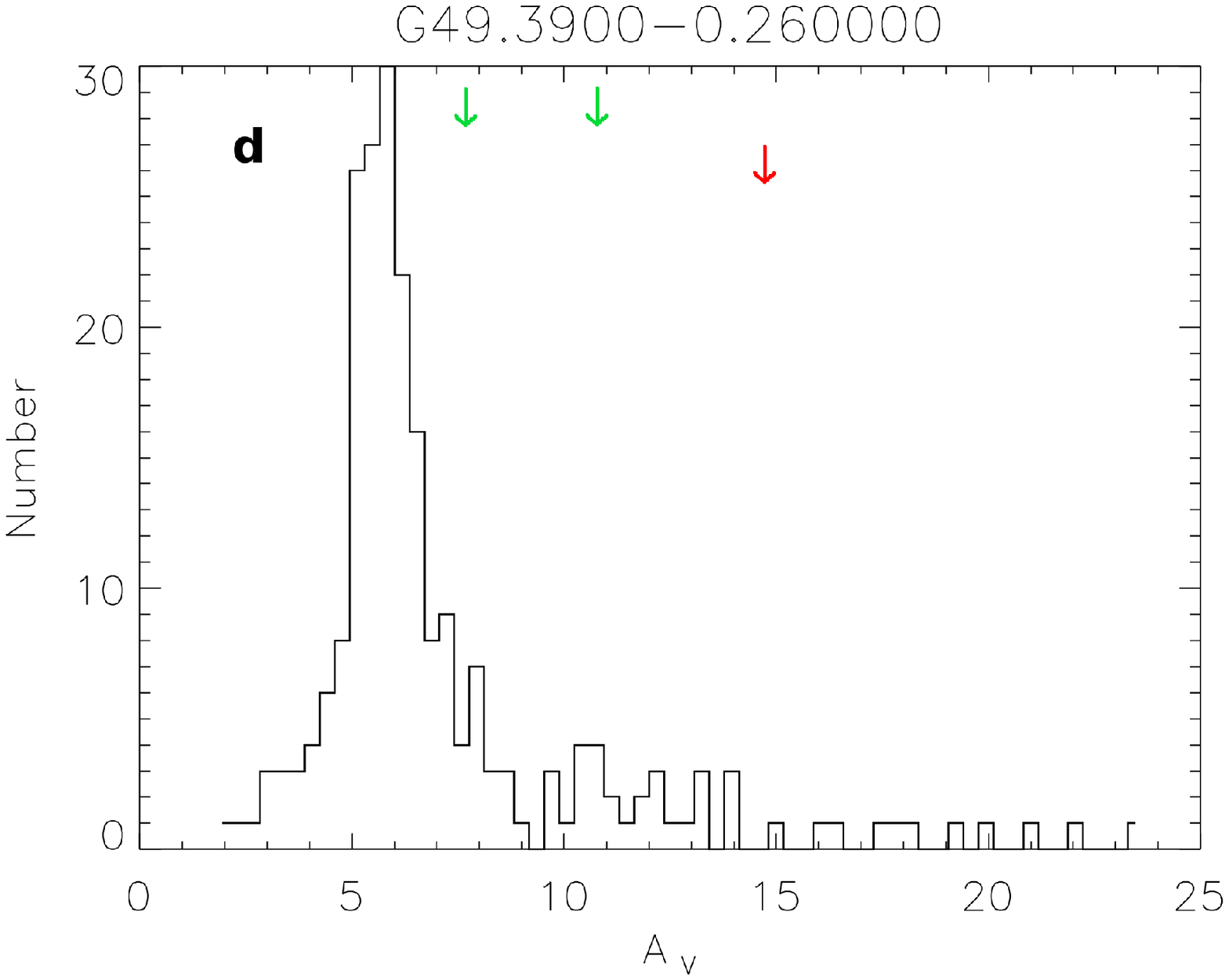}} \\
\resizebox{80mm}{!}{\includegraphics[angle=90]{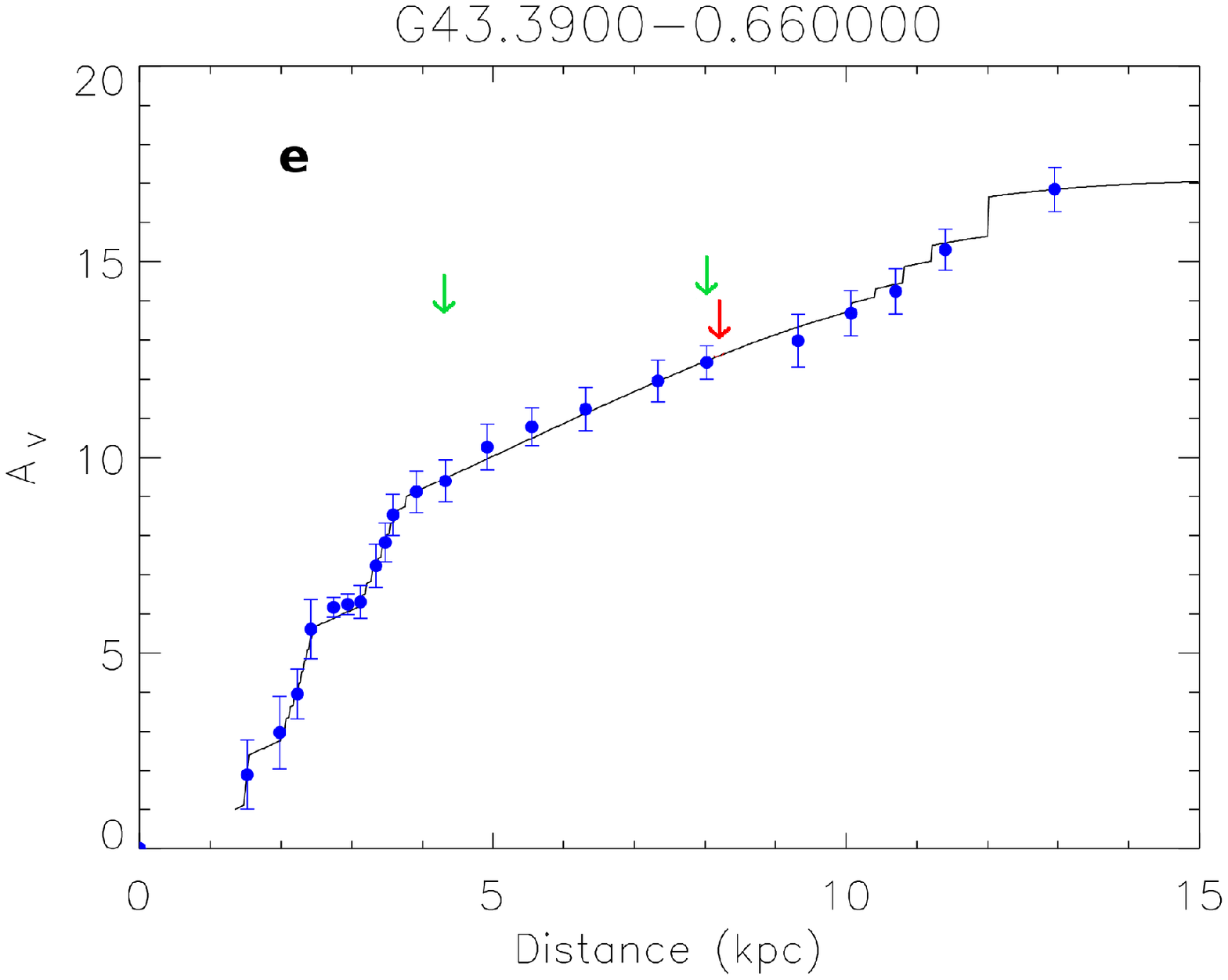}}&
\resizebox{80mm}{!}{\includegraphics[angle=90]{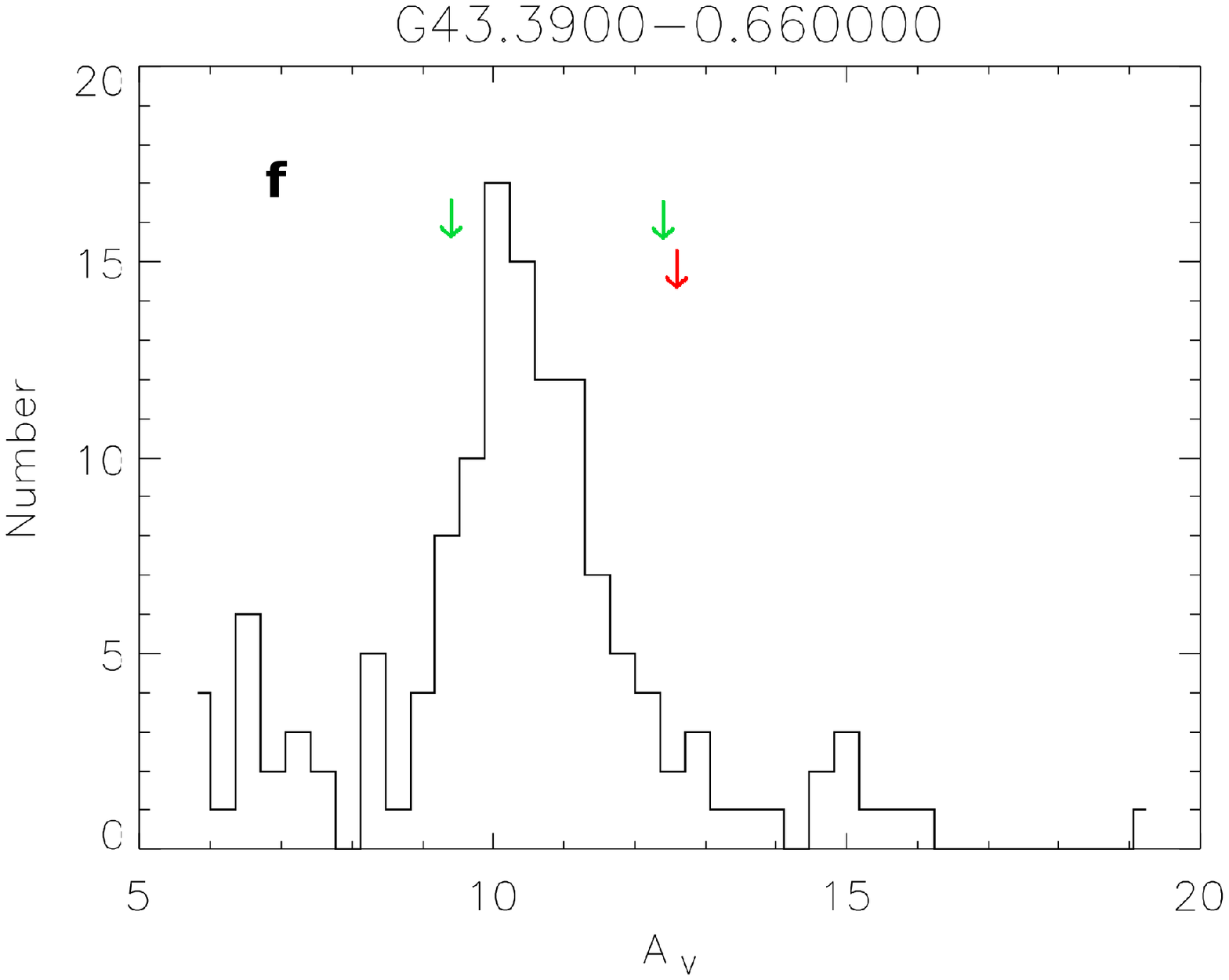}} 
    \end{tabular} 
\caption[]{\small a: A distance-A$_V$ plot containing the extinction distribution model (black line) created to reproduce the M06 data (blue error bars) along the line of sight of the GRS molecular cloud G042.04-0.01. The GRS and two RMS kinematic distances are plotted as red and green arrows respectively. b: An A$_V$ histogram of real stars from a small region centred on the cloud G042.04-0.01. As data are insufficient to exclude the far kinematic distance, but there is also no obvious cloud at the near distance, the cloud has been placed in the first subset of Group 2 (see text). c: Same as (a) for the cloud G049.39-0.26. d: Same as (b) for the cloud G049.39-0.26 except the data are insufficient to exclude both the near and far kinematic distances. For this reason the cloud has been placed in the second subset of Group 2. e: Same as (a) for the cloud G043.3900-0.66. f: Same as (b) for the cloud G043.39-0.66, except no obvious clouds are present in the A$_V$ histogram and so the cloud has been placed in Group 3.}
\label{fig:three_groups}
\end{center}
\end{figure*}

\begin{figure*}
  \begin{center}
    \begin{tabular}{cc}
      \resizebox{165mm}{!}{\includegraphics[angle=0]{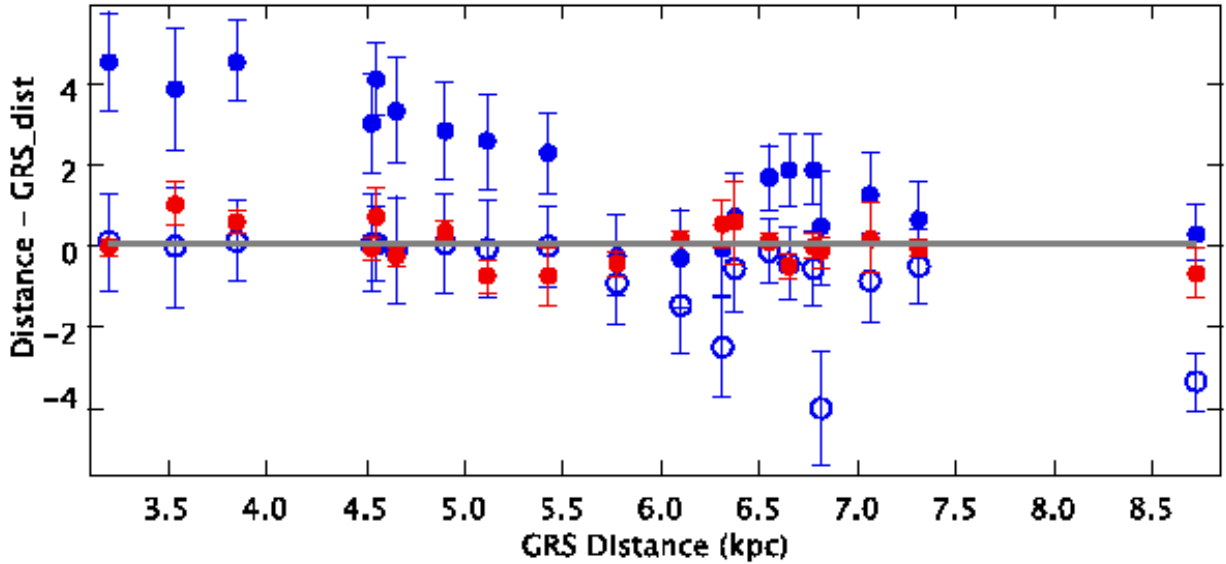}}
    \end{tabular}
    \caption{\small For the 20 molecular clouds in Group 1, the GRS distances, both RMS distances and the distances derived to each cloud in this paper have been subtracted from the GRS distance. The grey line crossing the origin therefore represents the GRS distance subtracted from itself. The open blue circles represent the near RMS distance and the closed blue circles represent the far RMS distance. The red error bars represent the distances derived in this paper.}
    \label{fig:GRSvsDist}
  \end{center}
\end{figure*}

The final 34 molecular clouds have each been sorted into one of three groups depending upon the features present in the subsequent A$_V$ histograms. For each of the final 34 clouds we have created both a distance-A$_V$ plot, containing the M06 data and an extinction distribution model (EDM), and an A$_V$ histogram. \\
$\indent$The first group contains 20 clouds and the corresponding A$_V$ histogram of each has a strong presence of one or more molecular clouds. We derive the distances of these molecular clouds in an attempt to distinguish between the near and far kinematic distances. Results are presented in Table \ref{tabz:results_one} and the table is split into two subgroups. The first subgroup contains regions where only one obvious molecular cloud is present in the A$_V$ histogram, such as the cloud containing G048.9897-0.2992 (Fig. \ref{fig:G48.9897_AvHist}). The second subgroup contains regions where multiple molecular clouds are present in the A$_V$ histogram, such as the cloud G041.04-0.664 (Fig. \ref{fig:G41.04_Av_dist}b). In cases where there are multiple molecular clouds, we use the cloud that appears to be associated with either the near or far distance, and address the presence of the additional clouds in the Appendix by looking for overlapping clouds in the GRS. The columns of Table \ref{tabz:results_one} are as follows: (1)	the molecular cloud name; (2) the near RMS distance; (3) the far RMS distance; (4) the GRS distance; (5) and (6) the distance and error derived in this paper.\\
$\indent$The second group contains 10 clouds and the corresponding A$_V$ histograms do not contain an adequate number of sources with large enough values of A$_V$ to exclude the far kinematic distance. Results are presented in Table \ref{tabz:results_two} and the table is split into two subgroups. In some cases it is still possible to exclude the near distance due to the apparent lack of a molecular cloud at the near distance. The first and second subgroups contain regions where we have and have not been able to exclude the near kinematic distance respectively, Fig. \ref{fig:three_groups} (b) gives an example of an A$_V$ histogram where data are insufficient to exclude the far kinematic distance, but also contains no obvious cloud at the near distance, thereby solving the near/far ambiguity through exclusion. Fig. \ref{fig:three_groups} (d) gives an example of an A$_V$ histogram where data are insufficient to exclude both the near and far kinematic distances. The columns of Table \ref{tabz:results_two} are the same Table \ref{tabz:results_one} except as follows; (5) a flag denoting if the near kinematic distance can be excluded. \\
$\indent$The third group contains 4 clouds and the corresponding A$_V$ histograms do not show the significant presence of a molecular cloud at either the near or far kinematic distance. Results are presented in Table \ref{tabz:results_three}. The columns of Table \ref{tabz:results_three} are the same Table \ref{tabz:results_one}. \citet{roman09} provide the $^{13}$CO luminosity (K km s$^{-1}$ pc$^2$) of each cloud in the GRS which can be used to estimate the A$_V$ of each cloud. This is done as follows; an average $^{13}$CO luminosity of each cloud is determined by estimating the area of each from the Galactic longitudinal and latitudinal FWHM and the GRS distance. This value is then converted to a $^{13}$CO column clump density, in cm$^{-2}$, using the equation N($^{13}$CO) = 8.75 x 10$^{14}$W($^{13}$CO), from \citet{simon01}. Finally this value is converted to A$_V$ using the equation A$_V$ = 4.24 x 10$^{-16}$N($^{13}$CO) + 1.67 from \citet{goodman09}. We do not expect this computed value of A$_V$ to agree particularly well with the value of A$_V$, used to determine the distance to the cloud, measured in each histogram. This is because the computed value of A$_V$ has been averaged over the entire molecular cloud and the resolution of the GRS data is many times greater than the approximate, 6$\arcmin$x6$\arcmin$ tiles used to construct each A$_V$ histogram. Furthermore, the spatial substructure of each cloud has not been considered. \\
$\indent$Three of the clouds in group 3 have the three lowest A$_V$ values, derived from the $^{13}$CO luminosities, in our sample of 34 clouds. This provides a likely explanation as to why the molecular clouds could not be detected above the noise in the A$_V$ histograms. For the remaining cloud, G30.79-0.06, the reverse is true. G30.79-0.06 has one of the largest A$_V$ values, derived from the $^{13}$CO luminosities, out of the sample of 34 clouds. Despite this however, it failed the goodness of fit test as the line of sight extinction is too large, from the combination of both the cloud G30.79-0.06 and a second molecular cloud, as shall be discussed in the Appendix, to detect an adequate amount of stars at the near side of the cloud. Fig. \ref{fig:three_groups} (f) contains an example A$_V$ histogram from group three. Distance-A$_V$ plots and A$_V$ histograms for the remaining regions, from all three groups, are available with the online material. \\
\begin{table*}
\centering
\caption{Clouds where sufficient data are present to examine near and far distances.}
\begin{tabular}{ l c c c c c }
\hline
\hline
	GRS Cloud	&	Near 	&	Far	&	GRS 	&	D	&	error	\\
		&	(kpc)	&	(kpc)	&	(kpc)	&	(kpc)	&		\\
		(1)	&	(2)	&	(3)	&	(4)	&	(5)	&	(6)\\
\hline
GRSMC G031.39+0.29	&	6.35	&	8.16	&	6.55	&	6.6 	&	0.5	\\
GRSMC G032.09+0.09	&	6.13	&	8.29	&	7.07	&	7.2 	&	1.1 	\\
GRSMC G039.34--0.31	&	4.55	&	8.59	&	4.55	&	5.3 	& 	0.7 	\\
GRSMC G040.34--0.26	&	5.34	&	7.63	&	5.43	&	4.5 	&	0.8 	\\
GRSMC G042.14--0.61	&	4.99	&	7.62	&	5.12	&	4.3 	&	0.5 	\\
GRSMC G042.44--0.26	&	4.87	&	7.68	&	4.90	&	5.1 	&	0.7 	\\
GRSMC G043.34--0.36	&	4.46	&	7.91	&	4.65	&	4.1 	&	0.7 	\\
GRSMC G045.14+0.14	&	4.53	&	7.48	&	4.53	&	4.4 	&	0.6 	\\
GRSMC G050.24--0.61	&	3.22	&	7.66	&	3.20	&	3.2 	&	0.2 	\\
GRSMC G050.84+0.24	&	3.43	&	7.32	&	3.53	&	4.5 	&	0.4 	\\
GRSMC G052.79--0.56	&	4.78	&	5.47	&	5.78	&	5.3 	&	0.6 	\\
GRSMC G054.14--0.06	&	3.76	&	6.21	&	6.32	&	7.0 	&	1.0 	\\
GRSMC G054.39--0.46	&	2.76	&	7.20	&	6.82	&	6.8 	&	0.8 	\\
\hline
GRSMC G029.89--0.06	&	6.14	&	8.60	&	6.78	&	6.5 	&	0.7 	\\
GRSMC G030.29--0.21	&	6.76	&	7.90	&	7.32	&	7.2 	&	0.7 	\\
GRSMC G031.04+0.29	&	6.16	&	8.44	&	6.65	&	6.3 	&	0.8	\\
GRSMC G033.04+0.04	&	5.30	&	8.96	&	8.73	&	8.1 	&	0.9 	\\
GRSMC G041.04--0.66	&	5.78	&	7.04	&	6.38	&	6.9 	&	0.5 	\\
GRSMC G043.89--0.81	&	3.92	&	8.33	&	3.85	&	4.4 	&	0.6 	\\
GRSMC G052.79+0.29	&	4.55	&	5.70	&	6.10	&	6.9 	&	1.1 	\\
\hline
\end{tabular}
\label{tabz:results_one}
  \end{table*}
\begin{table*}
\centering
\caption{Clouds where data are insufficient to exclude far distances.}
\begin{tabular}{ l c c c c }
\hline
\hline
GRS Cloud	&	Near 	&	Far	&	GRS 	&	Exclude Near?	\\
                &	(kpc)	&	(kpc)	&	(kpc)	&	\\
		(1)	&	(2)	&	(3)	&	(4)	&	(5)	 \\
\hline
GRSMC G030.44--0.26	&	6.62	&	8.04	&	7.30	&		Y	\\
GRSMC G041.34+0.09	&	4.12	&	8.64	&	8.55	&		Y	\\
GRSMC G042.04--0.01	&	3.87	&	8.74	&	8.60	&		Y	\\
GRSMC G043.19--0.51	&	4.15	&	8.24	&	8.27	&		Y	\\
GRSMC G044.29+0.04	&	4.33	&	7.83	&	8.02	&		Y	\\
GRSMC G044.34--0.21	&	5.33	&	6.90	&	6.80	&		Y	\\
GRSMC G045.49+0.04	&	4.71	&	7.22	&	7.45	&		Y	\\
\hline
GRSMC G028.59+0.04	&	6.16	&	8.76	&	6.60	&		N 	\\
GRSMC G037.69+0.09	&	6.18	&	7.20	&	6.70	&		N	\\
GRSMC G049.39--0.26	&	4.99	&	6.04	&	6.82	&		N	\\
\hline
\end{tabular}
\label{tabz:results_two}
  \end{table*}
\begin{table*}
\centering
\caption{Clouds where data are insufficient to detect the presence of a molecular cloud.}
\begin{tabular}{ l c c c c }
\hline
\hline
GRS Cloud	&	Near 	&	Far	&	GRS 	\\
                &	(kpc)	&	(kpc)	&	(kpc)	\\
	(1)	&	(2)	&	(3)	&	(4) \\
\hline
GRSMC G032.99+0.59	&	5.91	&	8.36	&	8.05		\\
GRSMC G030.79--0.06	&	5.77	&	8.84	&	6.22		\\
GRSMC G043.39--0.66	&	4.32	&	8.01	&	8.18		\\
GRSMC G043.49--0.71	&	3.32	&	8.95	&	2.83		\\
\hline
\end{tabular}
\label{tabz:results_three}
  \end{table*}
$\indent$To draw a comparison between our results and both the RMS and GRS kinematic distances, Fig. \ref{fig:GRSvsDist} contains a plot of the GRS distances, both RMS distances and the distances derived to each of the 20 clouds in Group 1, subtracted from the GRS distance. The grey line crossing the origin therefore represents the GRS distance subtracted from itself. The open blue circles represent the near RMS distance and the closed blue circles represent the far RMS distance. The red error bars represent the 20 distances derived in this paper from group 1. \\
$\indent$The near/far distance ambiguity has been resolved for 27 clouds out of the sample of 34 molecular clouds. Distances have been derived to 20 clouds that are independent to kinematic methods. Of the 20 distances derived, 19 agree the GRS distance to within the 2$\sigma$ level, and 18 of the 20 agree to within the 1$\sigma$ level, however this does not take into account the error on the GRS distance. One cloud, G050.84+0.24 with a GRS distance D = 3.5 kpc and a distance derived in this paper D = 4.5$\pm$0.4, agrees only within the 3$\sigma$ level. There is a second overlapping cloud with a GRS distance of 5.3 kpc, however there is no obvious sign of a second molecular cloud in the A$_V$ histogram, furthermore the distance derived in this paper correctly resolves the near/far ambiguity. \\
$\indent$In many cases the GRS distance agrees within errors with one of the RMS kinematic distances. However, as discussed previously, due to the difference in choice of the Galactic rotation models used to derive both the RMS and GRS kinematic distances, there are some differences between the two sets of kinematic distances. Error bars for the GRS distances have not been plotted as \citet{roman09} do not include specific errors in each measurement, however they will be very similar to the size of the error bars on the RMS distances. Taking this into account, all distances measured in this paper agree well with the GRS resolved kinematic distances.  \\
$\indent$There are often large differences between kinematic distances and those derived by independent means. For example, there are differences of $\sim$2 kpc between kinematic distances and distances derived using trigonometric parallax in the Perseus spiral arm \citep{xu06}. Of the 18 masers previously mentioned that were studied by \citet{reid09}, 3 lie within the same Galactic longitude range used in this paper. The differences between the kinematic distances and trigonometric parallaxes of these 3 masers, presented by \citet{reid09}, are in each case less than $\sim$0.4 kpc. This value is therefore smaller than both the average error of the distances derived in this paper, and the average error of the RMS distances.\\
$\indent$One final limitation to this method is that the most reddened stars are predominantly M class giants, and as the most reddened stars, they are required to probe the most distant molecular clouds. It has been noted previously that the M0III stars should be dereddened along specific M0III reddening tracks. This is even more important for the most reddened stars where the difference between reddening tracks is very apparent \citep{stead09}. Although the error caused by using a K0III reddening track has been accounted for, the error created is a systematic one, affecting all results in the same manner. It may be possible however to use the $\Bes$ data to estimate the position on a CCD where the dominant spectral type changes and adjust the choice of reddening track accordingly. The error caused by the inappropriate choice of reddening track is $\sim$0.03 mag in each photometric band, therefore similar to the error cut applied to the UKIDSS data in this paper.

\section{Conclusions}
\label{sec:conclusion}

Of the 34 molecular clouds studied in this paper, we have directly resolved the near/far kinematic distance ambiguity for 20 clouds by deriving a distance to each cloud. These distances have been derived using near-infrared photometry and an existing line of sight specific extinction-distance relationship. Therefore the distances presented in this paper are independent to kinematic methods typically used to determine distances to molecular clouds. These distances have been used to resolve the kinematic distance ambiguity associated with each cloud, and from comparison to the work of \citet{roman09}, all 20 clouds have been resolved correctly.\\
$\indent$For the remaining 14 clouds that we were unable to derive a distance to, 3 of the clouds were too optically shallow to detect above the noise in subsequent A$_V$ histograms. One cloud had several high extinction clouds along the same line of sight and so, due to a lack of foreground stars, we were unable to constrain a distance to the cloud. Finally, the method failed for 10 clouds as the extinction along the line of sight was too large to detect stars at, and beyond, the far distance. However, for 7 of these 10 clouds, we could not detect the presence of a molecular cloud at the near distance, thereby solving the near/far ambiguity through exclusion. In all 7 occurrences our results agreed with those of \citet{roman09}. In total, 27 clouds had their near/far ambiguity resolved. \\
$\indent$Although it was not possible to observe stars beyond the far RMS kinematic distance assigned to 10 of the 34 molecular clouds, it may still be possible to extend our sample to include clouds further than the original 9 kpc cut applied. The majority of derived distances are below 7.5 kpc, however the method fails when the extinction along the line of sight is too large to detect stars at, and beyond, the far distance. For this reason the method failed on some clouds, as close as 6.6 kpc, and succeeded on others as far away as 8.7 kpc. Success of the method depends on the total extinction along the line of sight. Although UKIDSS GPS data would allow the detection of stars suffering up to A$_V$$\sim$25 using the error cuts detailed in this paper, from inspection of the A$_V$ histograms presented in this paper we can determine distances to clouds that suffer below A$_V$$\sim$18 along the line of sight, including the visual extinction of the cloud itself. It may be possible to increase the depth by relaxing the error cut in the J band, the photometric band that suffers from the effect of extinction the most, without contaminating the resultant A$_V$ histograms significantly. Furthermore, for regions where the extinction along the line of sight exceeds 18 magnitudes of visual extinction, it may still be possible to confidently exclude the near distance.\\
$\indent$In this paper the GRS data provide a way to obtain the Galactic location and spatial extent of each molecular cloud studied. However the GRS only covers a Galactic longitude range from l=18$^o$ and l=55.7$^o$, between $|b|<$1$^o$. To extend the work in this paper to regions outside of this range, molecular cloud extinction maps (section \ref{sec:Xmaps}) could be used to identify molecular clouds and measure their spatial extent. This process would work best at larger Galactic longitudes where the average line of sight extinction is typically lower, making it easier to detect the presence of molecular clouds above the background extinction level. We will be able to extend and improve the method to clouds in the southern hemisphere when the deeper VISTA data become available \citep{minniti09}.

\section*{Acknowledgements}

We thank the anonymous referee for suggestions that improved the results of this paper. This work is based in part on data obtained as part of the UKIRT Infrared Deep Sky Survey. This publication makes use of data products from the Two Micron All Sky Survey, which is a joint project of the University of Massachusetts and the Infrared Processing and Analysis Centre/California Institute of Technology, funded by the National Aeronautics and Space Administration and the National Science Foundation. We made use of the VizieR service (http://vizier.u-strasbg.fr/viz-bin/VizieR) to obtain 2MASS data and the M06 extinction distributions. This publication makes use of molecular line data from the Boston University-FCRAO Galactic Ring Survey (GRS). The GRS is a joint project of Boston University and Five College Radio Astronomy Observatory, funded by the National Science Foundation under grants AST-9800334, AST-0098562, and AST-0100793.

\bibliographystyle{mn2e}
\bibliography{Joey}{}

\appendix

\section{Regions with multiple molecular clouds}

\begin{figure*}
  \begin{center}
    \begin{tabular}{cc}
      \resizebox{80mm}{!}{\includegraphics[angle=90]{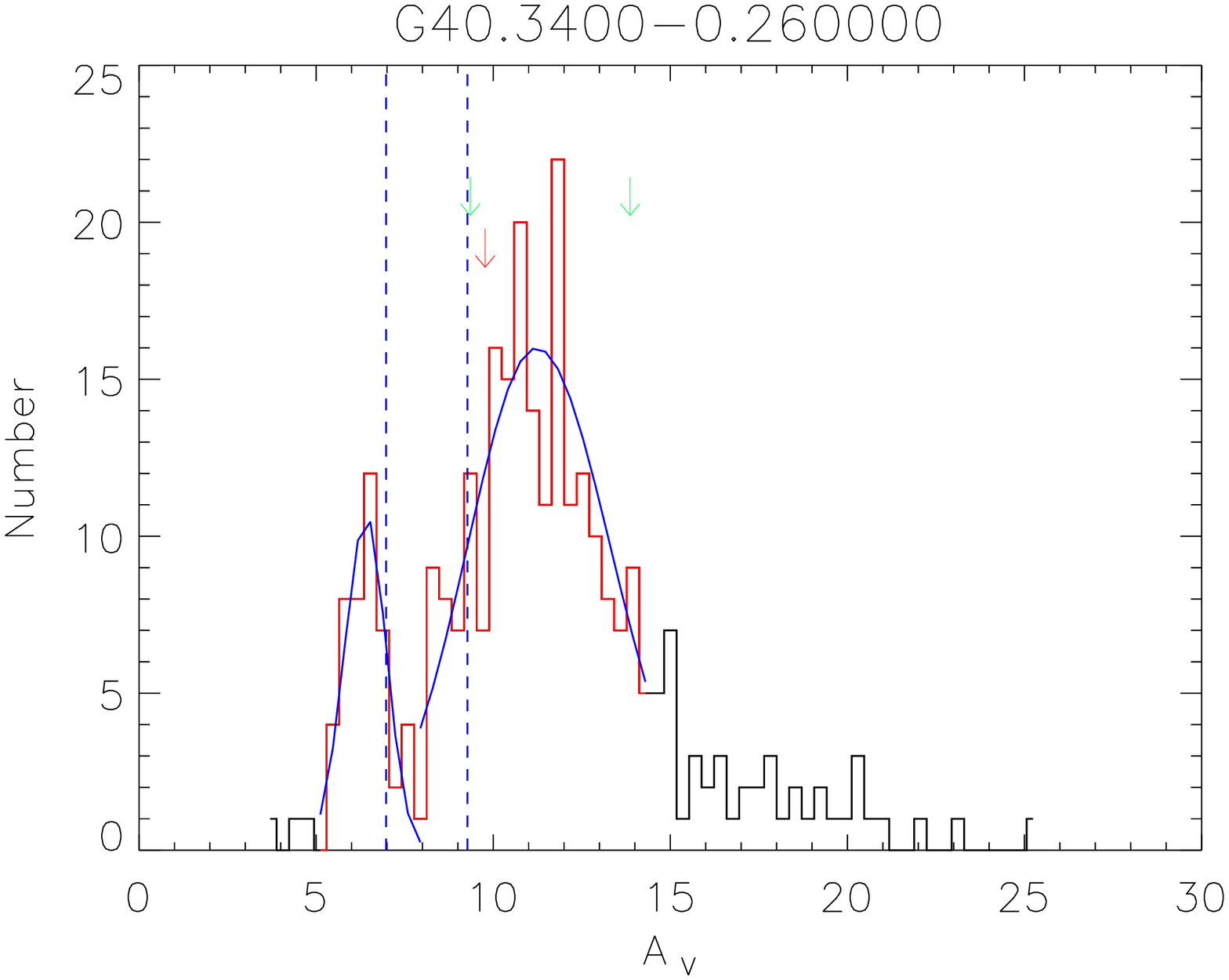}} &
      \resizebox{80mm}{!}{\includegraphics[angle=90]{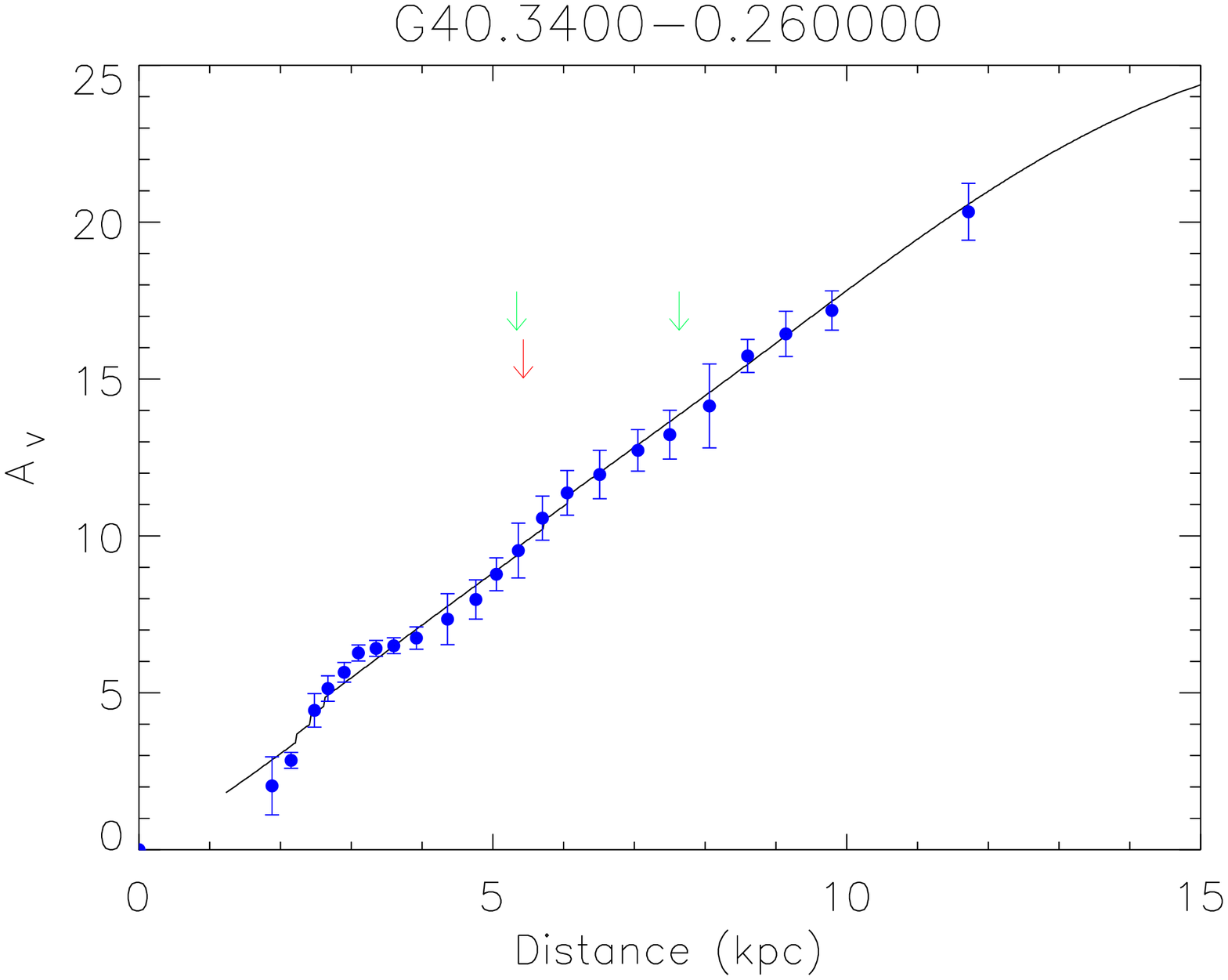}}\\
      \resizebox{80mm}{!}{\includegraphics[angle=90]{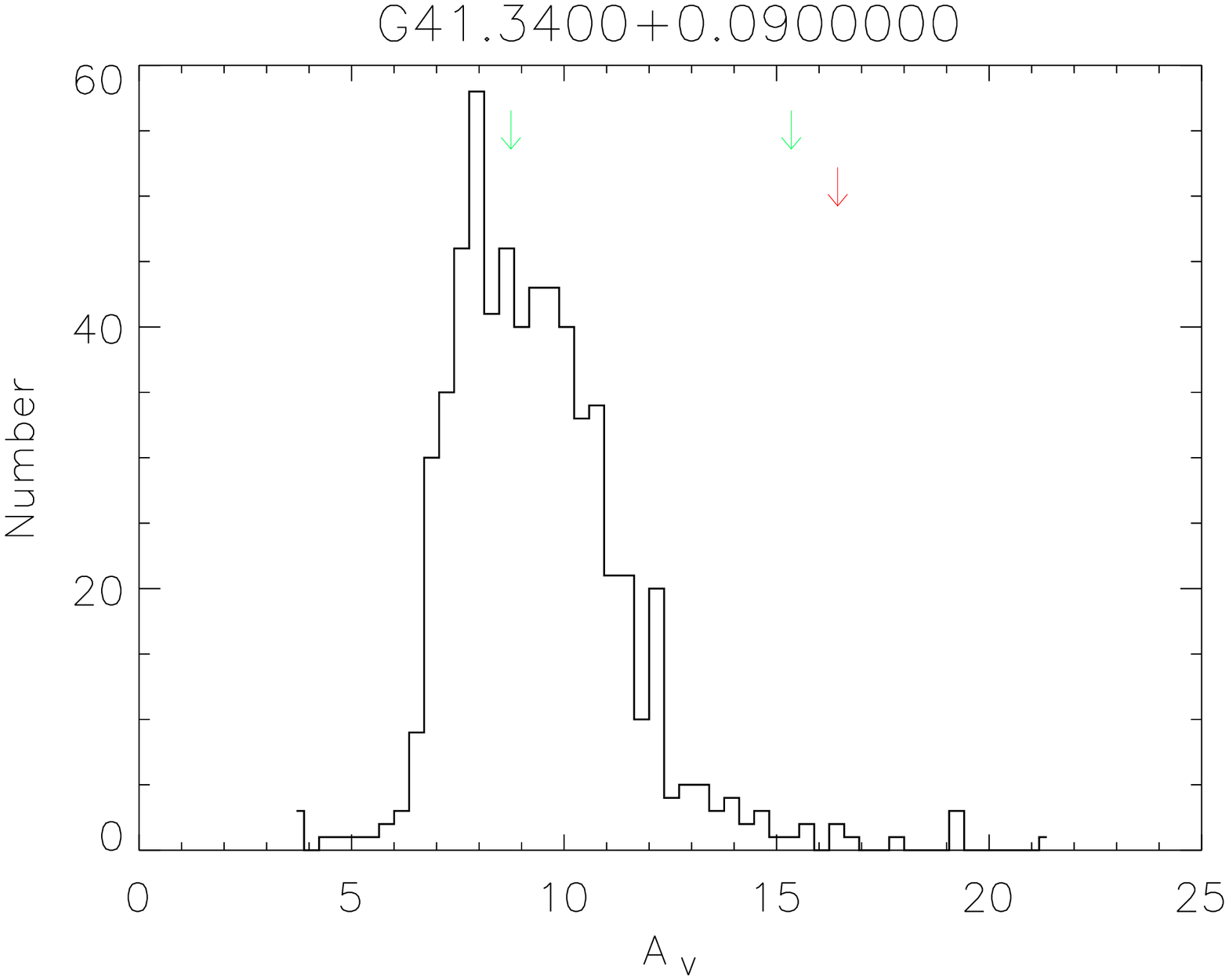}} &
      \resizebox{80mm}{!}{\includegraphics[angle=90]{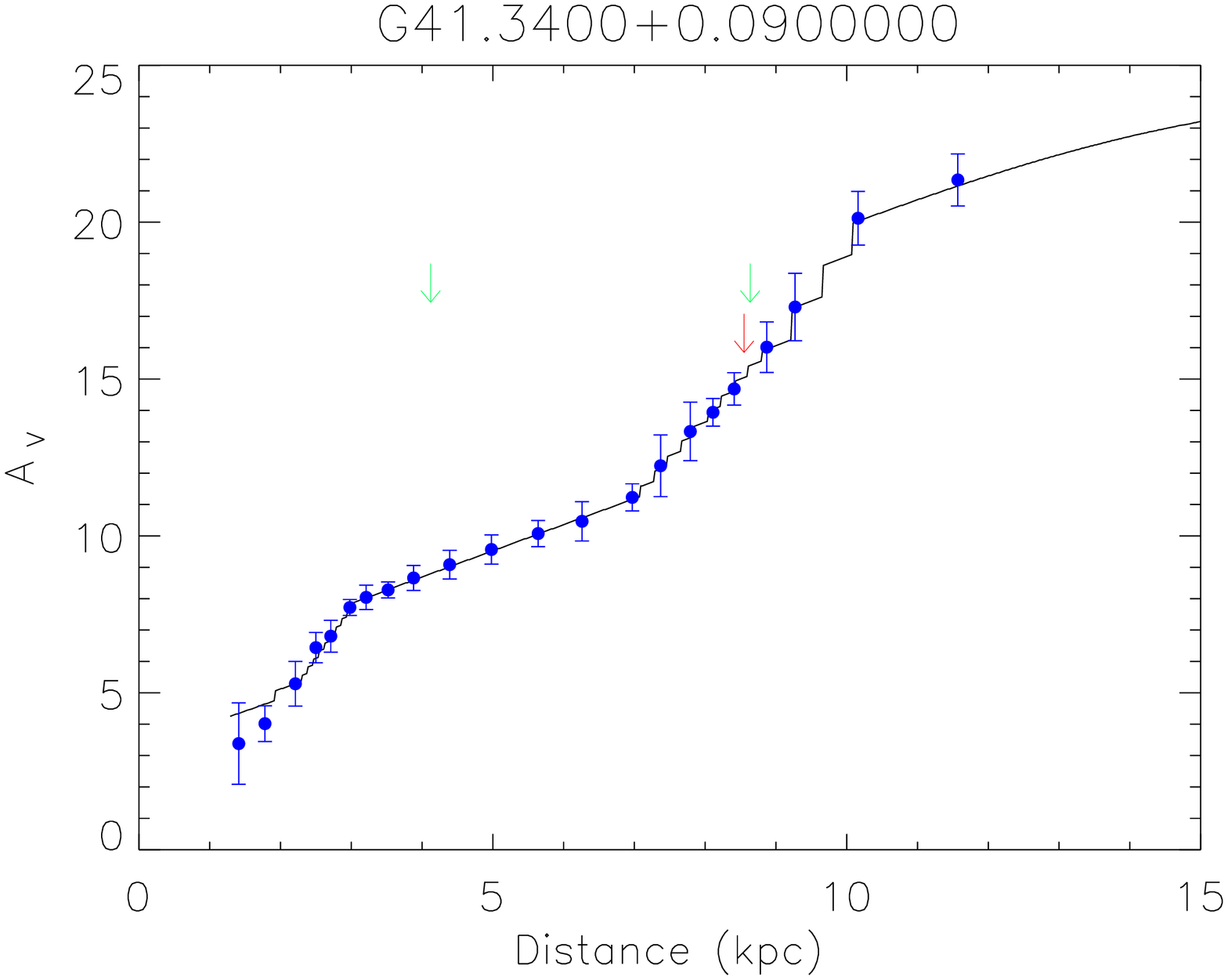}}\\
      \resizebox{80mm}{!}{\includegraphics[angle=90]{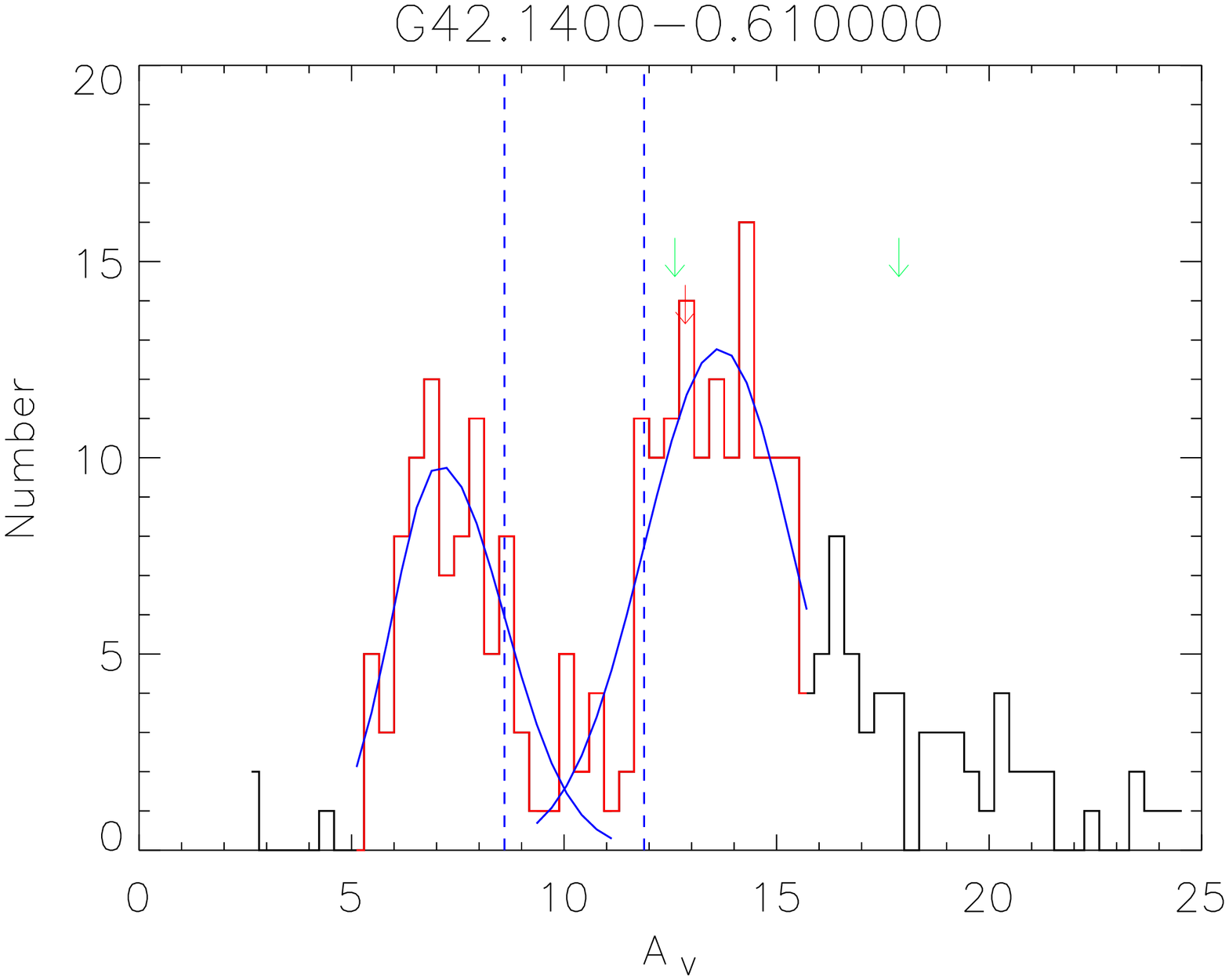}} &
      \resizebox{80mm}{!}{\includegraphics[angle=90]{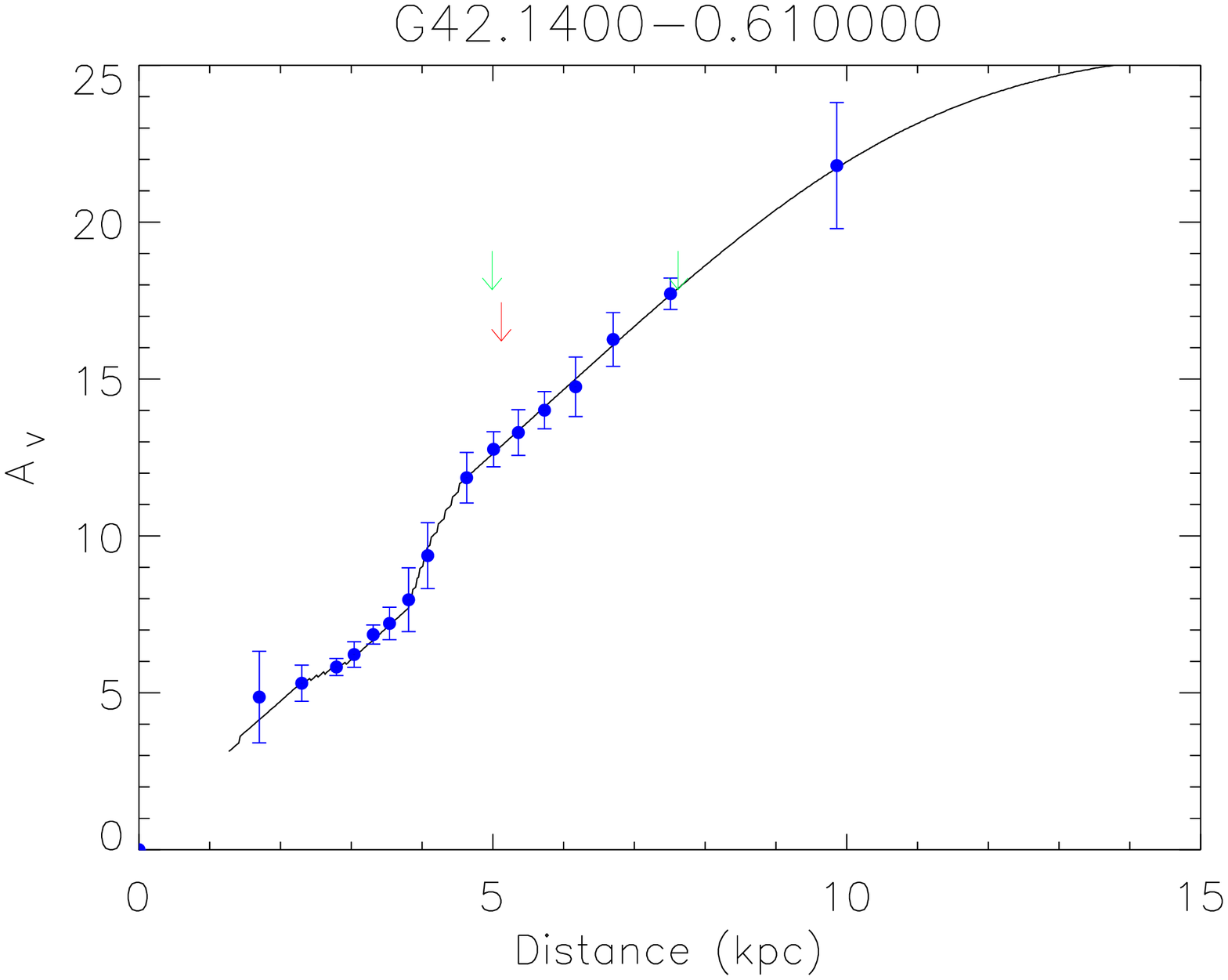}}\\
      \resizebox{80mm}{!}{\includegraphics[angle=90]{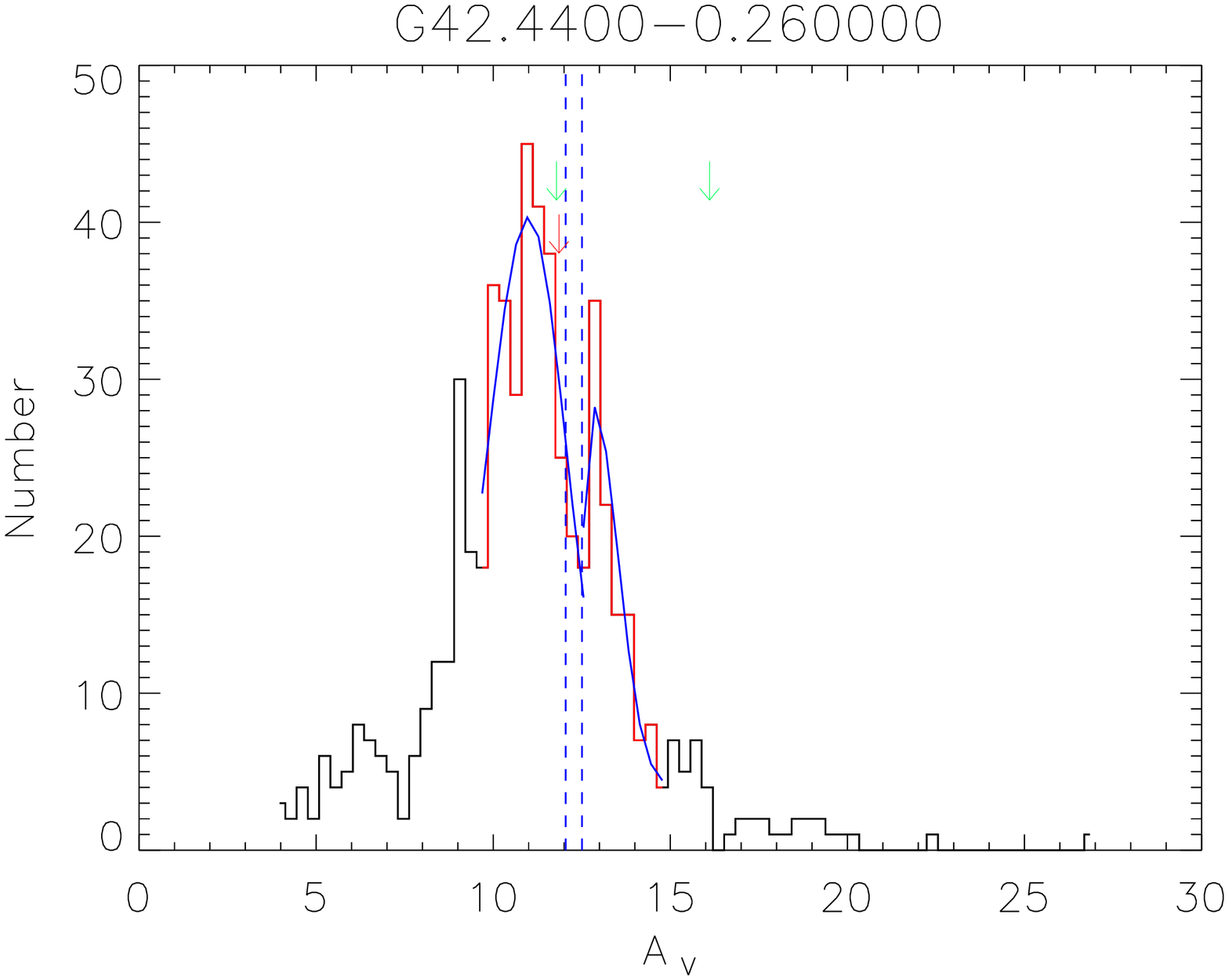}} &
      \resizebox{80mm}{!}{\includegraphics[angle=90]{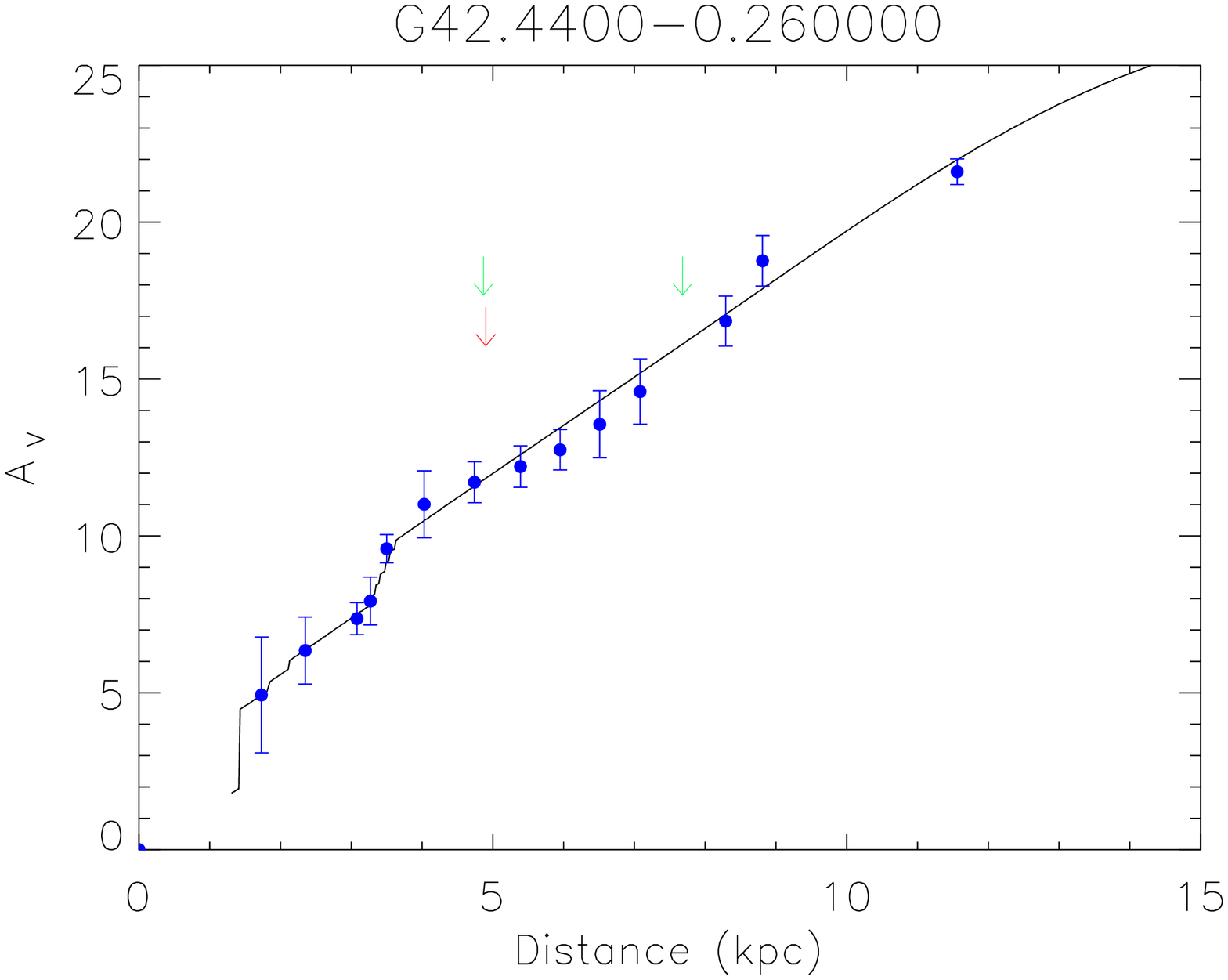}}
    \end{tabular} 
    \caption[]{\small }
    \label{fig:a4}
  \end{center}
\end{figure*}

\begin{figure*}
\begin{center}
    \begin{tabular}{cc}
      \resizebox{80mm}{!}{\includegraphics[angle=90]{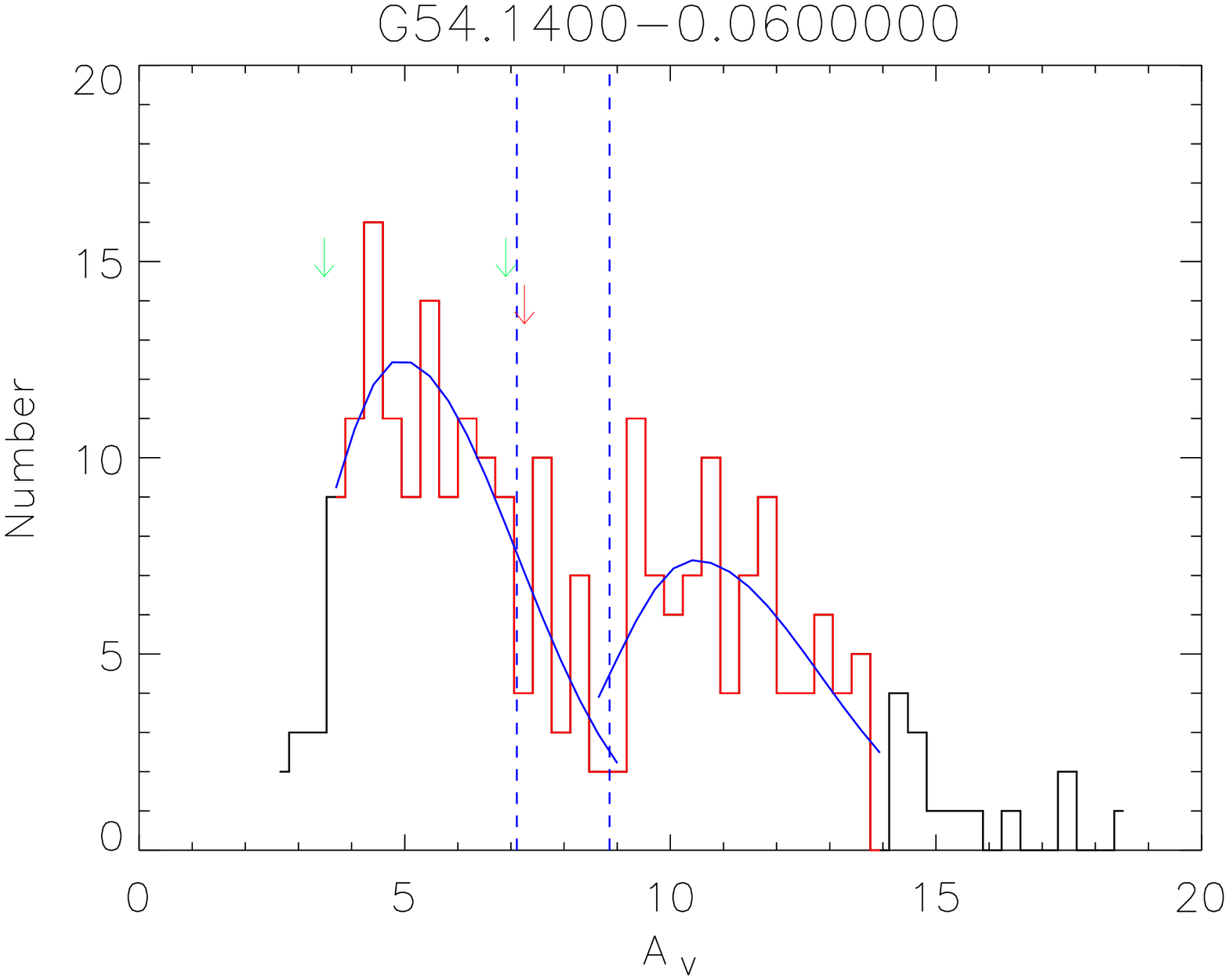}} &
      \resizebox{80mm}{!}{\includegraphics[angle=90]{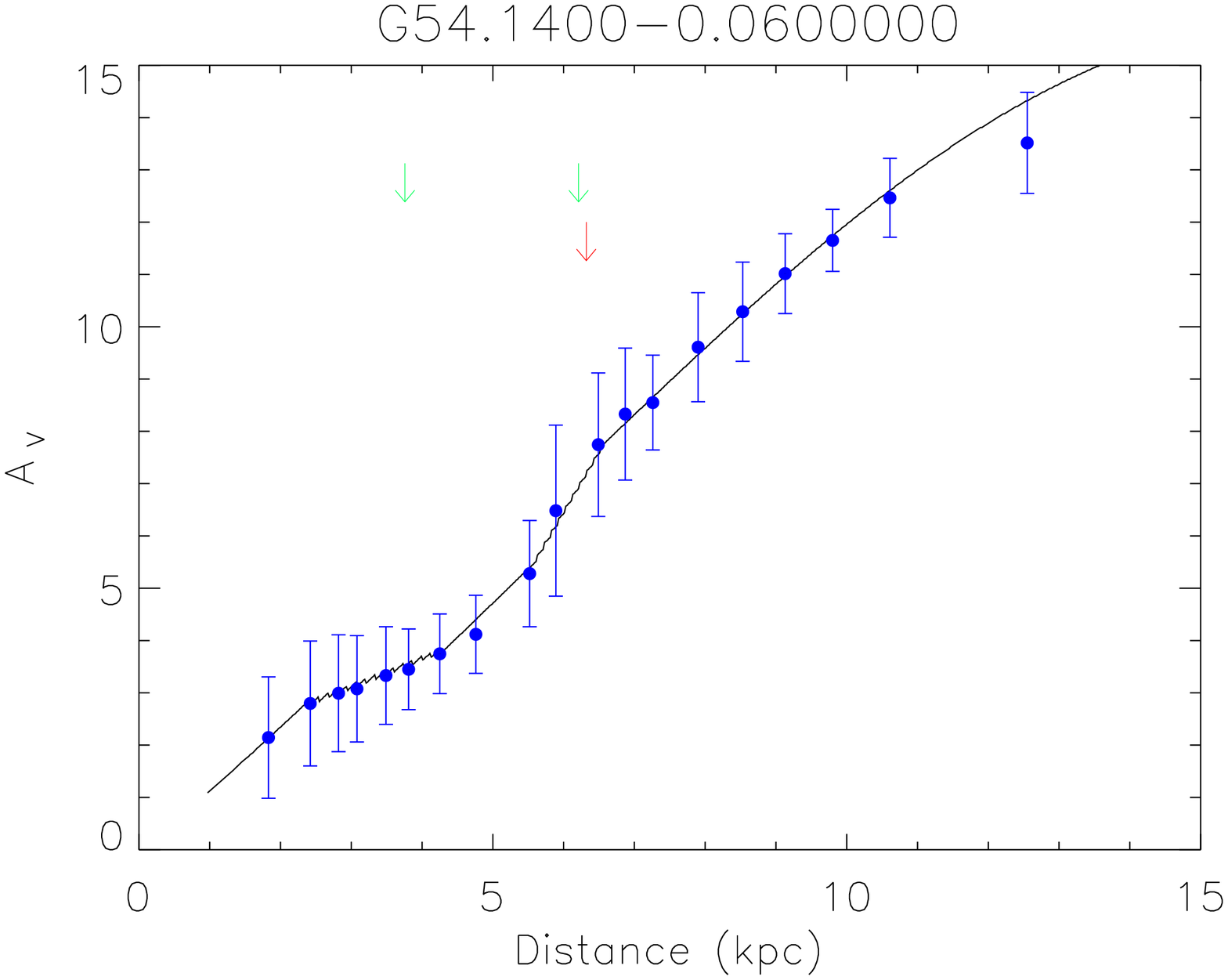}}\\
      \resizebox{80mm}{!}{\includegraphics[angle=90]{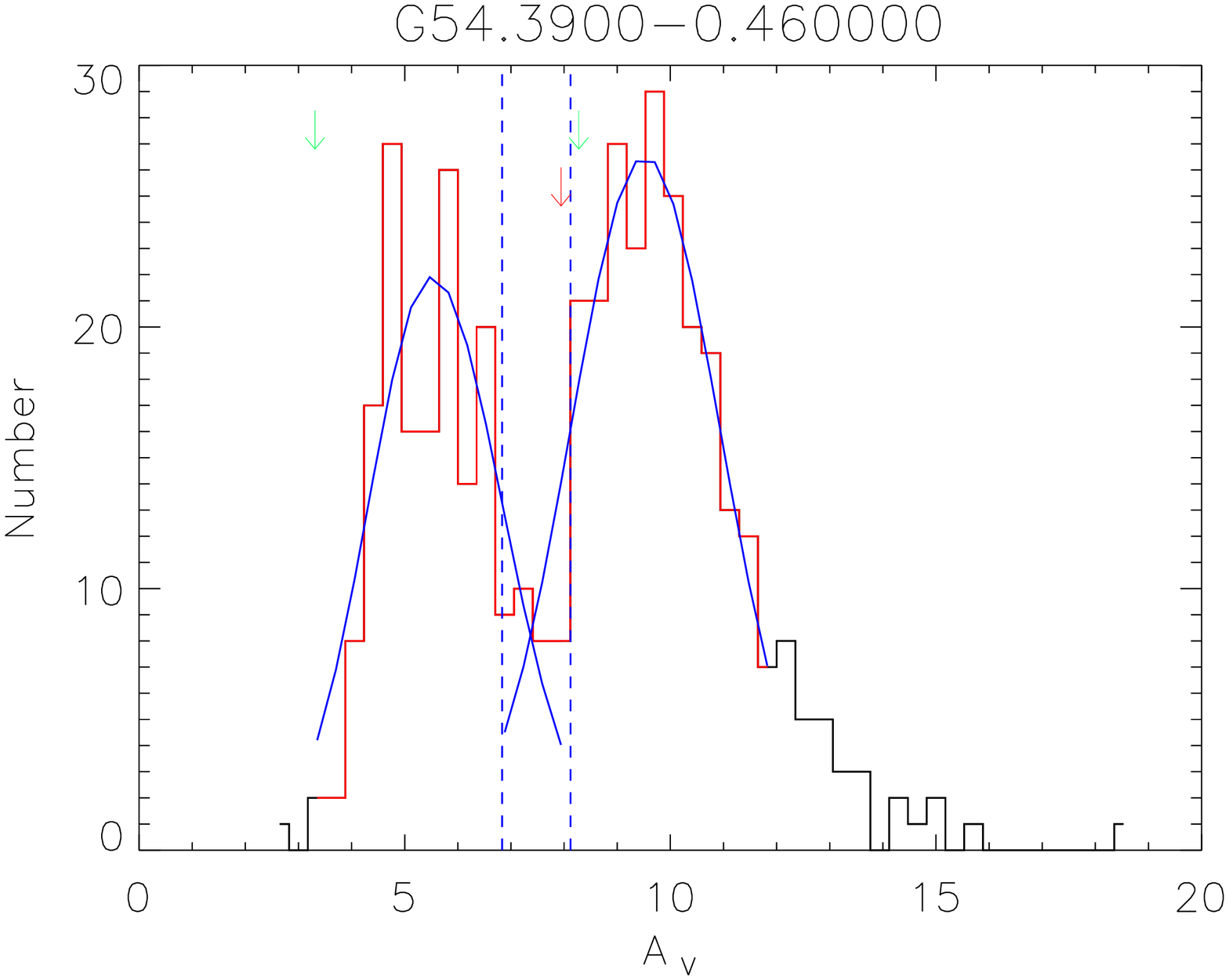}} &
      \resizebox{80mm}{!}{\includegraphics[angle=90]{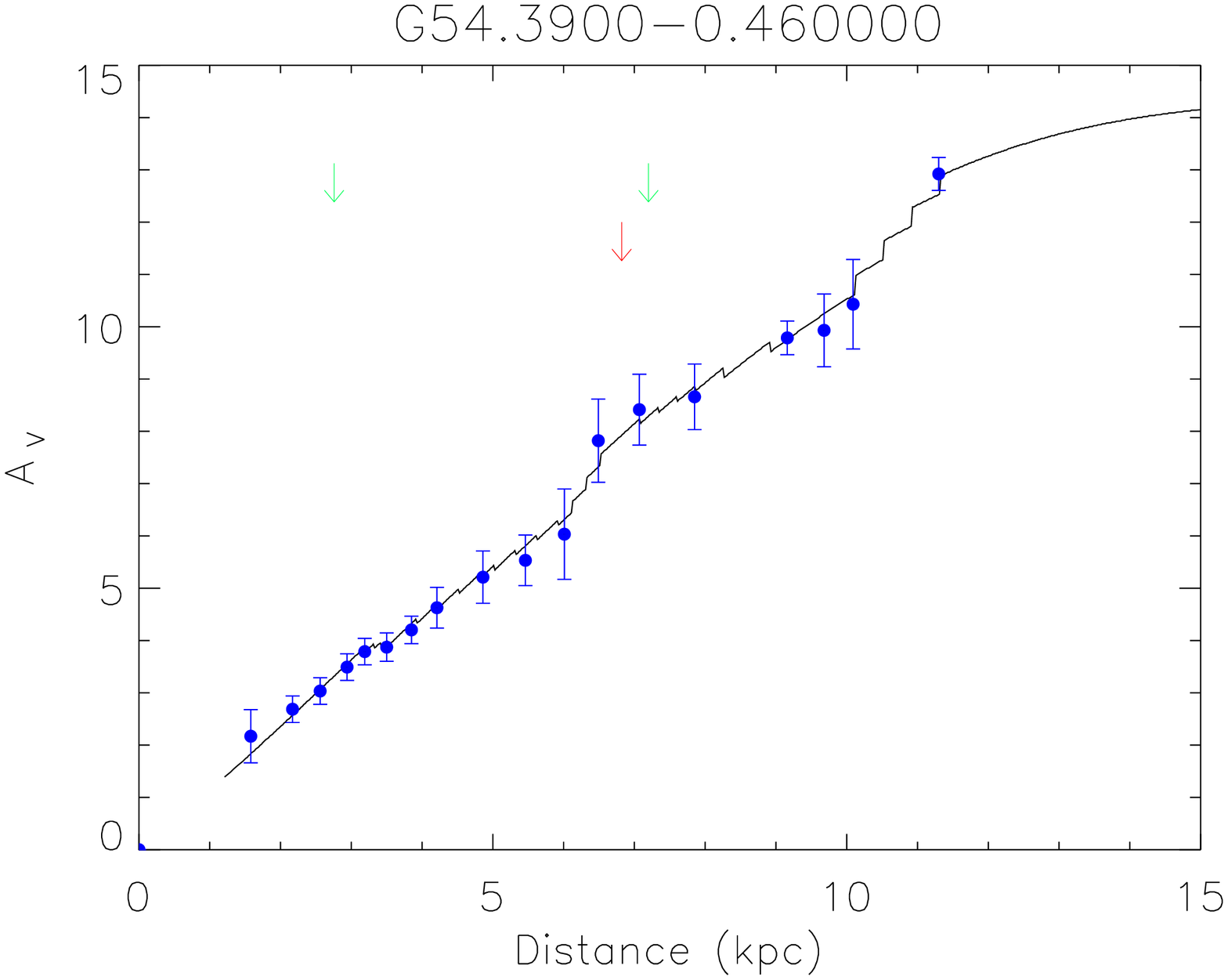}}
    \end{tabular} 
\caption[]{\small }
\label{fig:a8}
\end{center}
\end{figure*}

Distance-A$_V$ plots and A$_V$ histograms for all the remaining regions are available with the online material in Figures \ref{fig:a1} to \ref{fig:a8}. In all cases the distance-A$_V$ plots contain the extinction distribution model (black line) created to reproduce the M06 data (blue error bars) along the line of sight of each GRS cloud. The GRS and two RMS kinematic distances are plotted as red and green arrows respectively. The A$_V$ histograms contain the same red and green arrows corresponding to the GRS and both RMS kinematic distances. In cases where a molecular cloud was present in the A$_V$ histogram, the two Gaussians used to estimate the extinction of, and from that the distance to, each cloud are shown as blue curves. The 1$\sigma$ width of each skewed Gaussian has been indicated with blue dashed lines.

\subsection{G029.89-0.06}

The molecular cloud G029.89-0.06 has RMS distances of 6.1/8.6 kpc and a GRS distance of 6.8 kpc. The distance derived in this paper is D = 6.5$\pm$0.7. Fig. \ref{fig:a1} appears to contain a second molecular cloud at A$_V$$\sim$12. The derived distance to this second cloud is D = 8.3$\pm$0.7. This value agrees with the far RMS distance at the 1$\sigma$ level and so a certain level of ambiguity remains. \\
$\indent$The extracted region overlaps with a second molecular cloud G029.79-0.21. This second cloud, G029.89-00.26, has a high $^{13}$CO luminosity and a GRS distance of 8.45 kpc, therefore providing a likely explanation for the presence of the second cloud.

\subsection{G030.29-0.21}

The A$_V$ histogram, created from data extracted from around the molecular cloud, G030.29-0.21, appears to contain two A$_V$ dips of similar size (Fig. \ref{fig:a1}). Measuring the size of the second dip, situated at A$_V$$\sim$7, reveals a distance of D = 4.6$\pm$1.0 kpc. This distance is inconsistent with both RMS distances, 6.8/7.9 kpc, and the GRS distance, 7.3 kpc, assigned to this cloud. \\
$\indent$The extracted region overlaps with the molecular cloud G030.14-0.06 that has been associated with the RMS source G030.3830-00.1099. This particular RMS source has a near/far ambiguity of 5.4/9.3 kpc. As this far distance is larger than our 9 kpc selection criterion, it explains why the RMS source was not present in our final sample. The derived GRS kinematic distance of the molecular cloud is D = 5.6 kpc and so the presence of two molecular clouds in the A$_V$ histogram has been accounted for.

\subsection{G031.04+0.29}

The molecular cloud G031.04+0.29 has RMS distances of 6.2/8.4 kpc and a GRS distance of 6.7 kpc. The distance derived in this paper is D = 6.3$\pm$0.8. Fig. \ref{fig:a2} appears to contain a second molecular cloud at A$_V$$\sim$9. The derived distance to this second cloud is D = 7.5$\pm$0.6, therefore excluding the far RMS distance. \\
$\indent$The extracted region overlaps with several molecular clouds G030.89+00.14, G031.24+00.09 and G031.29+00.09 that have GRS distances of 7.28, 7.25 and 7.25 kpc respectively. All three clouds have high $^{13}$CO luminosities providing a possible explanation for the presence of a second cloud.

\subsection{G033.04+0.04}

\begin{figure*}
  \begin{center}
    \begin{tabular}{cc}
      \resizebox{80mm}{!}{\includegraphics[angle=90]{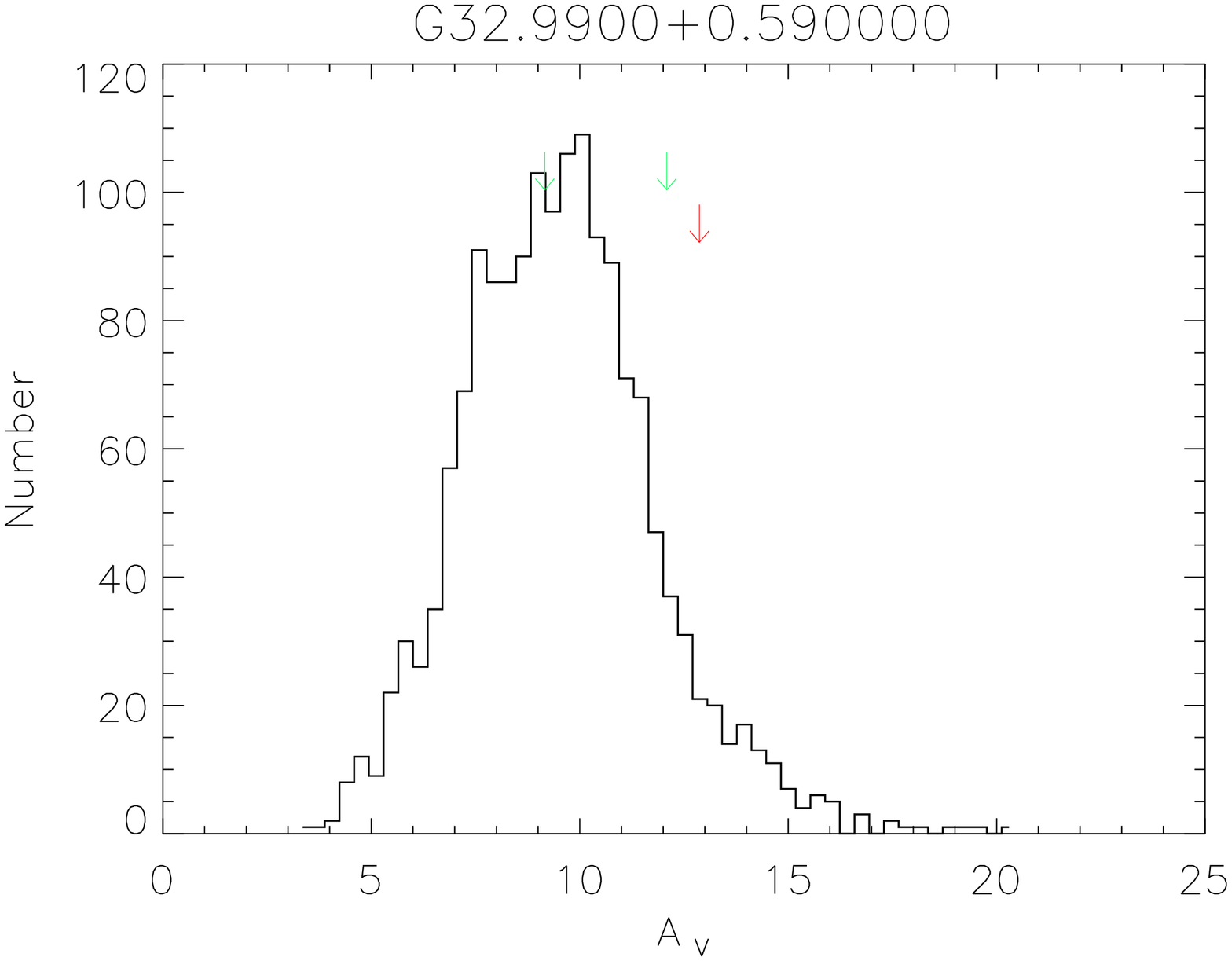}} &
      \resizebox{80mm}{!}{\includegraphics[angle=90]{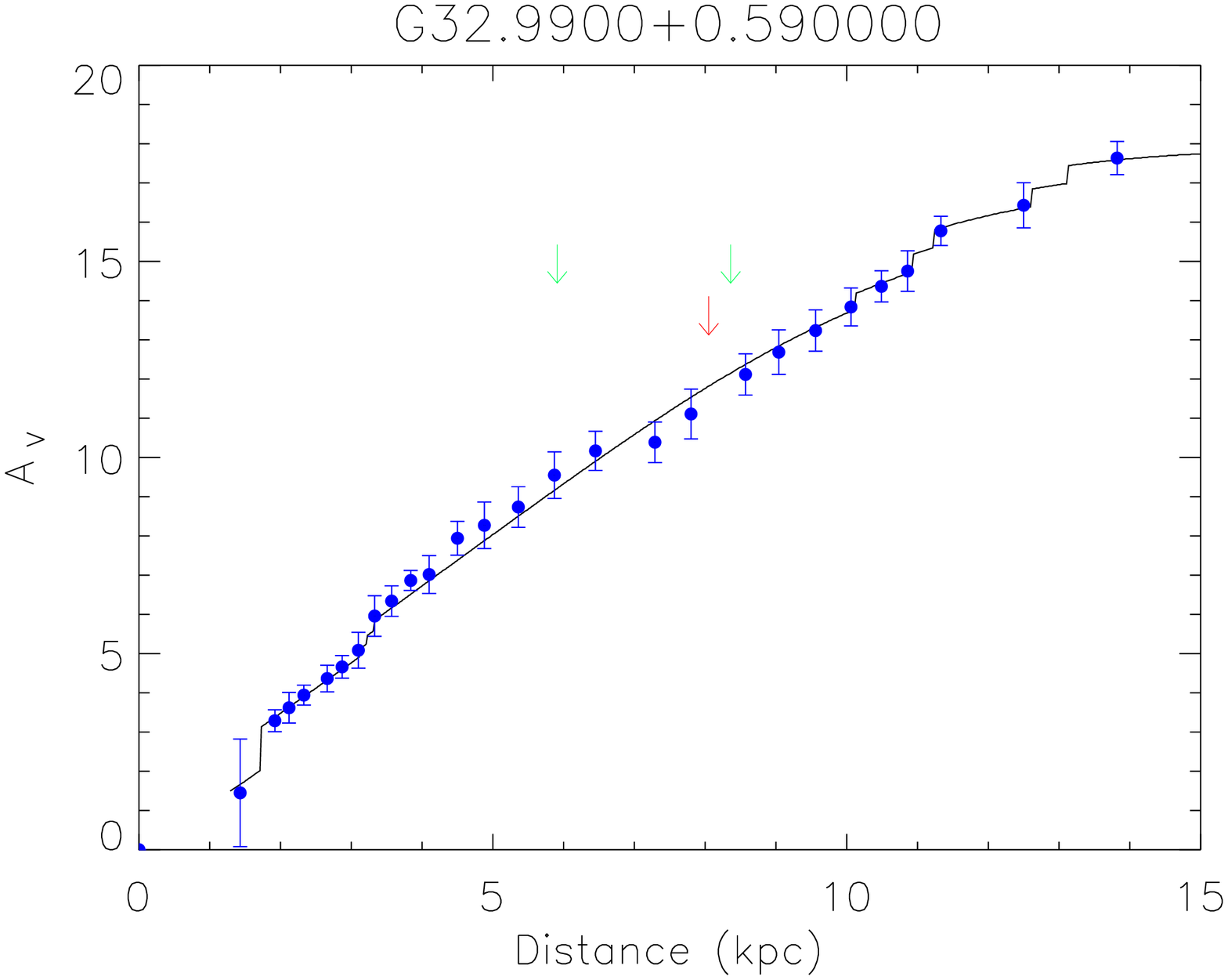}}\\
      \resizebox{80mm}{!}{\includegraphics[angle=90]{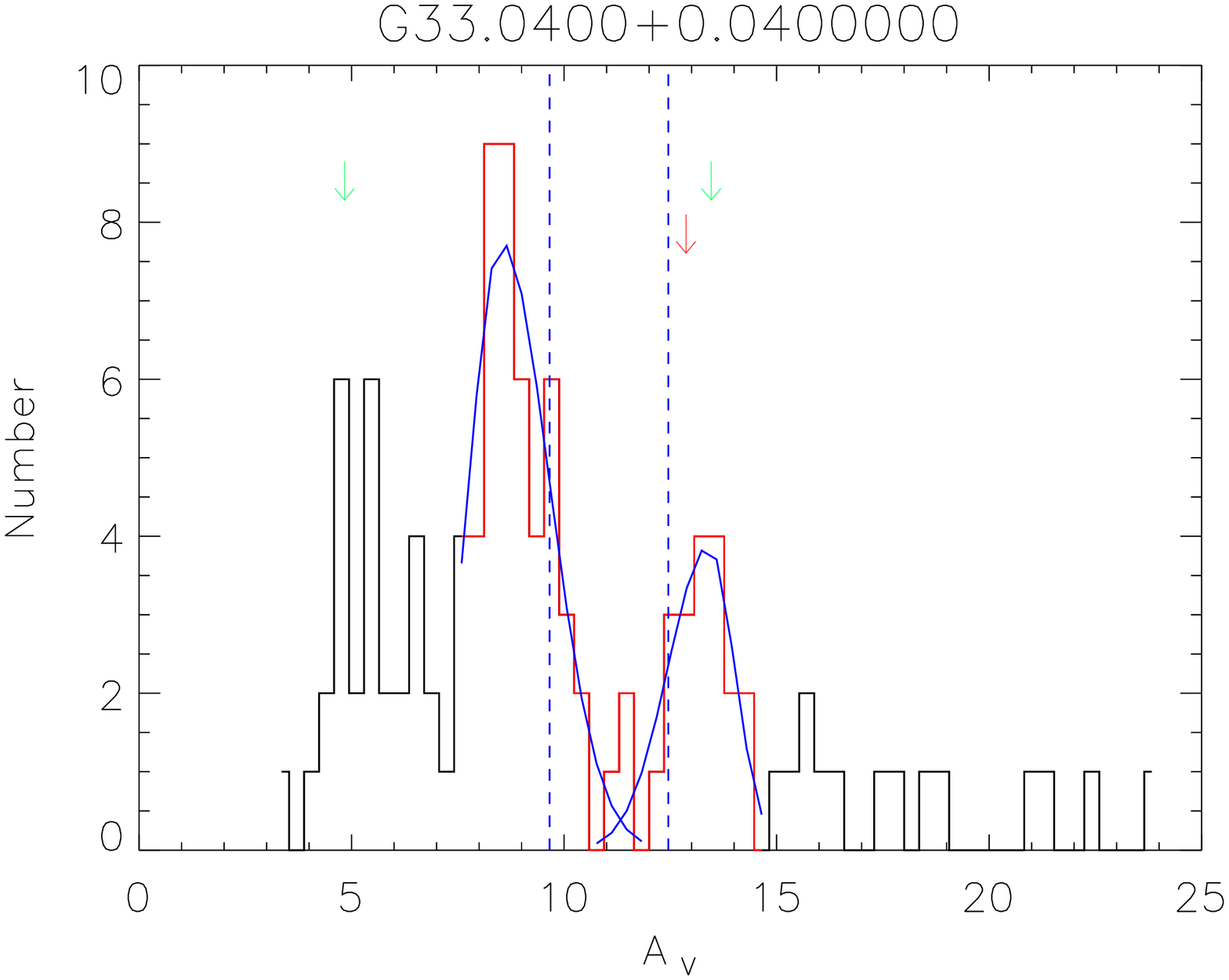}} &
      \resizebox{80mm}{!}{\includegraphics[angle=90]{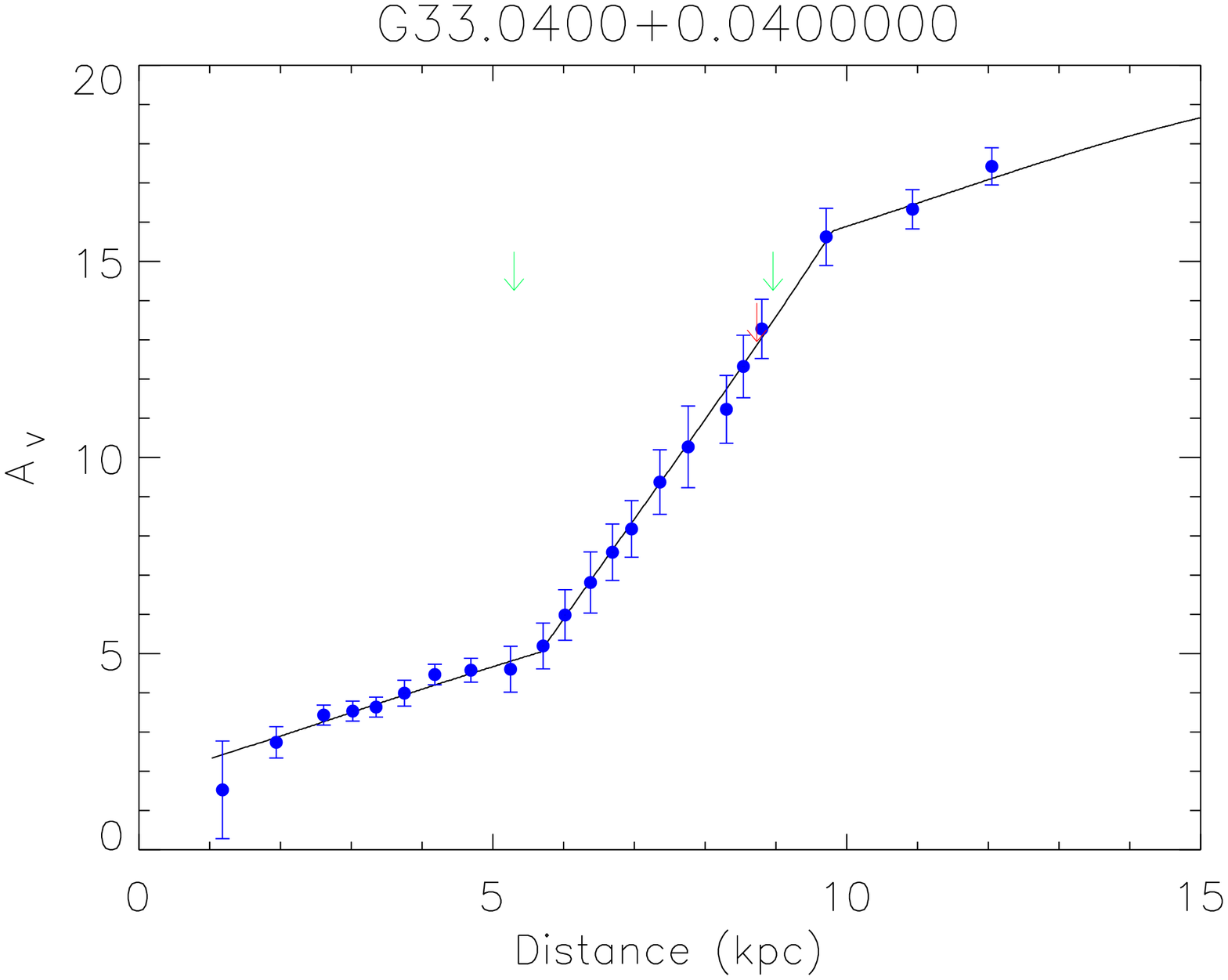}}\\
      \resizebox{80mm}{!}{\includegraphics[angle=90]{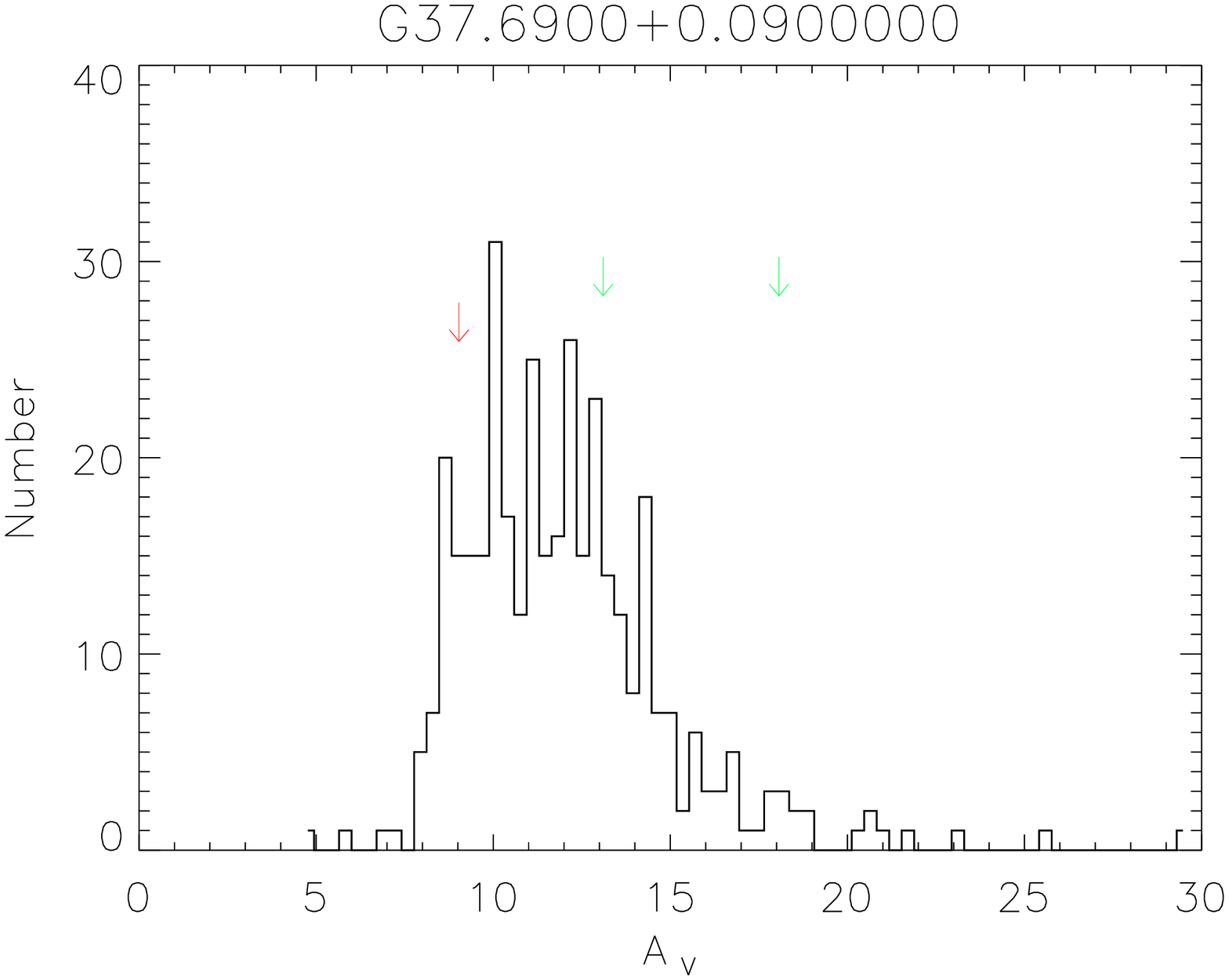}} &
      \resizebox{80mm}{!}{\includegraphics[angle=90]{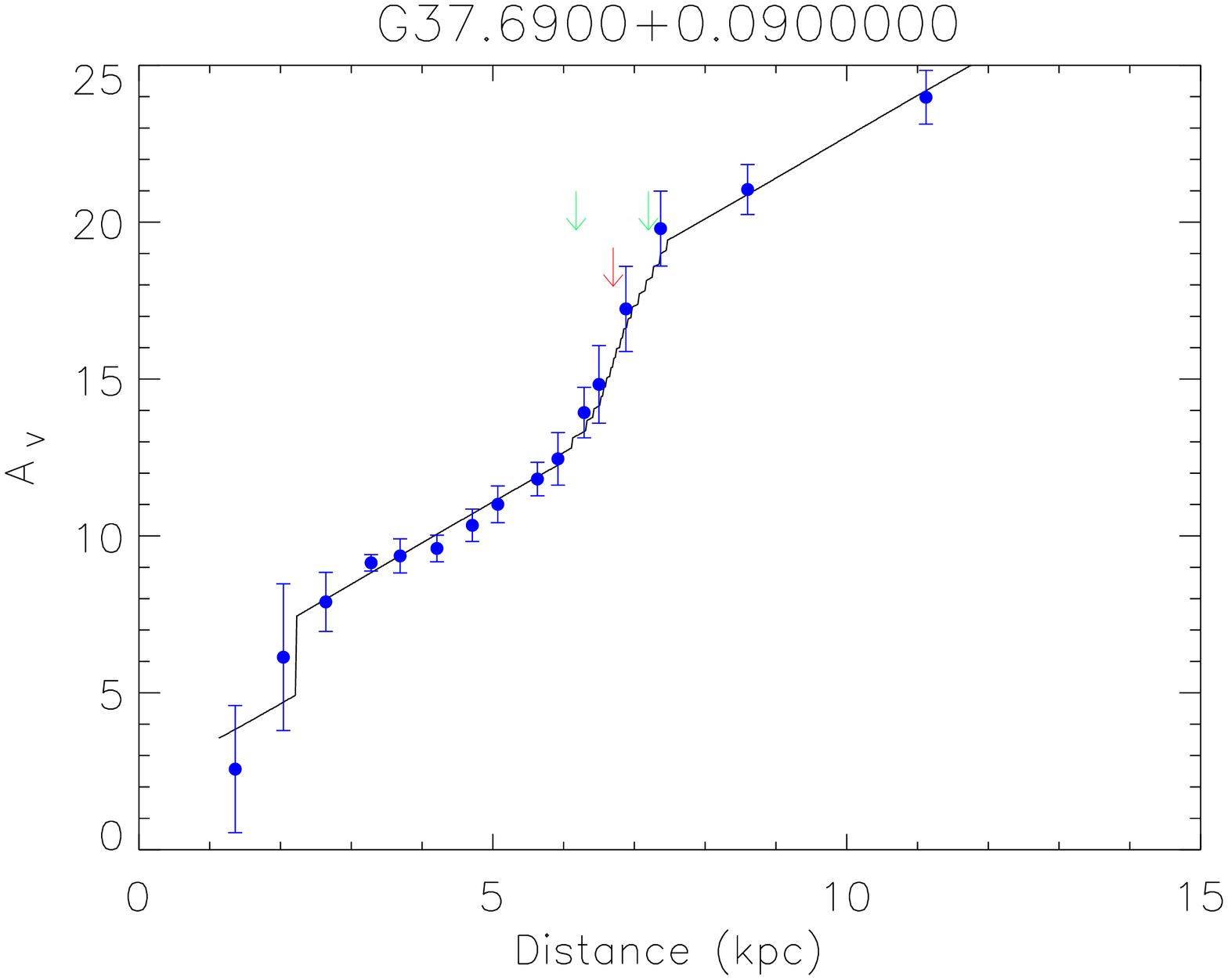}}\\
      \resizebox{80mm}{!}{\includegraphics[angle=90]{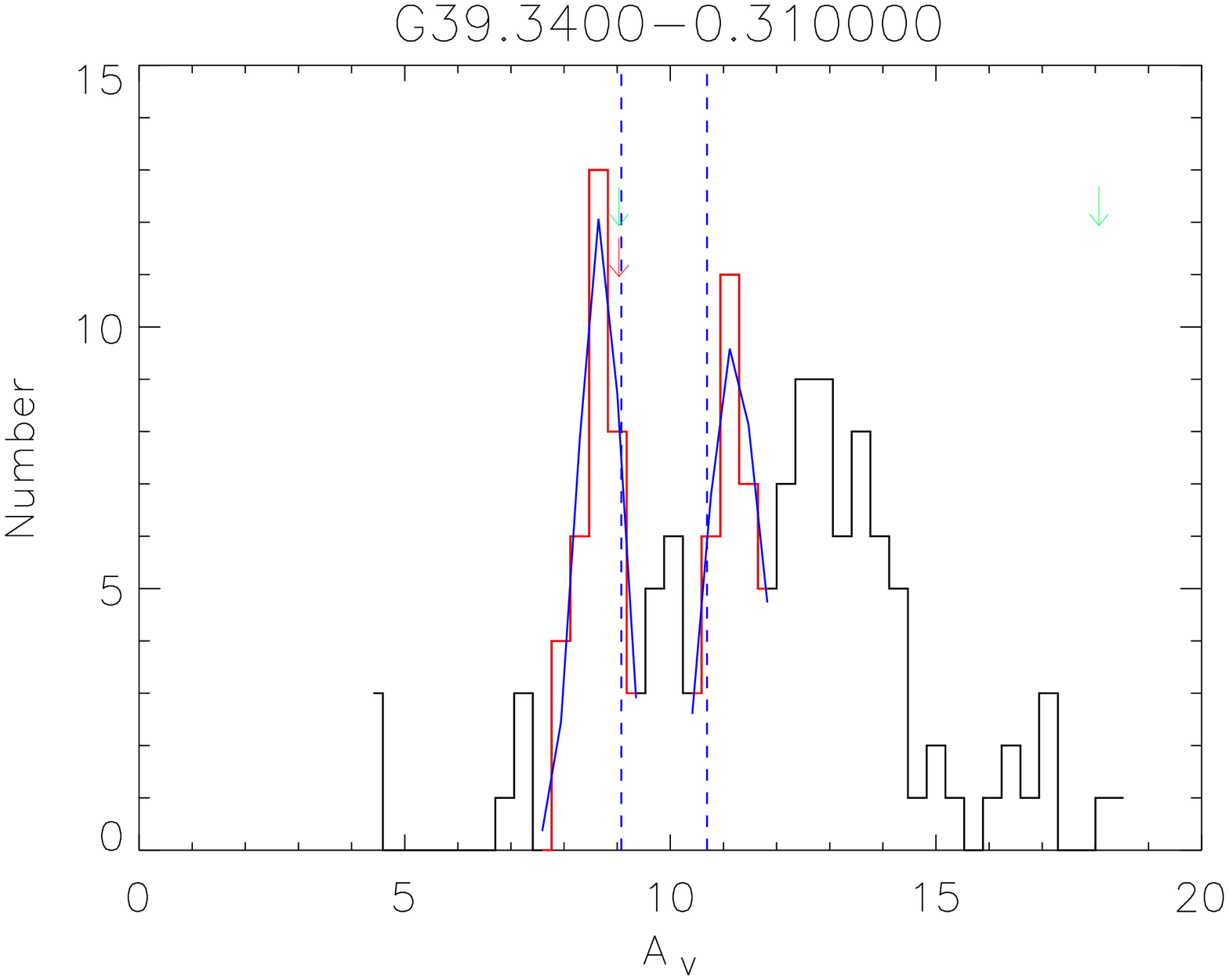}} &
      \resizebox{80mm}{!}{\includegraphics[angle=90]{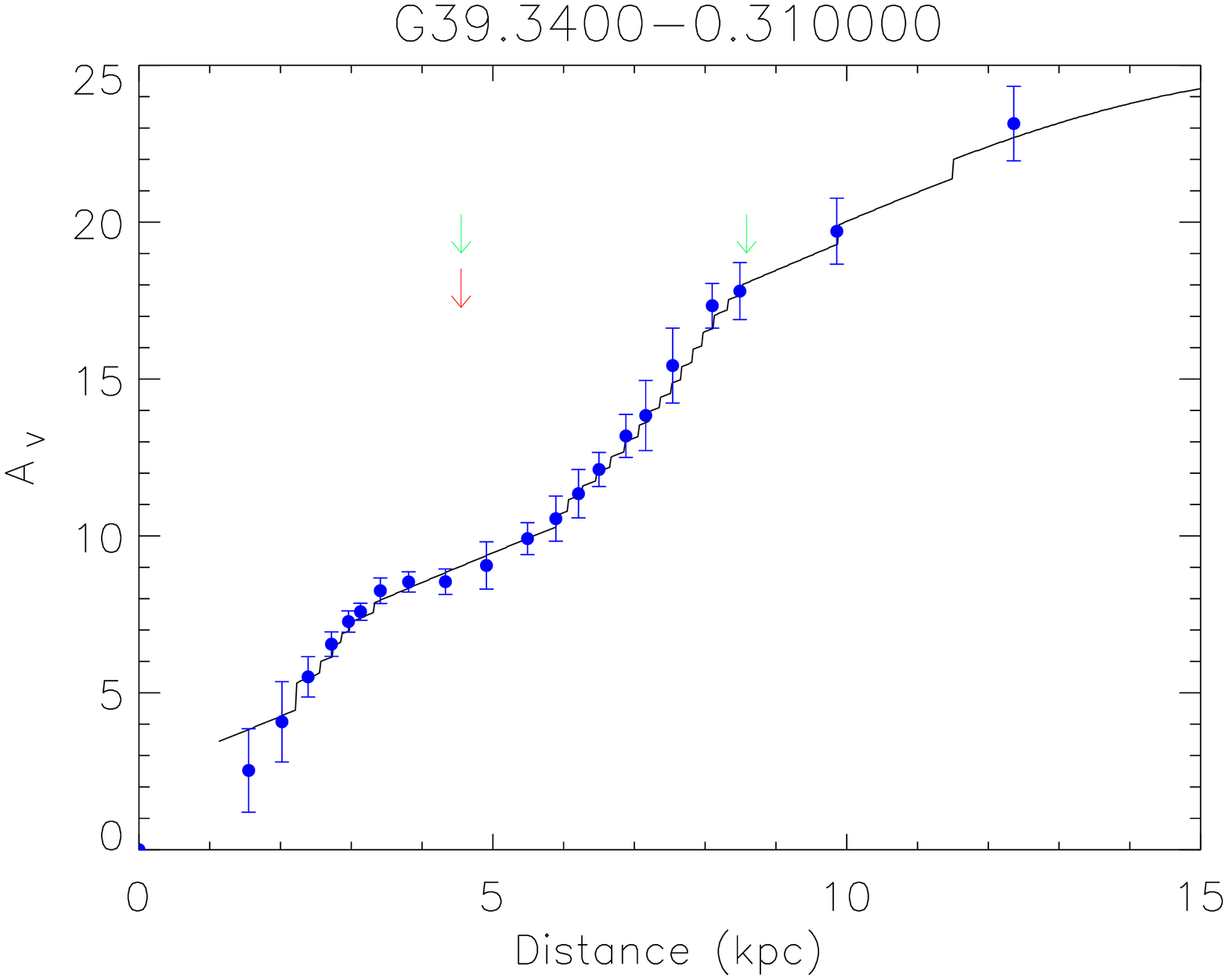}}
    \end{tabular} 
    \caption[]{\small }
    \label{fig:a3}
  \end{center}
\end{figure*}

The molecular cloud G033.04+0.04 has RMS distances of 5.3/9.0 kpc and a GRS distance of 8.7 kpc. The distance derived in this paper is D = 8.1$\pm$0.9. The second molecular cloud present in the A$_V$ histogram, at A$_V$$\sim$7 (Fig. \ref{fig:a3}), has a derived distance D = 6.3$\pm$0.8. This excludes the near RMS distance at the 1$\sigma$ level. There are four other molecular clouds nearby, all with GRS distances of 7.1 kpc, which could account for the presence of a second molecular cloud in the A$_V$ histogram.

\subsection{G043.89-0.81}

\begin{figure*}
  \begin{center}
    \begin{tabular}{cc}
      \resizebox{80mm}{!}{\includegraphics[angle=90]{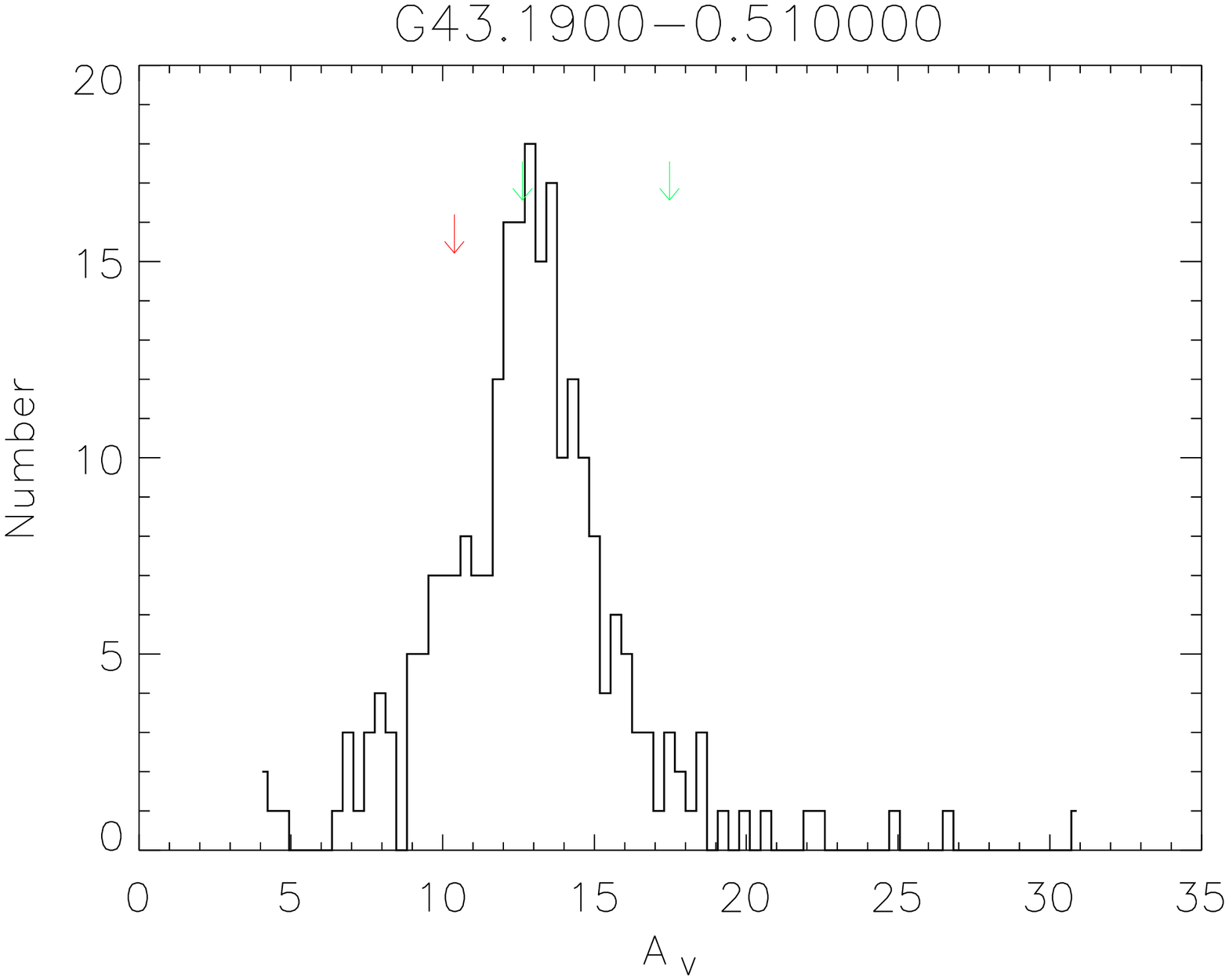}} &
      \resizebox{80mm}{!}{\includegraphics[angle=90]{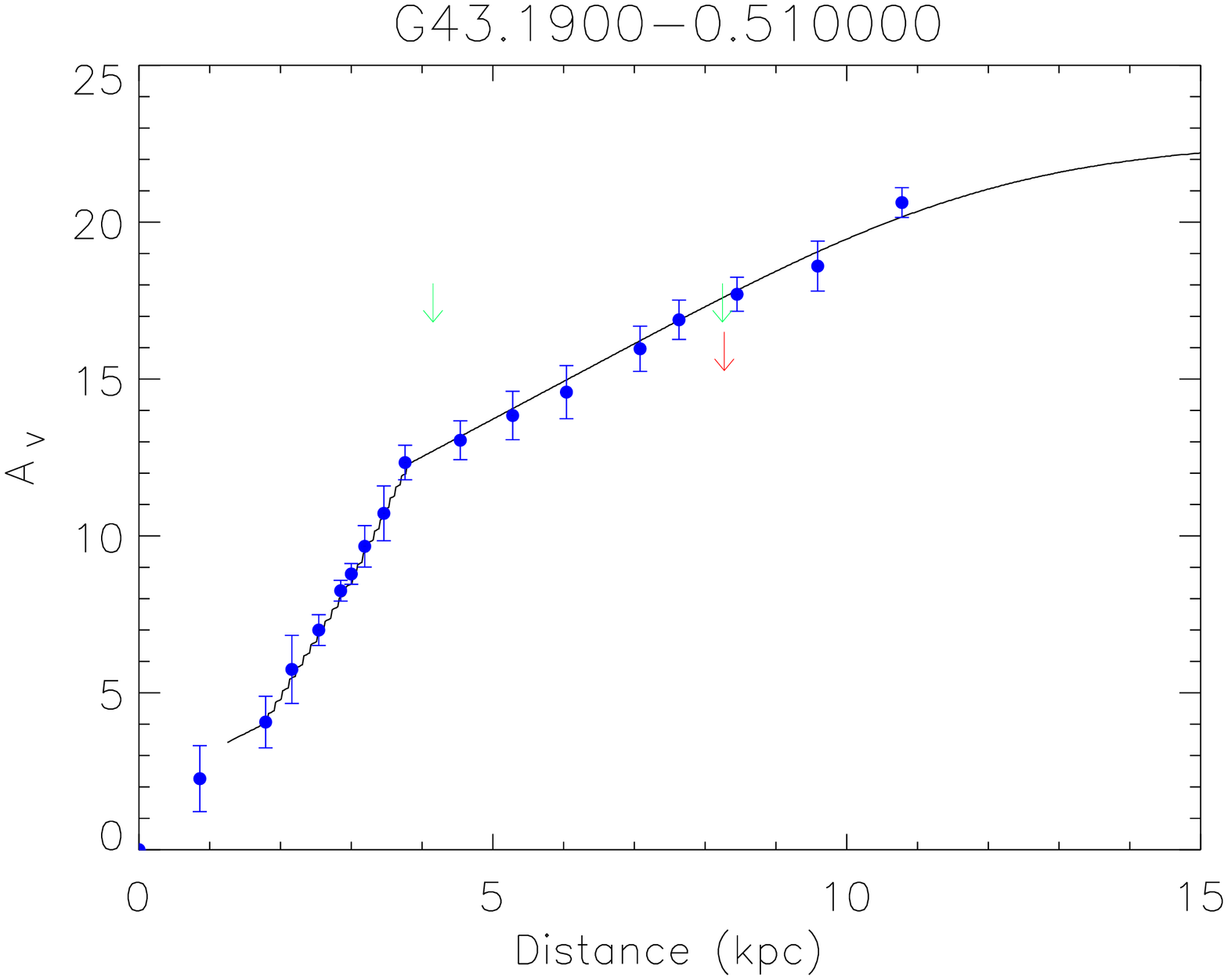}}\\
      \resizebox{80mm}{!}{\includegraphics[angle=90]{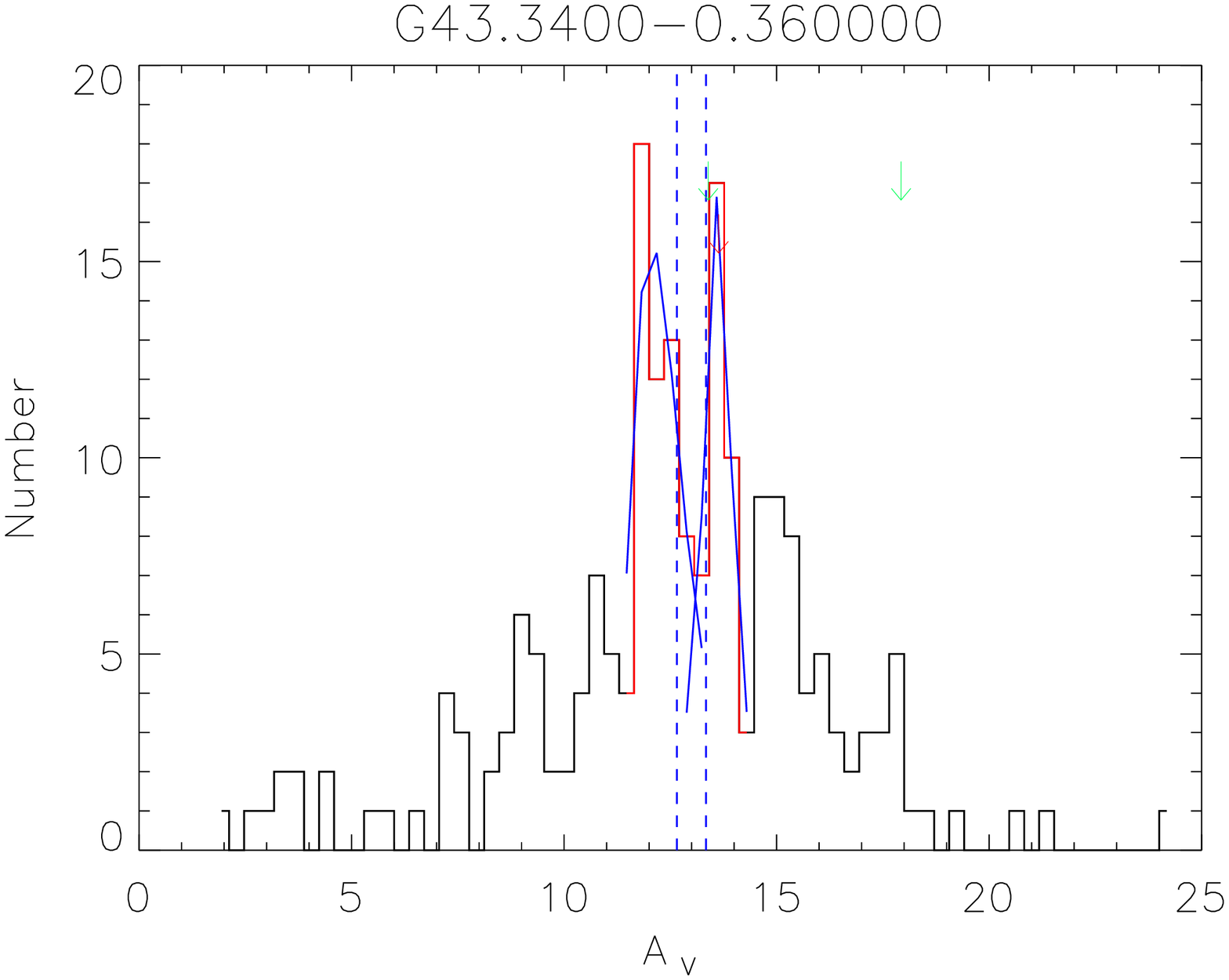}} &
      \resizebox{80mm}{!}{\includegraphics[angle=90]{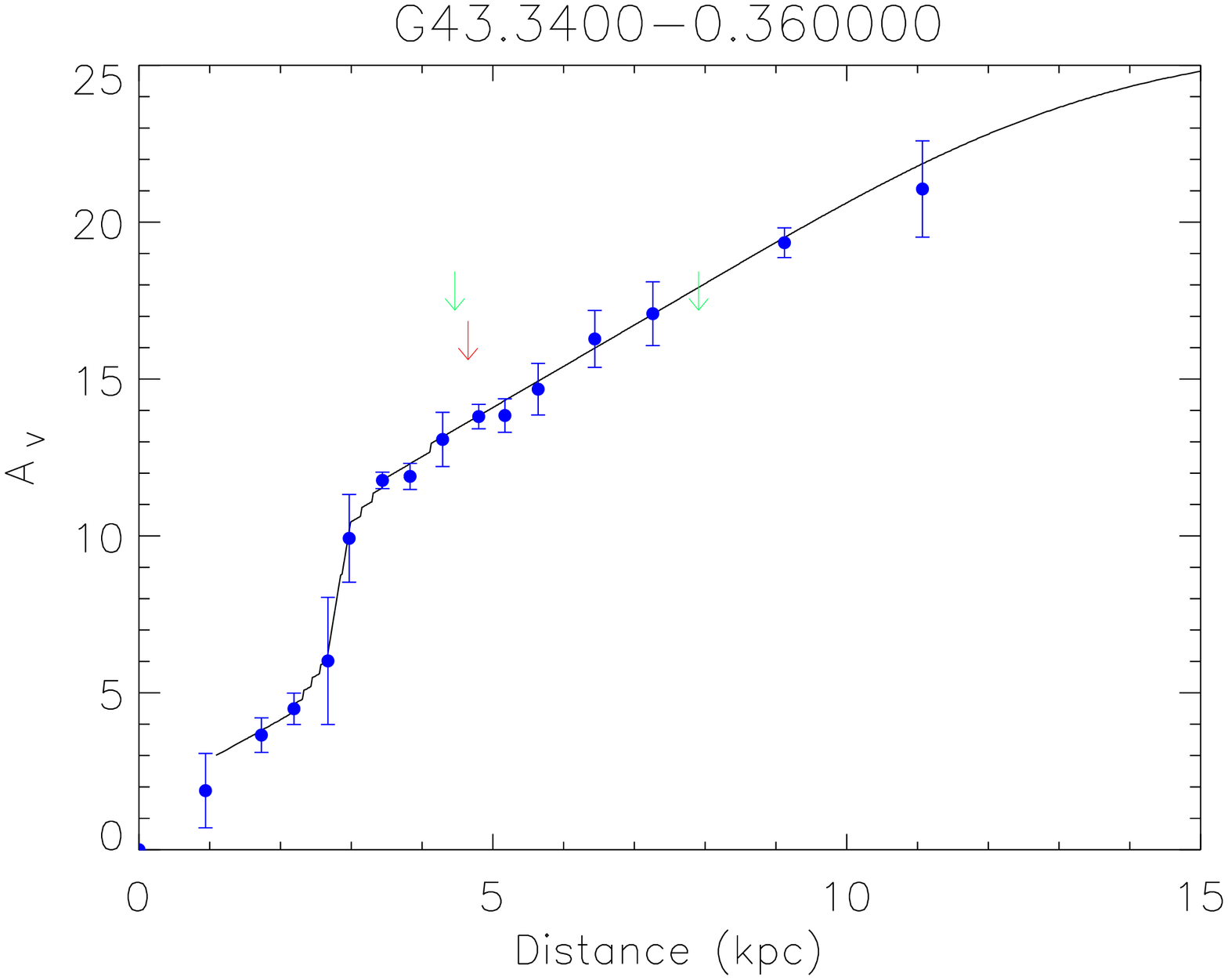}}\\
      \resizebox{80mm}{!}{\includegraphics[angle=90]{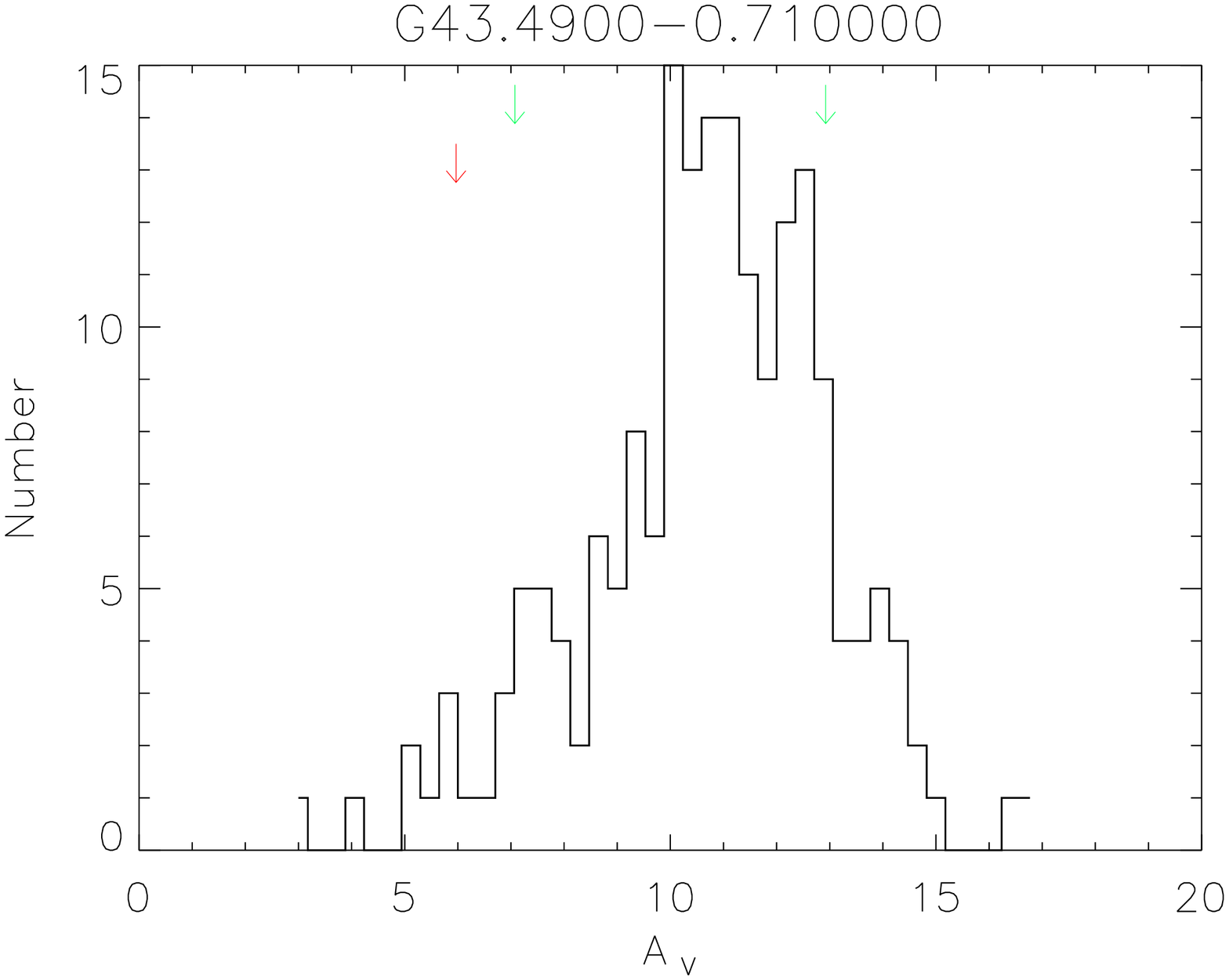}} &
      \resizebox{80mm}{!}{\includegraphics[angle=90]{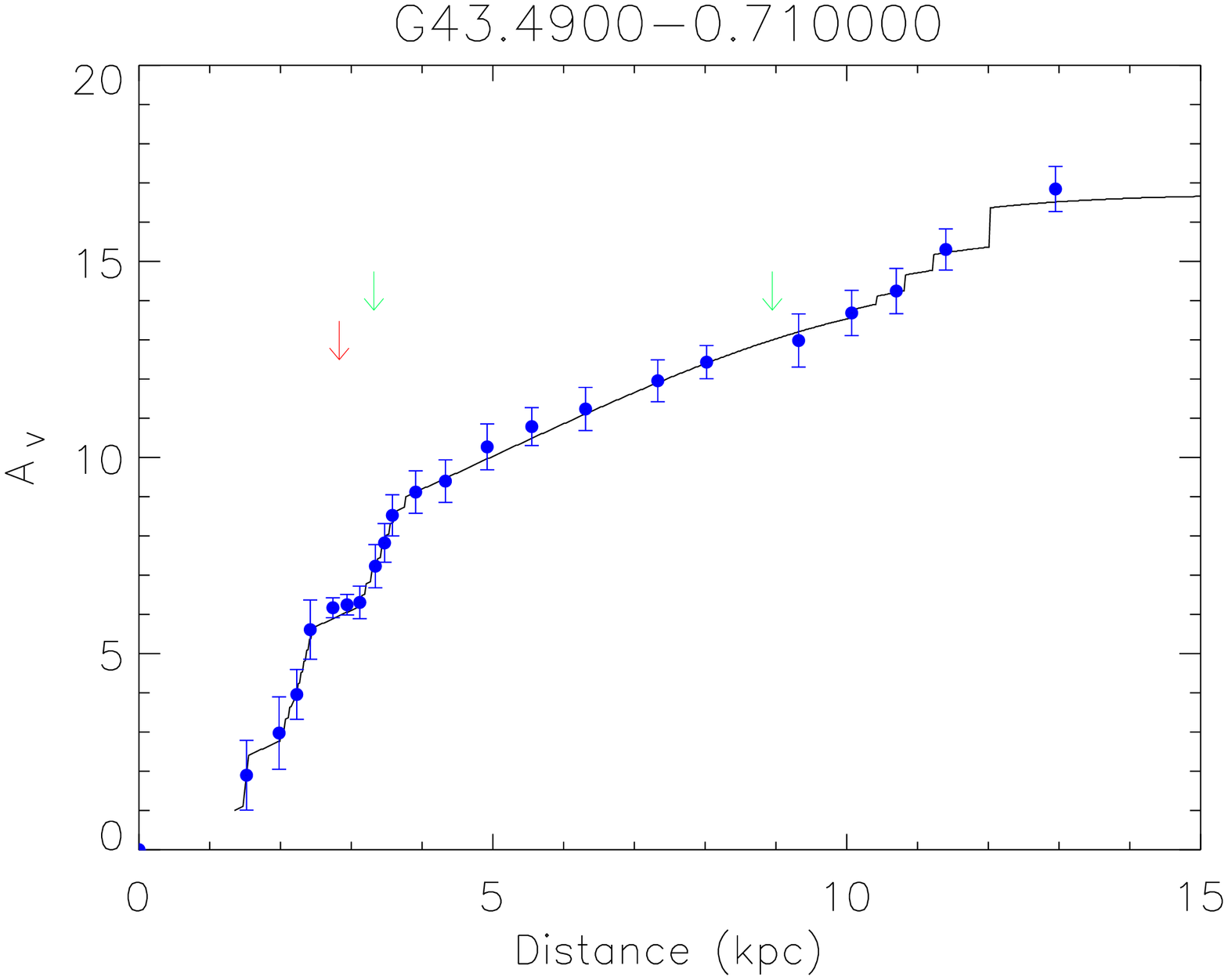}}\\
      \resizebox{80mm}{!}{\includegraphics[angle=90]{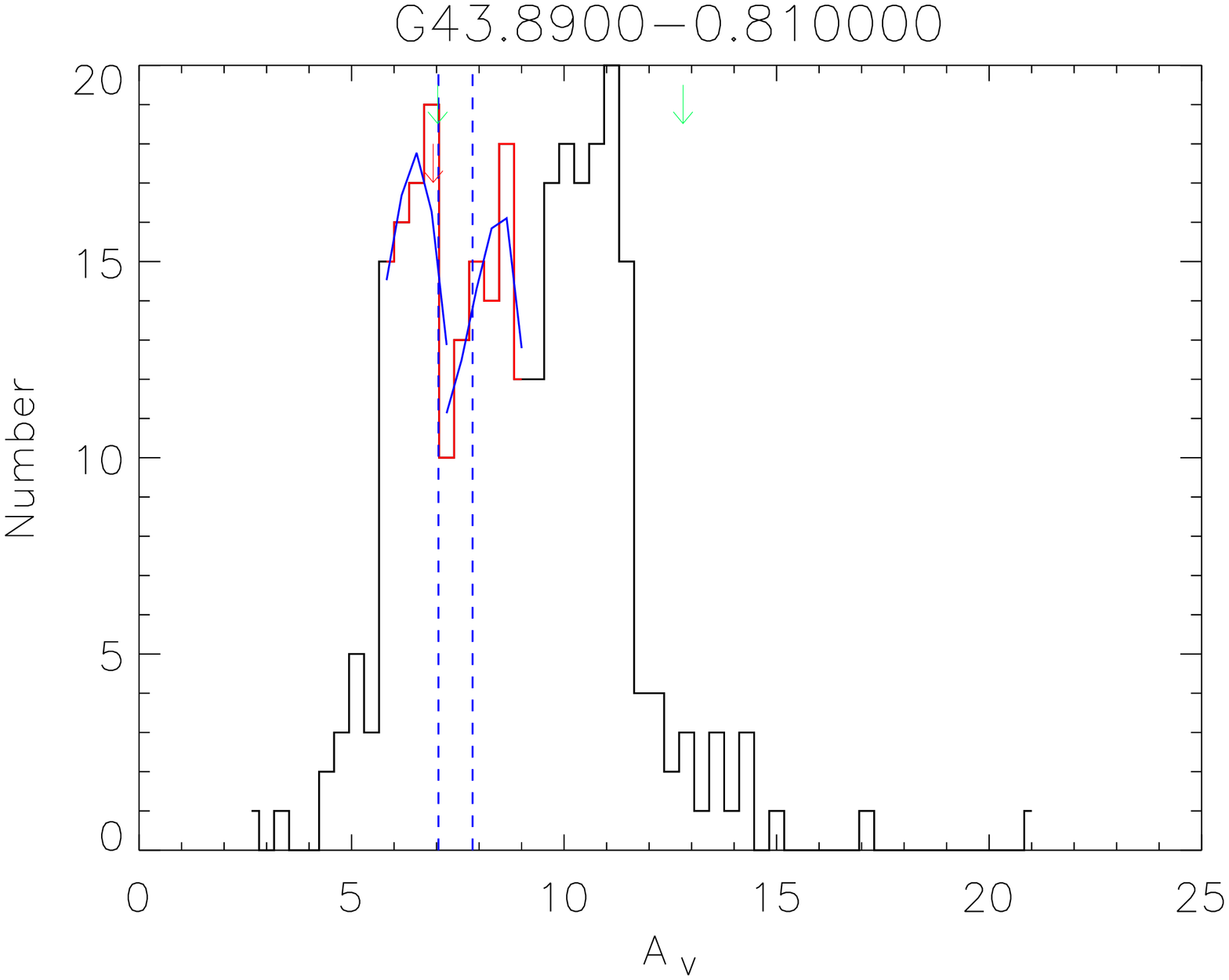}} &
      \resizebox{80mm}{!}{\includegraphics[angle=90]{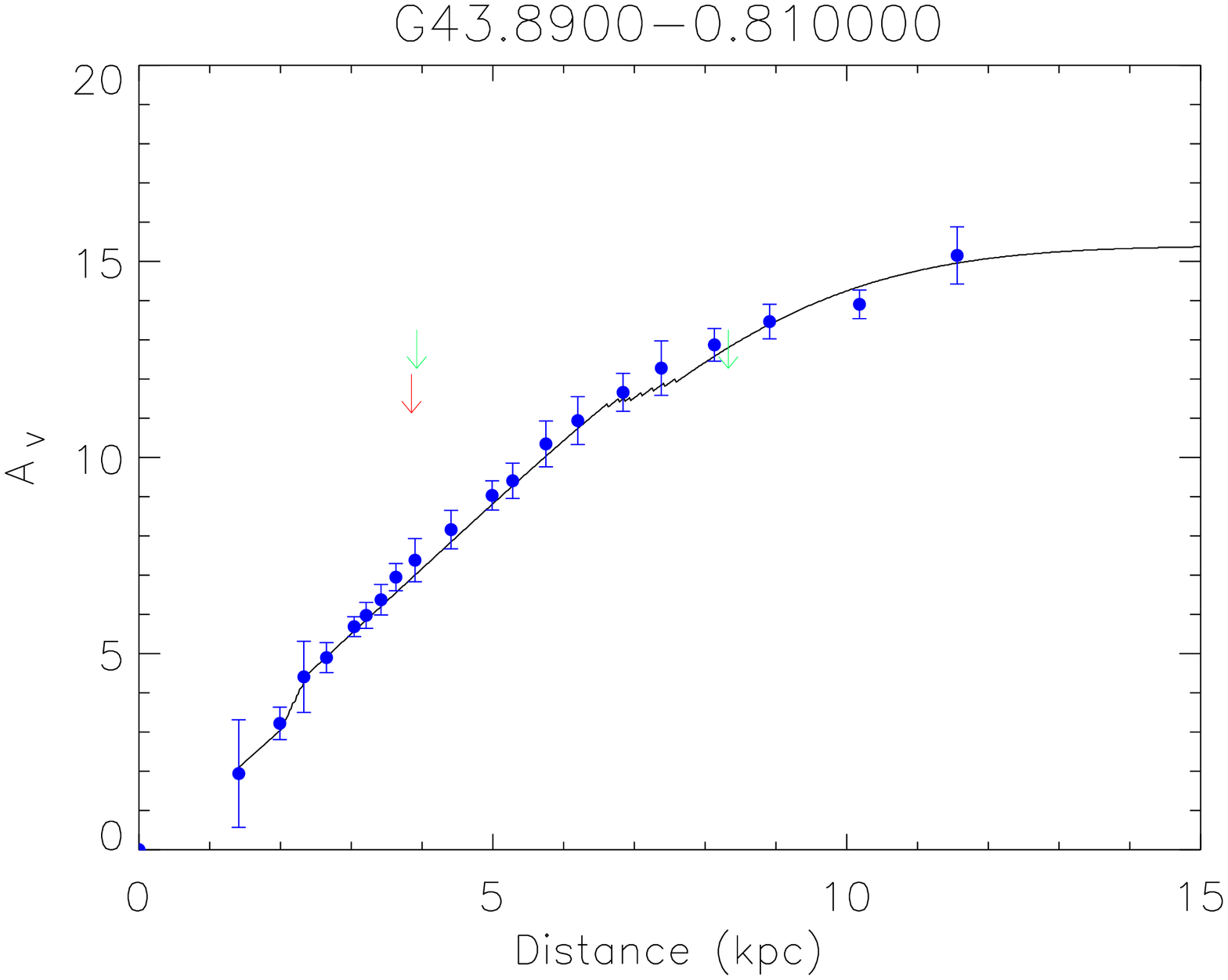}}
    \end{tabular} 
    \caption[]{\small }
    \label{fig:a5}
  \end{center}
\end{figure*}

The molecular cloud G043.89-0.81 has RMS distances of 3.9/8.3 kpc and a GRS distance of 3.9 kpc. The distance derived in this paper is D = 4.4$\pm$0.6. Fig. \ref{fig:a5} appears to contain a second molecular cloud at A$_V$$\sim$9. The derived distance to this second cloud is D = 5.3$\pm$0.6, therefore excluding the far RMS distance. \\
$\indent$The extracted region overlaps with a second molecular cloud G044.34-00.81 that has a GRS distance of 4.73 kpc, therefore providing a possible explanation for the presence of the second cloud.

\subsection{G052.79+0.29}

\begin{figure*}
  \begin{center}
    \begin{tabular}{cc}
      \resizebox{80mm}{!}{\includegraphics[angle=90]{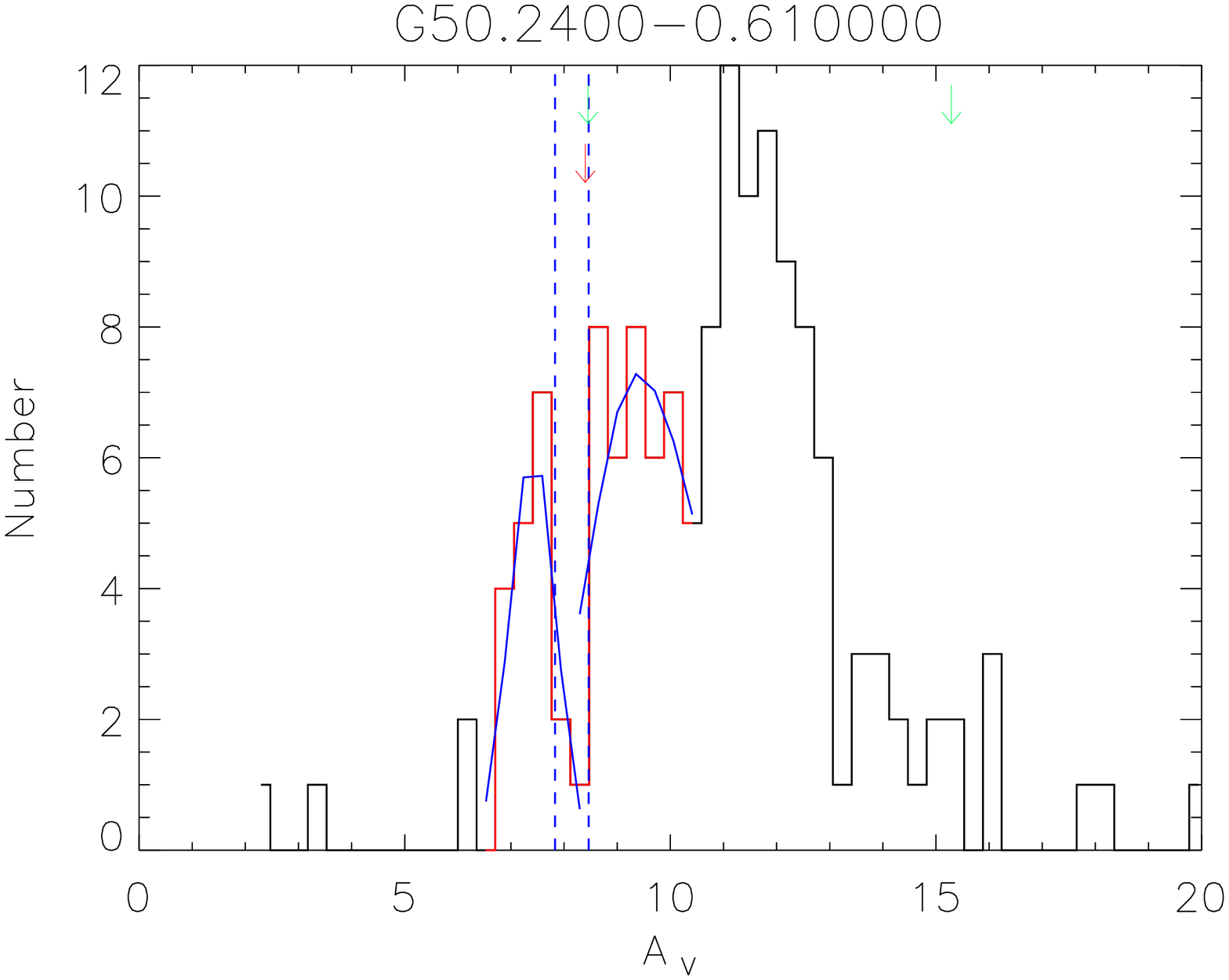}} &
      \resizebox{80mm}{!}{\includegraphics[angle=90]{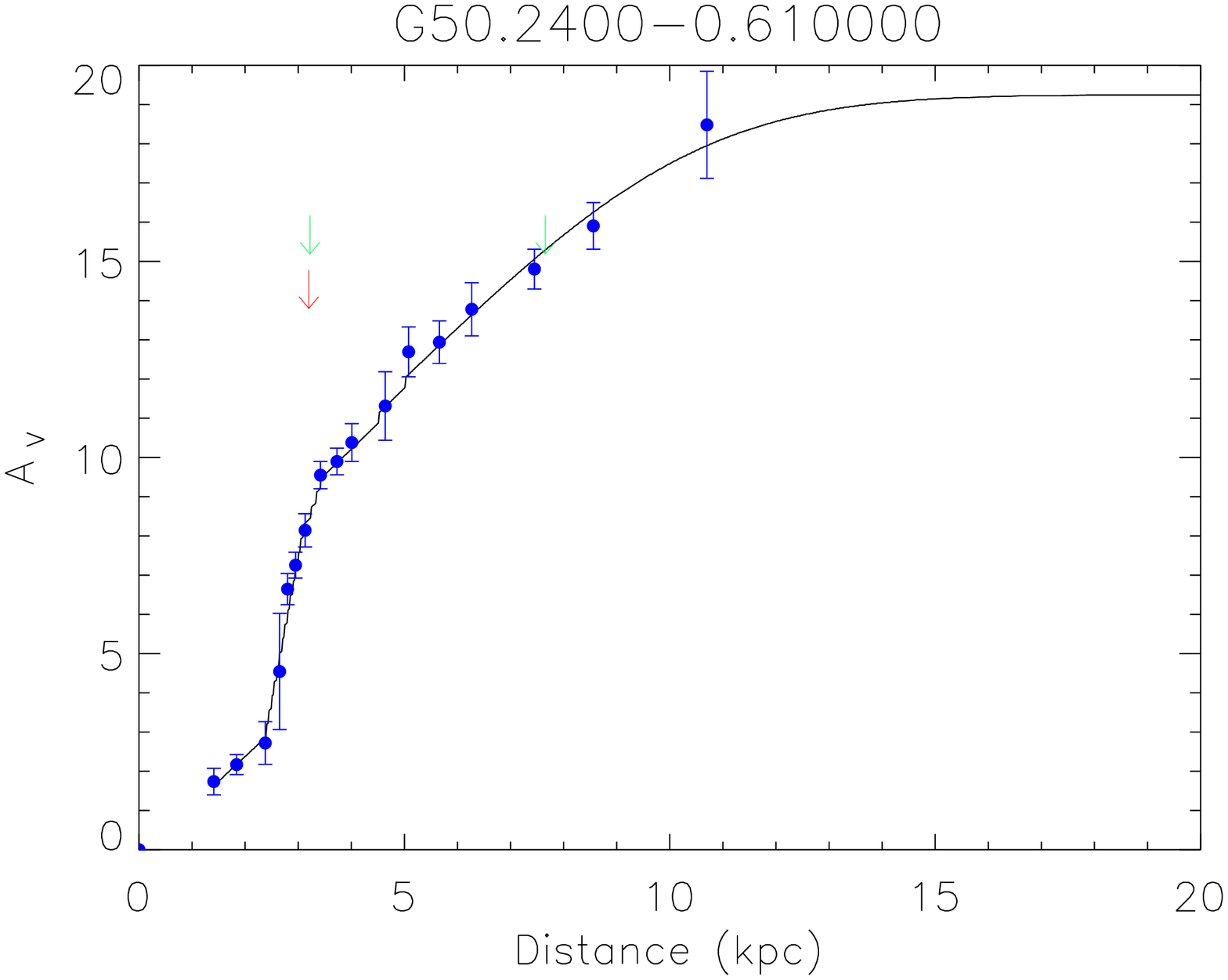}}\\
      \resizebox{80mm}{!}{\includegraphics[angle=90]{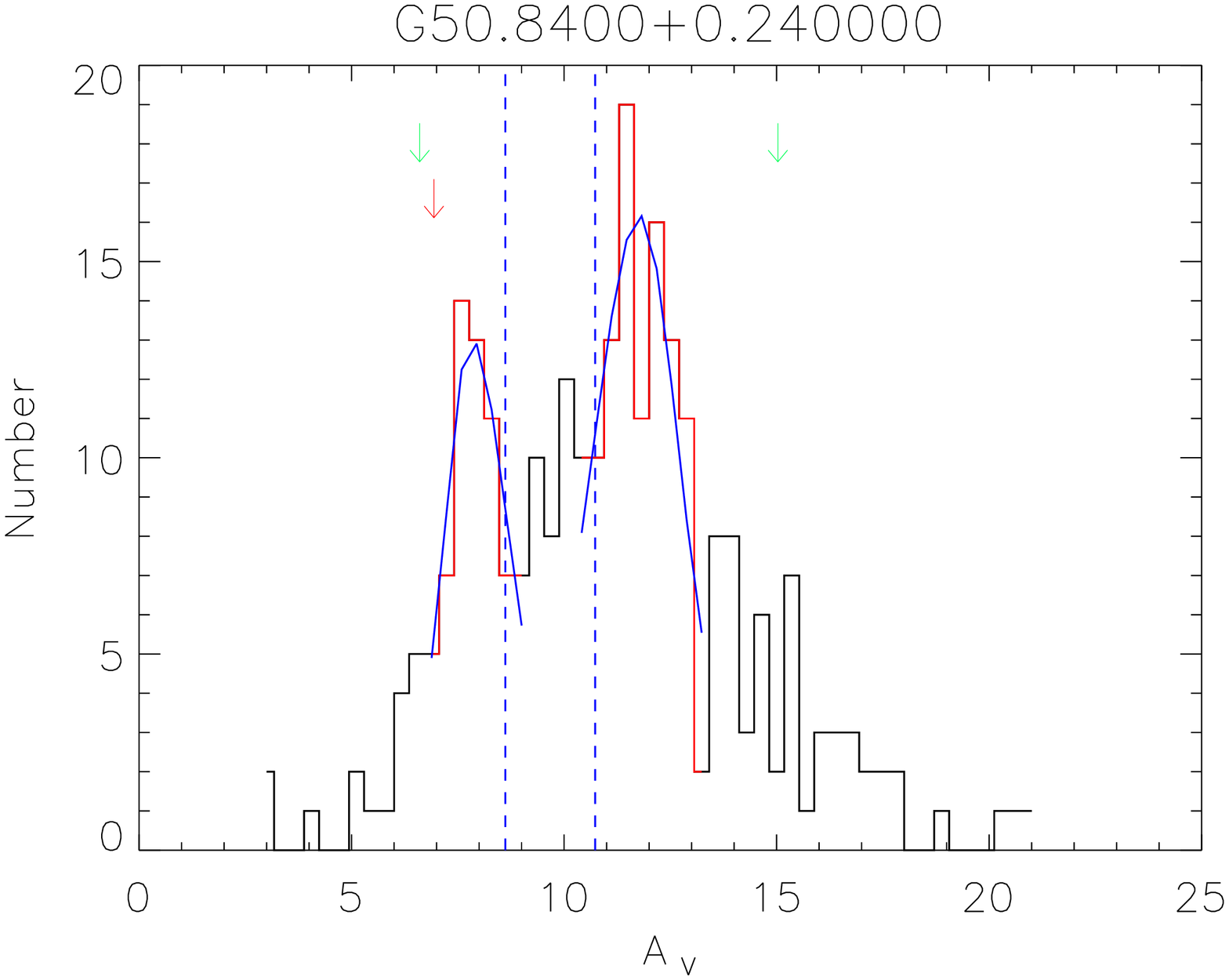}} &
      \resizebox{80mm}{!}{\includegraphics[angle=90]{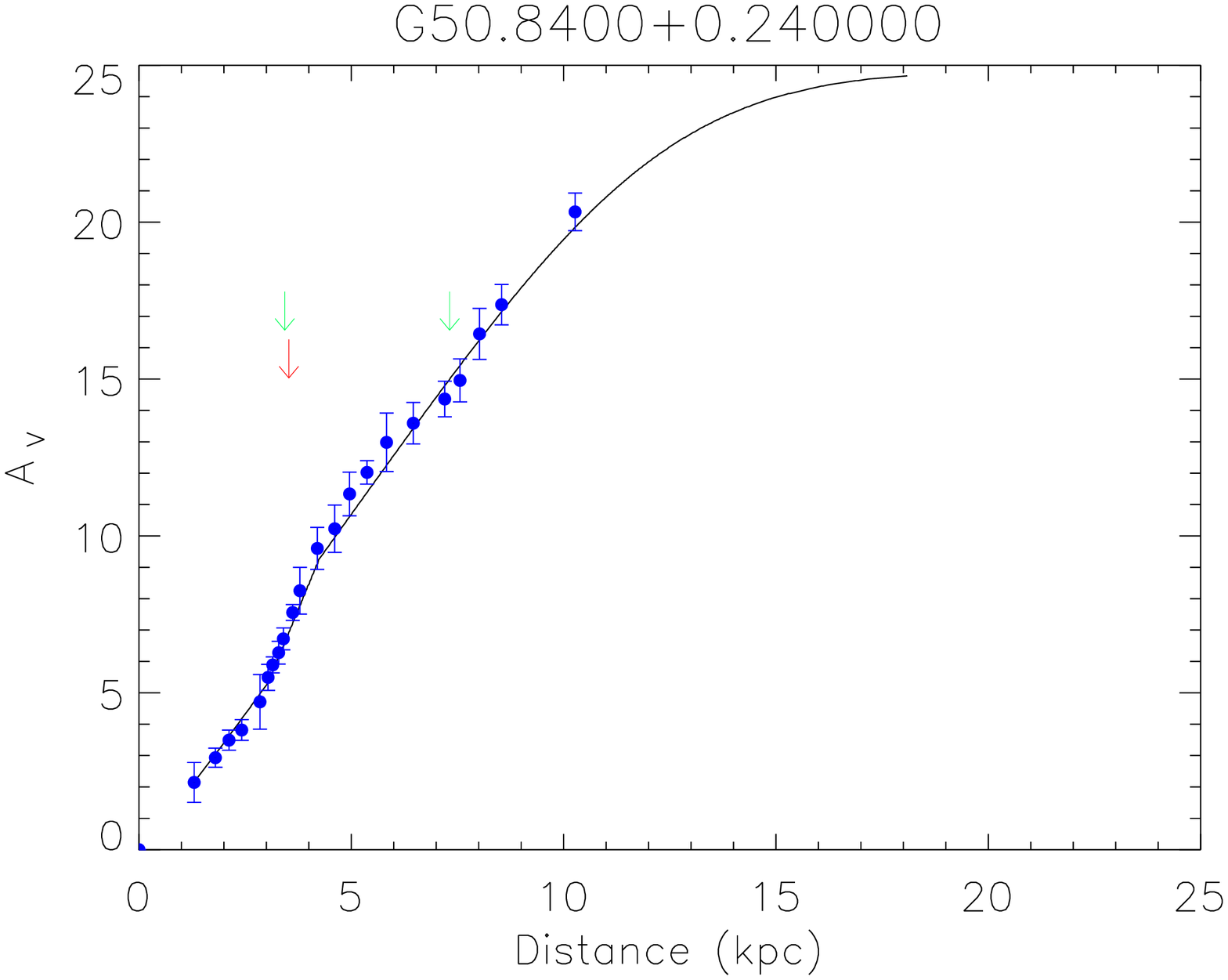}}\\
      \resizebox{80mm}{!}{\includegraphics[angle=90]{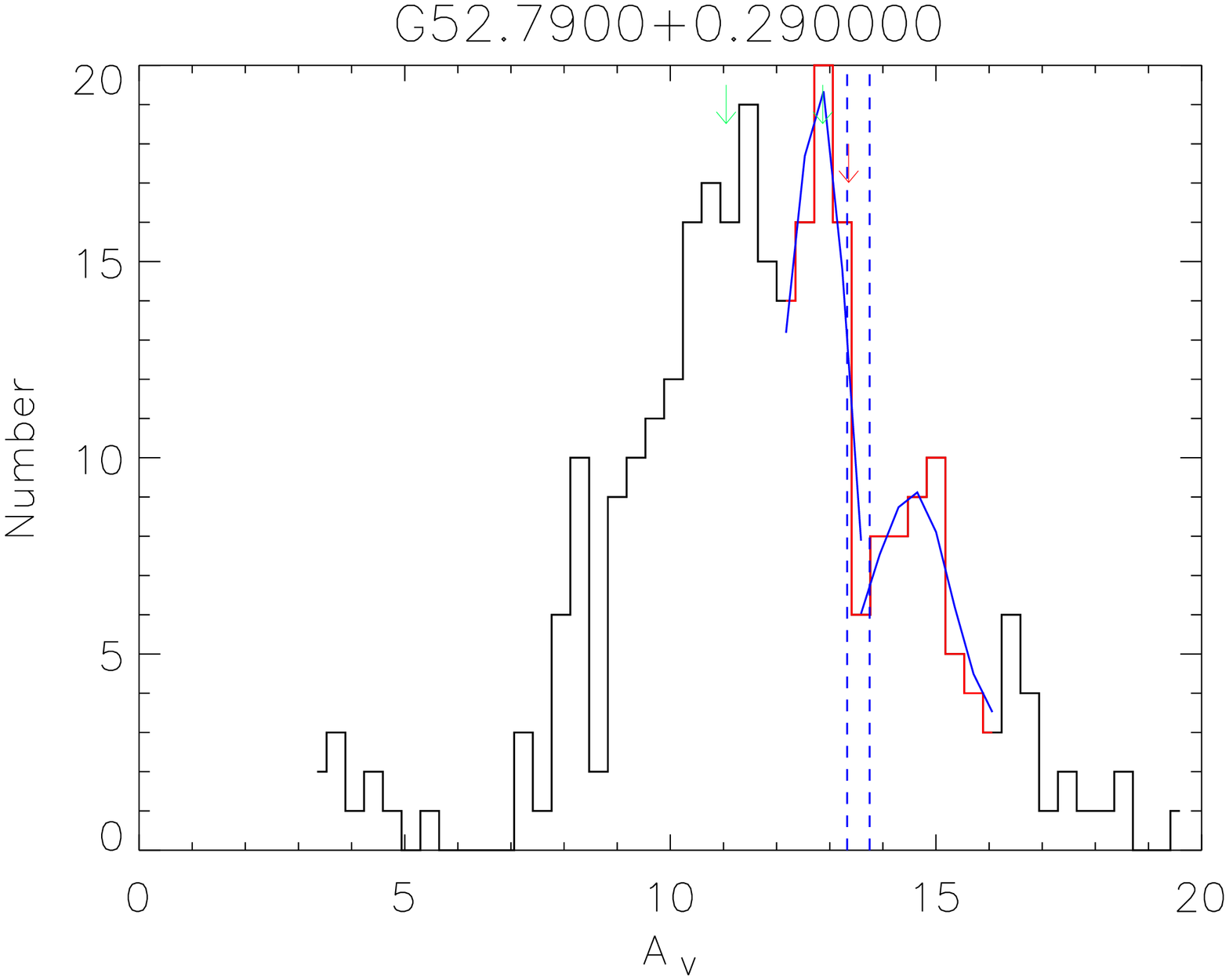}} &
      \resizebox{80mm}{!}{\includegraphics[angle=90]{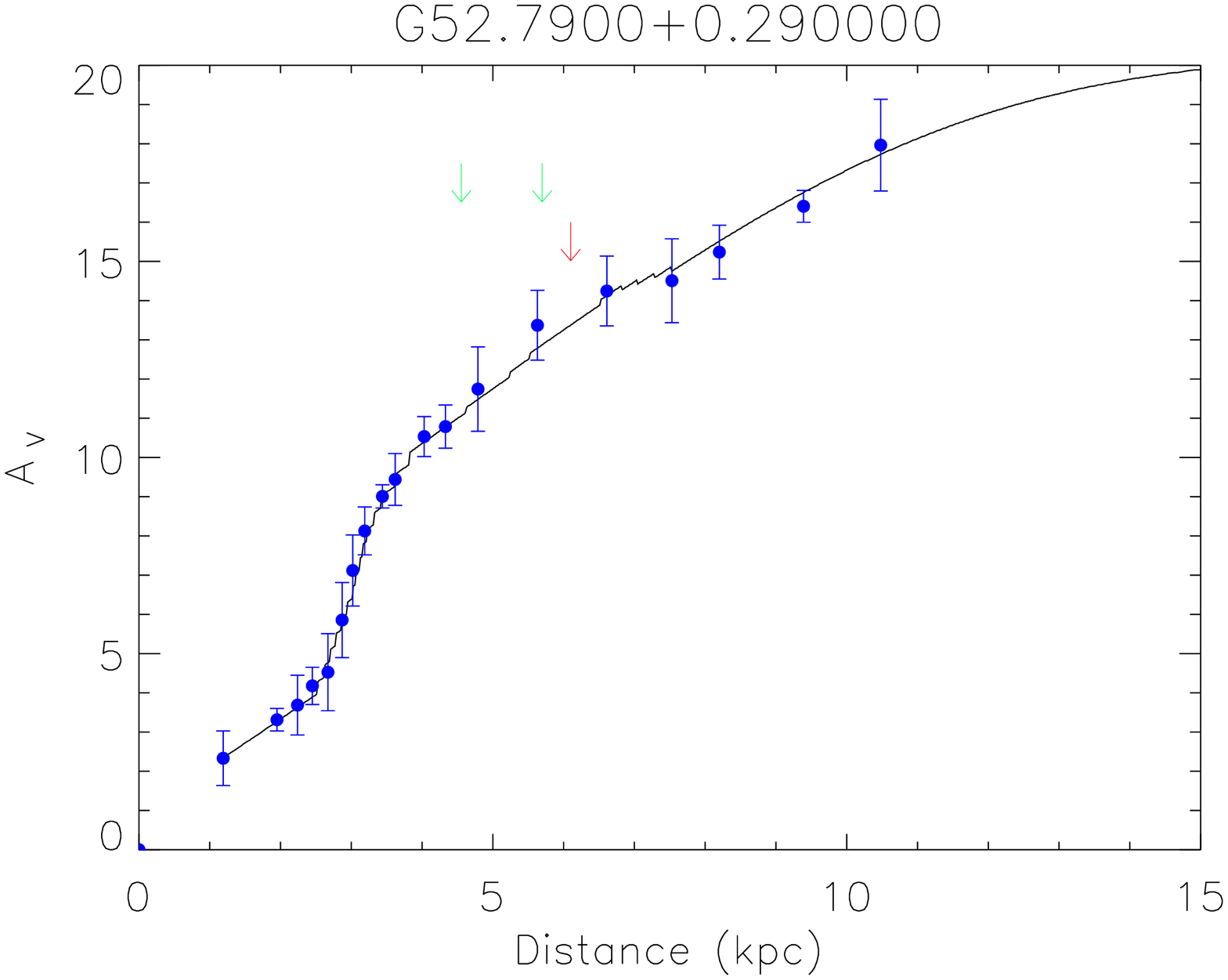}}\\
      \resizebox{80mm}{!}{\includegraphics[angle=90]{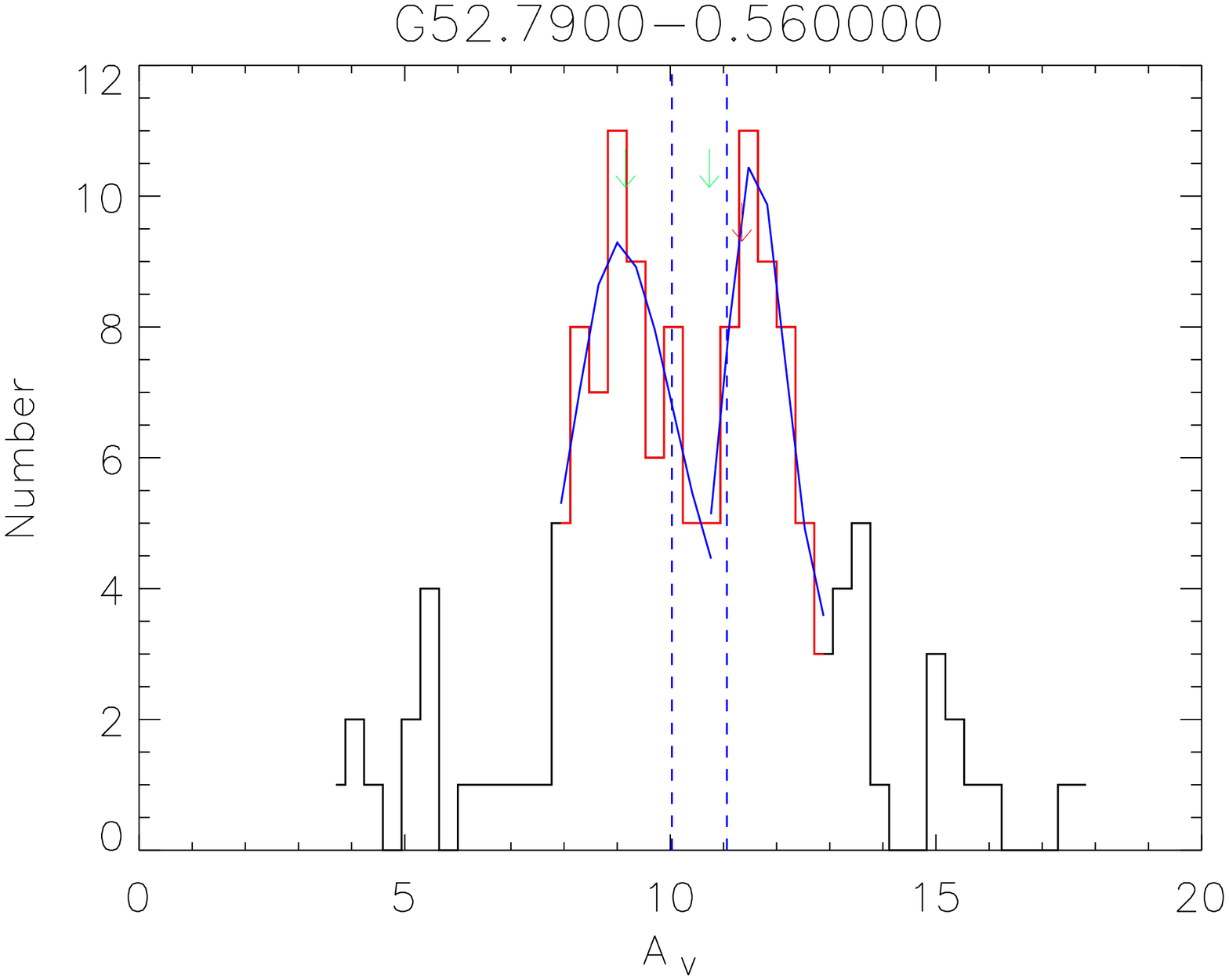}} &
      \resizebox{80mm}{!}{\includegraphics[angle=90]{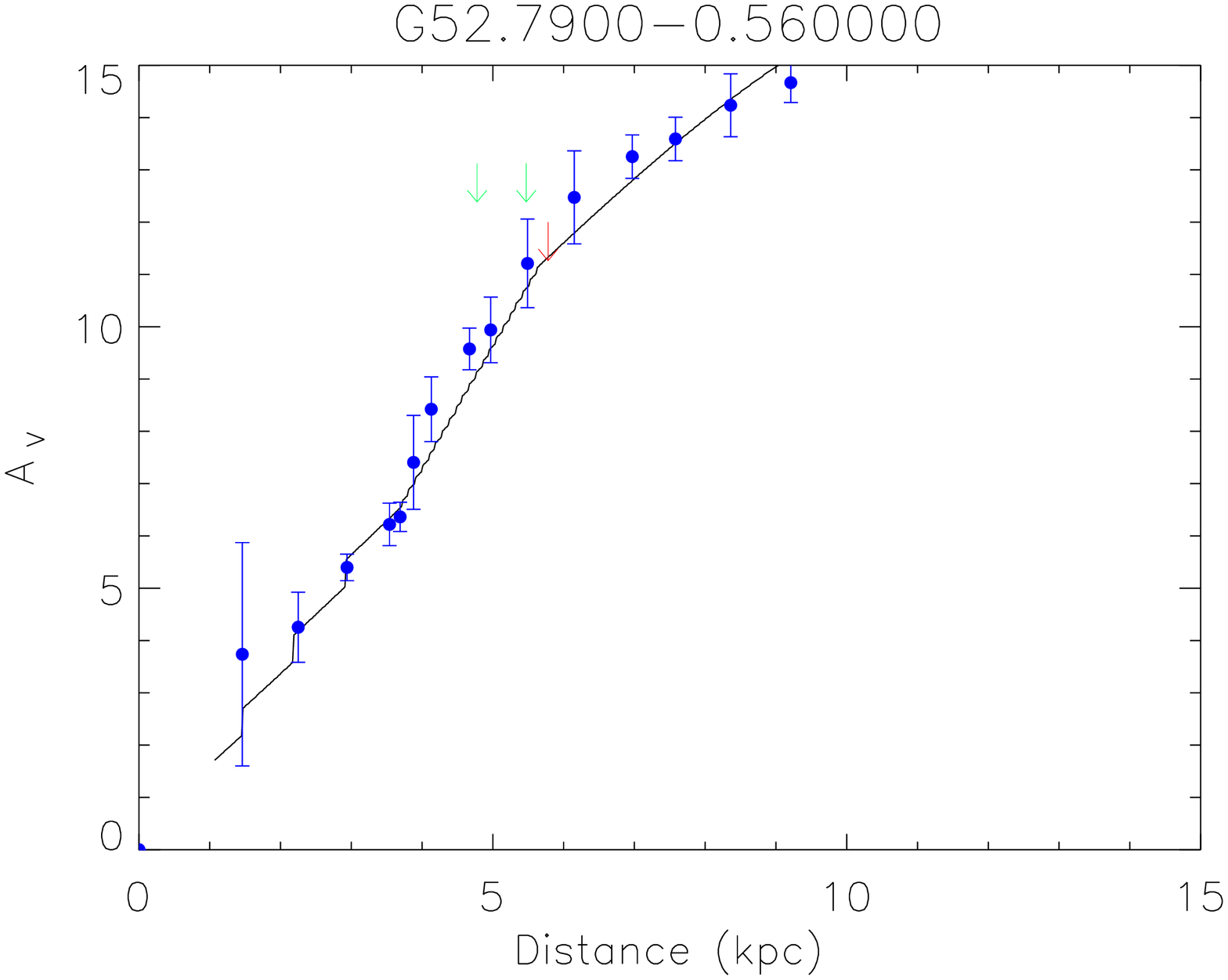}}
    \end{tabular} 
    \caption[]{\small }
    \label{fig:a7}
  \end{center}
\end{figure*}

The molecular cloud, G052.79+0.29, has RMS distances of 4.6/5.7 kpc and a GRS distance of 6.1 kpc. The distance derived in this paper is D = 6.9$\pm$1.1 kpc. Fig. \ref{fig:a7} appears to contain a second molecular cloud at A$_V$$\sim$12. The derived distance to this second cloud is D = 5.4$\pm$0.5, similar to the far RMS distance. There is however an overlap with a second cloud, G052.64+0.14, at a distance D = 5.1kpc providing a possible explanation for the presence of the second cloud.

\subsection{G028.59+0.04}

\begin{figure*}
\begin{center}
    \begin{tabular}{cc}
      \resizebox{80mm}{!}{\includegraphics[angle=90]{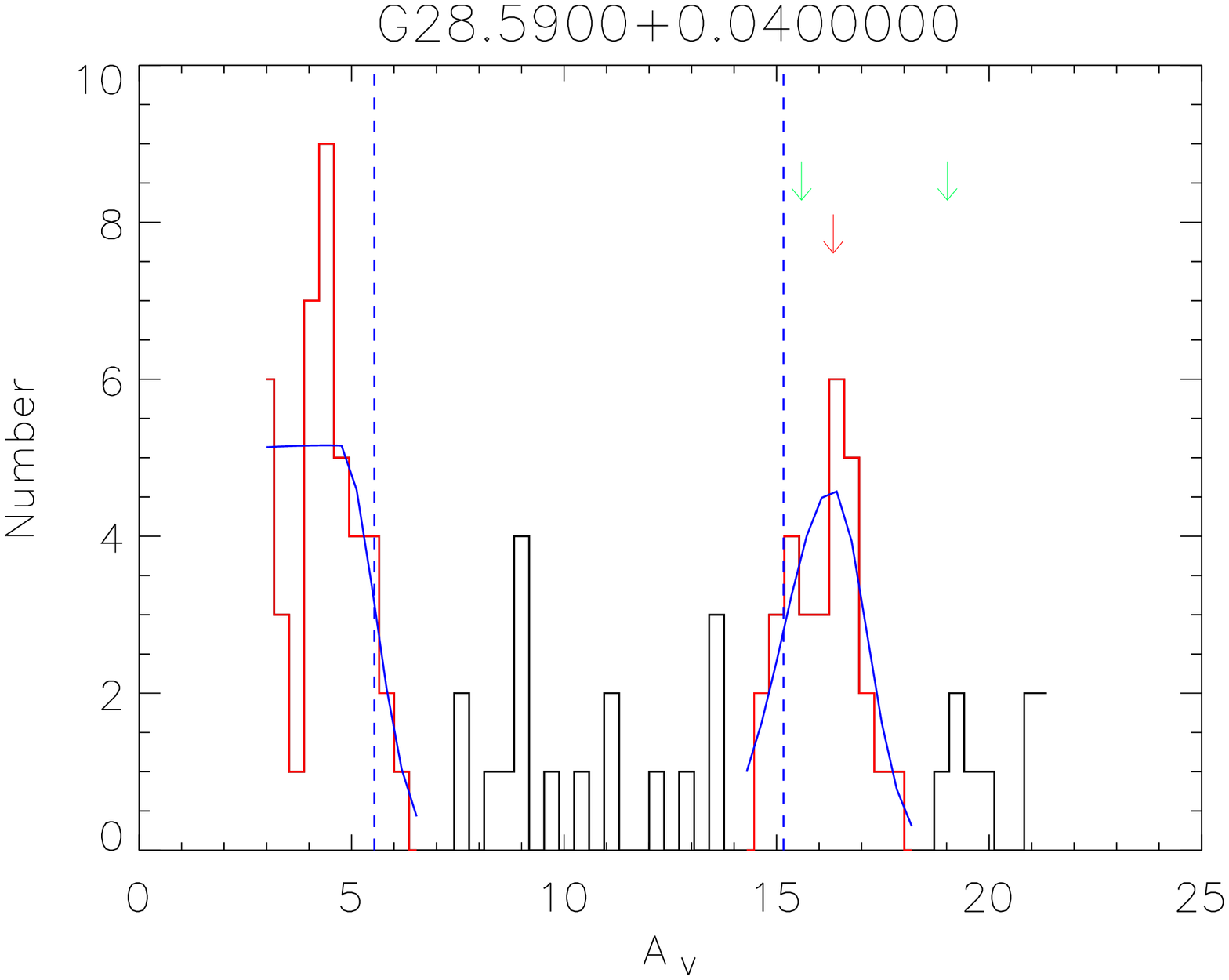}} &
      \resizebox{80mm}{!}{\includegraphics[angle=90]{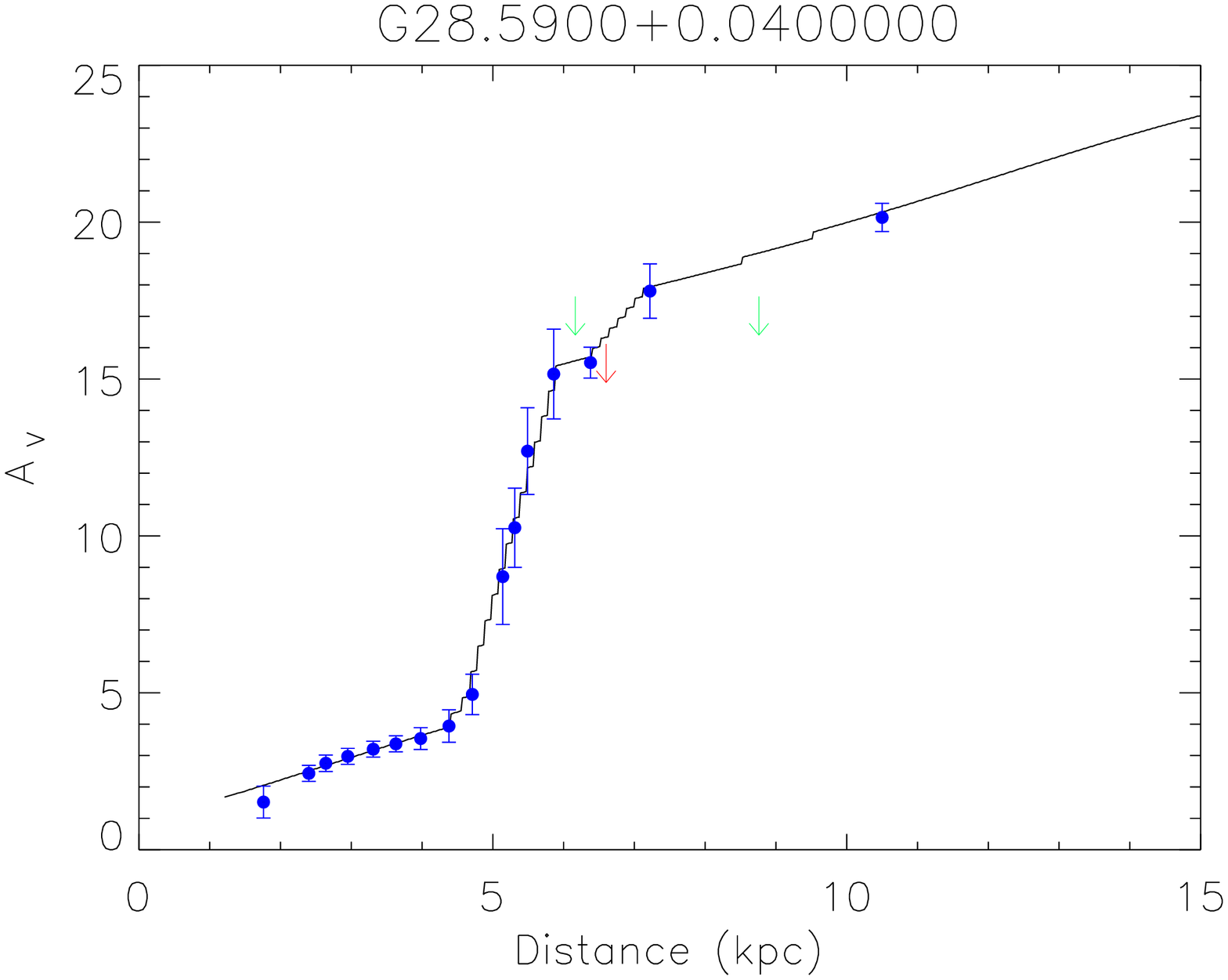}}\\
      \resizebox{80mm}{!}{\includegraphics[angle=90]{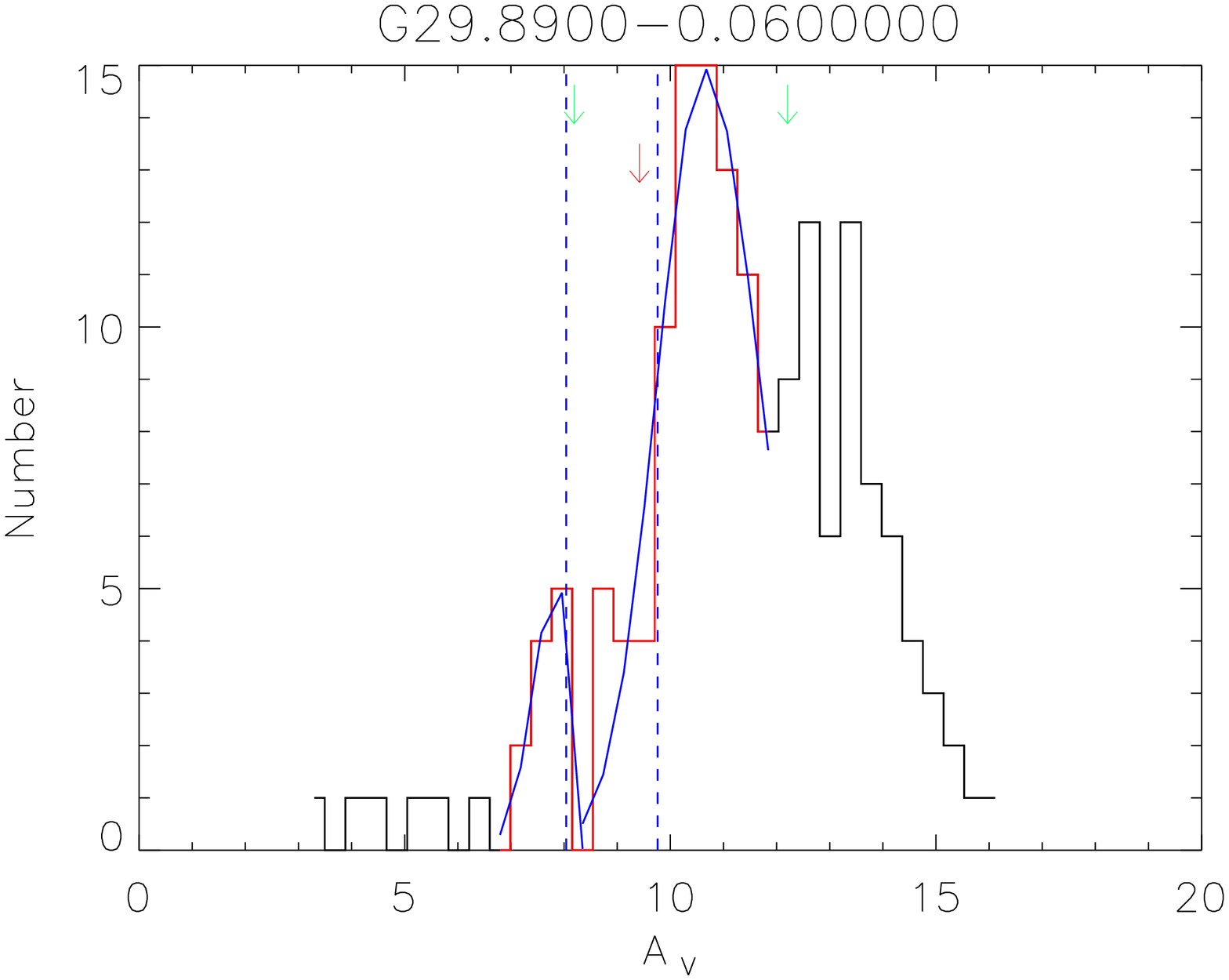}}&
      \resizebox{80mm}{!}{\includegraphics[angle=90]{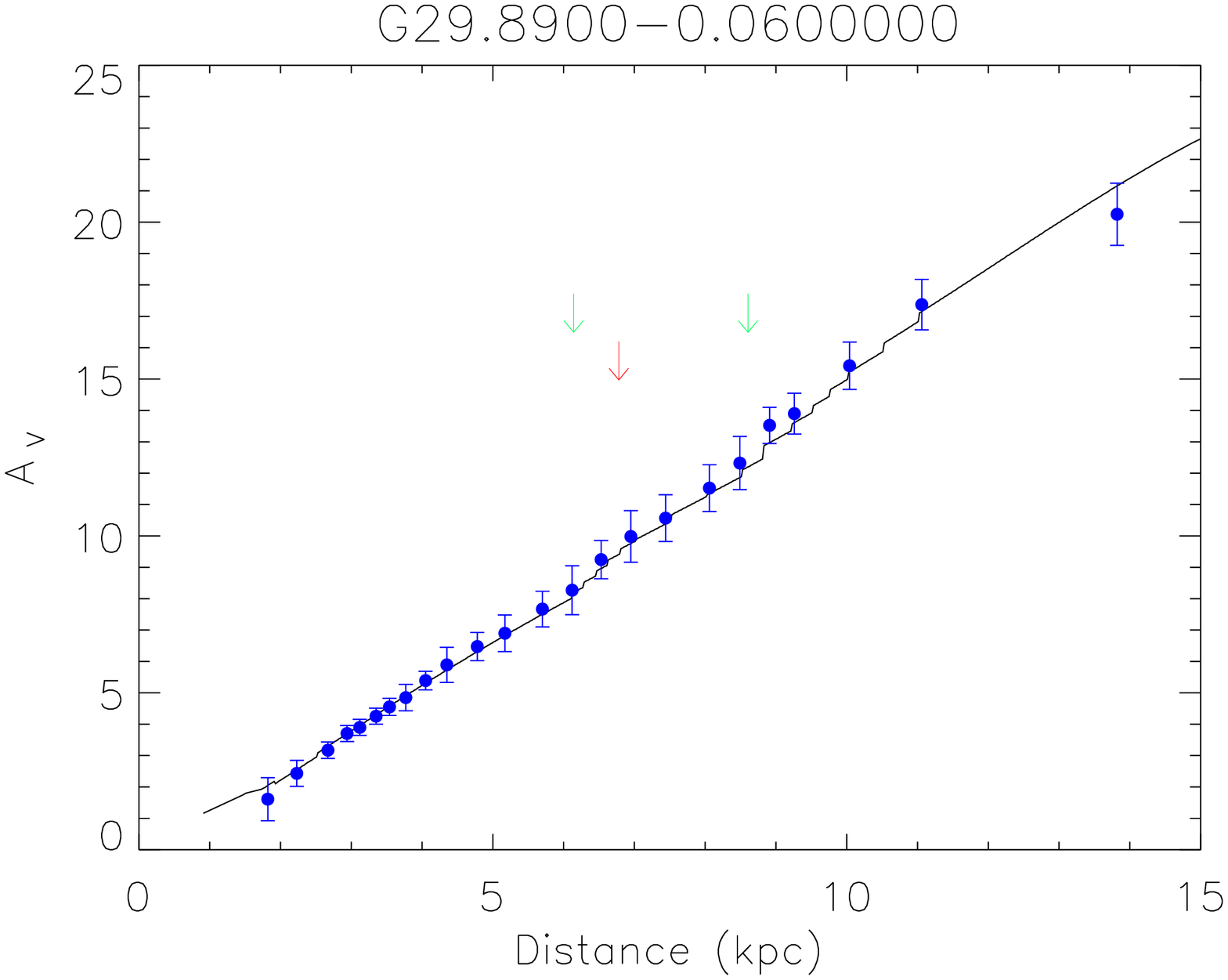}}\\
      \resizebox{80mm}{!}{\includegraphics[angle=90]{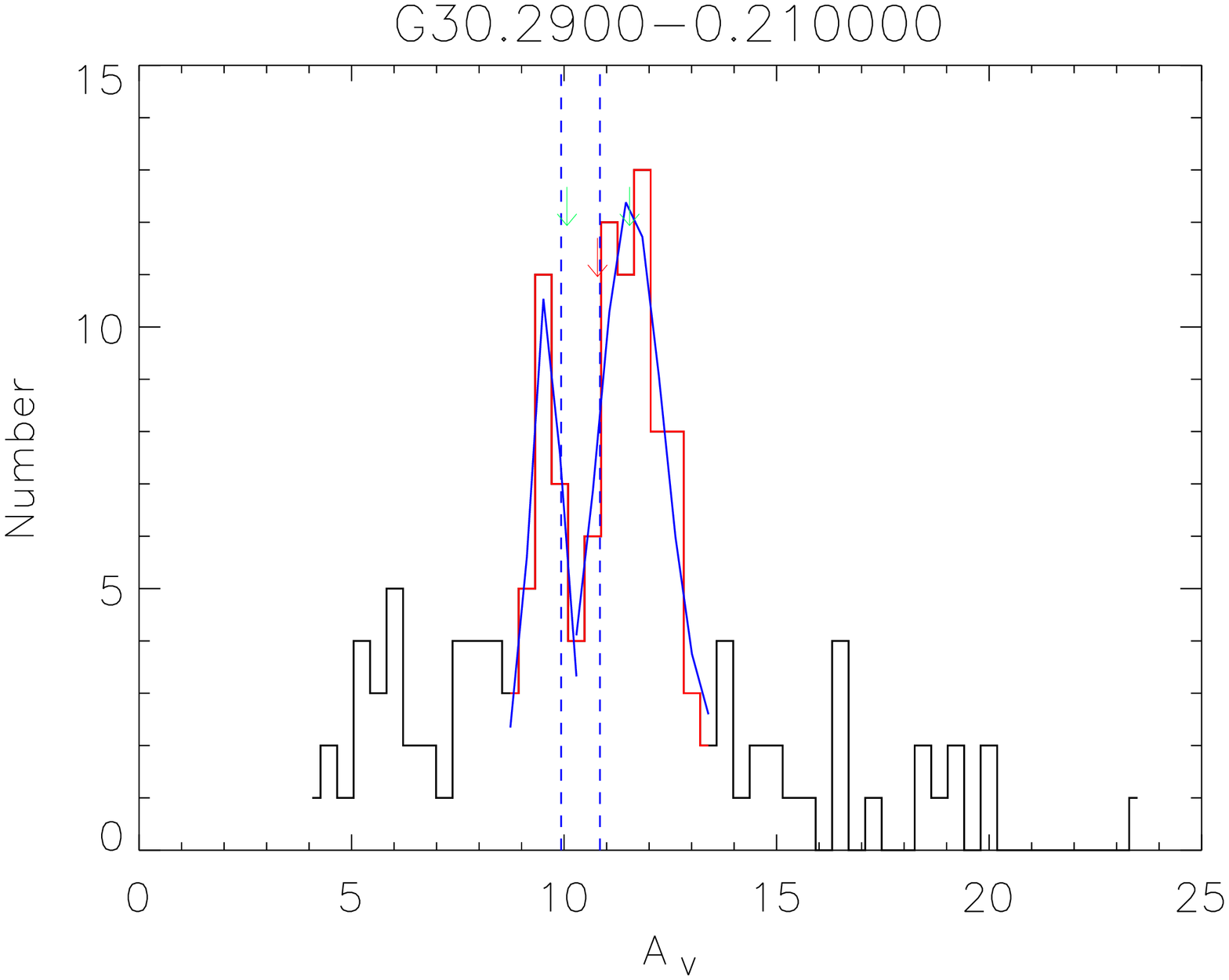}} &
      \resizebox{80mm}{!}{\includegraphics[angle=90]{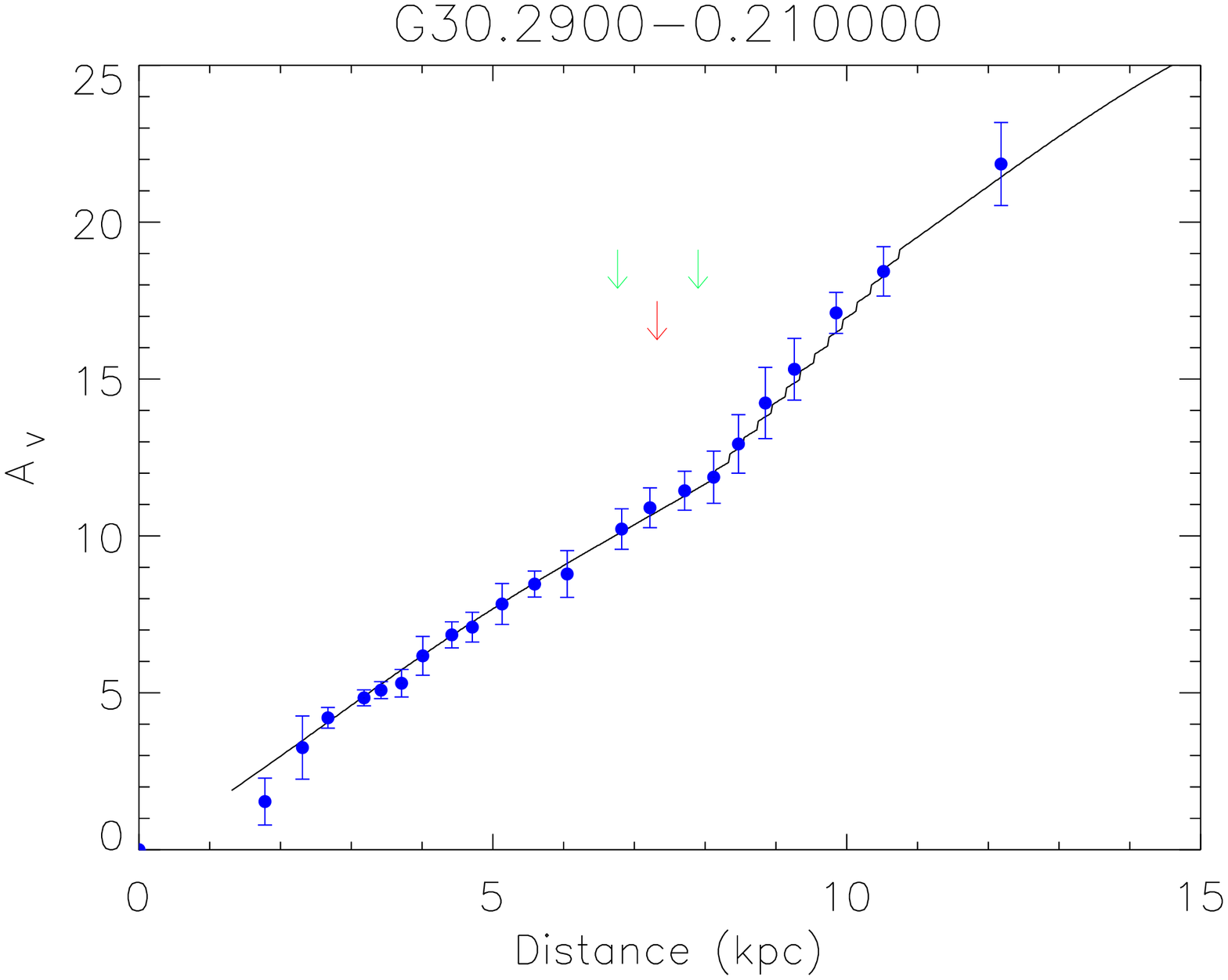}}\\
      \resizebox{80mm}{!}{\includegraphics[angle=90]{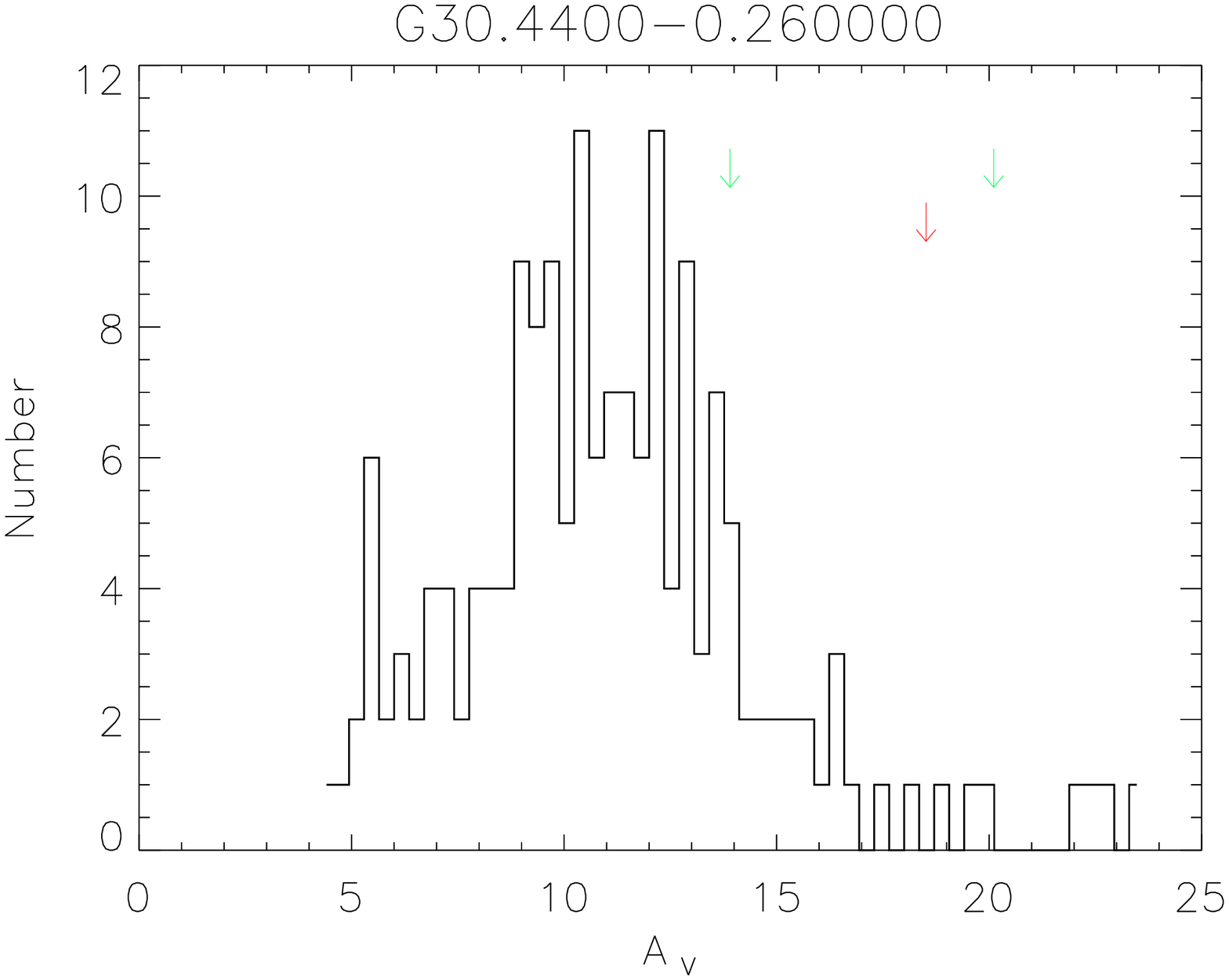}} &
      \resizebox{80mm}{!}{\includegraphics[angle=90]{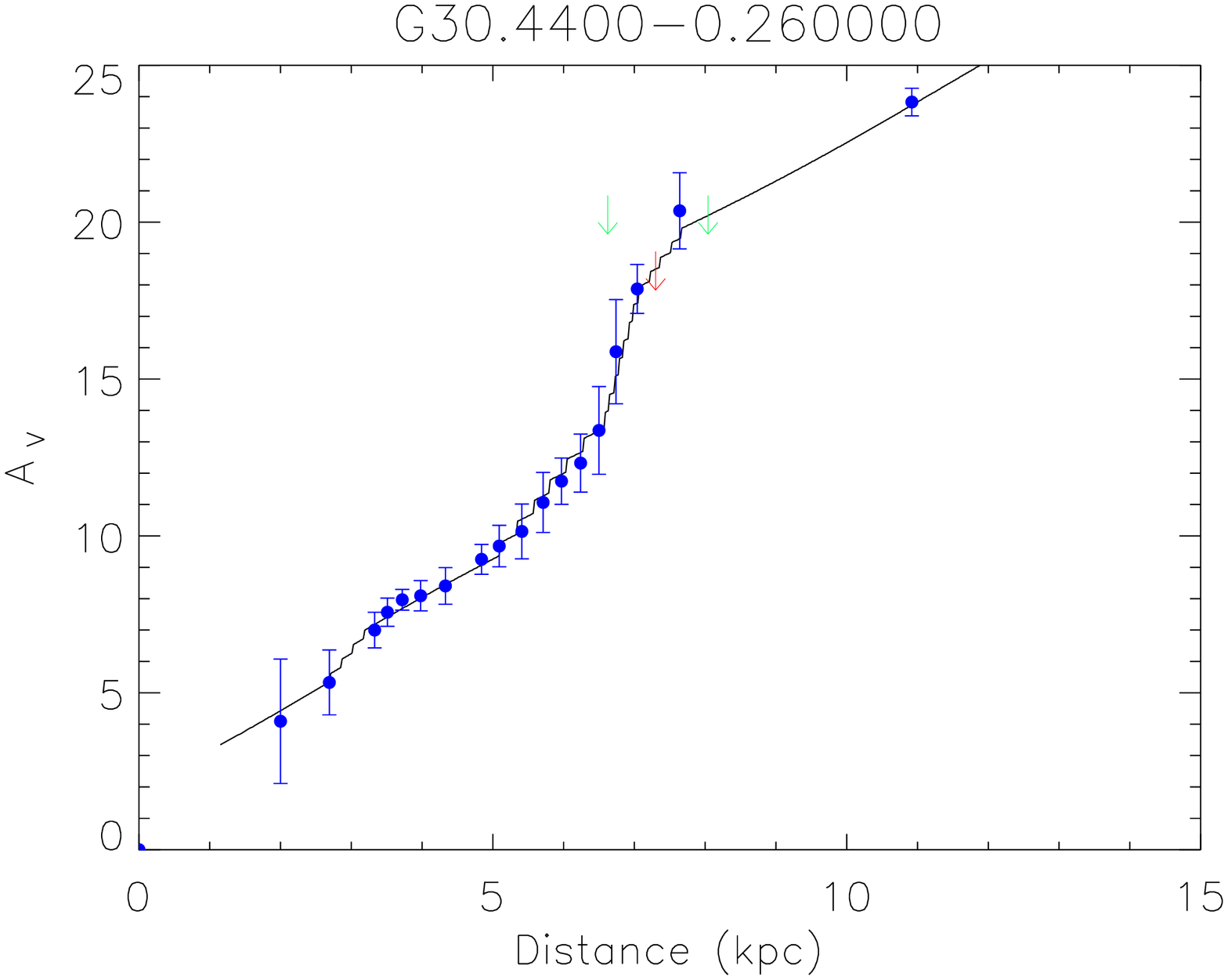}}
    \end{tabular} 
    \caption[]{\small }
    \label{fig:a1}
\end{center}
\end{figure*}

The molecular cloud G028.59+0.04 has RMS distances of 6.2/8.8 kpc and a GRS distance of 6.6 kpc. The A$_V$ histogram does not contain an adequate amount of data to detect a molecular cloud at the far RMS distance and so it has therefore been placed in group 2. There is however a very large dip in source counts in the A$_V$ histogram between A$_V$$\sim$6 and $\sim$14 (Fig. \ref{fig:a1}). The distance assigned to this dip is D = 5.3$\pm$0.9. The error is much smaller than might first be expected for such a large range of A$_V$, however upon inspection of the EDM  there is a very large increase in A$_V$ beginning at $\sim$4.5 kpc. Although the distance does agree with the near RMS source at the 1$\sigma$ level, there are several other overlapping clouds  covering a range of distances from $\sim$3.5 to $\sim$7.5 kpc.

\subsection{G045.49+0.04}

\begin{figure*}
  \begin{center}
    \begin{tabular}{cc}
      \resizebox{80mm}{!}{\includegraphics[angle=90]{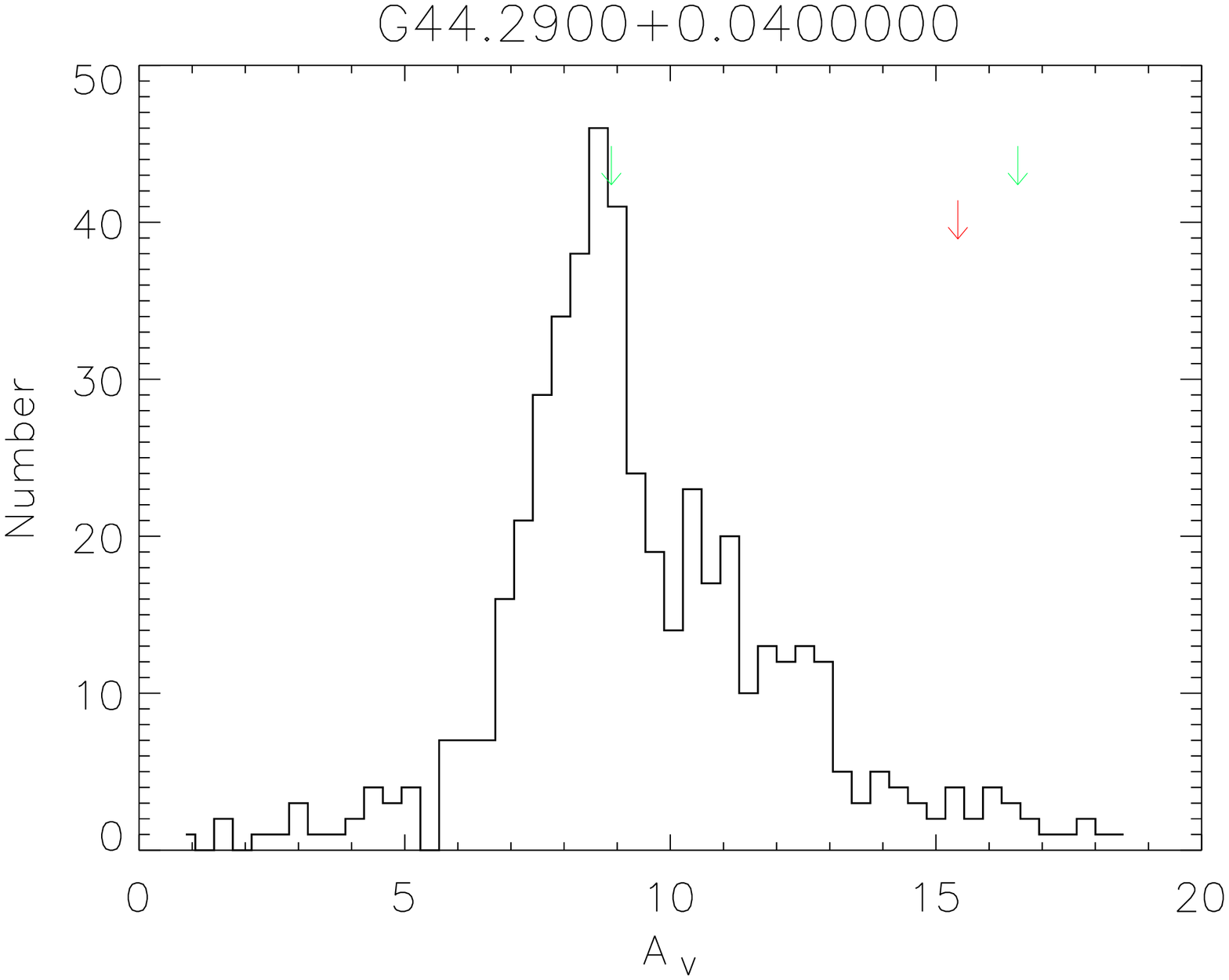}} &
      \resizebox{80mm}{!}{\includegraphics[angle=90]{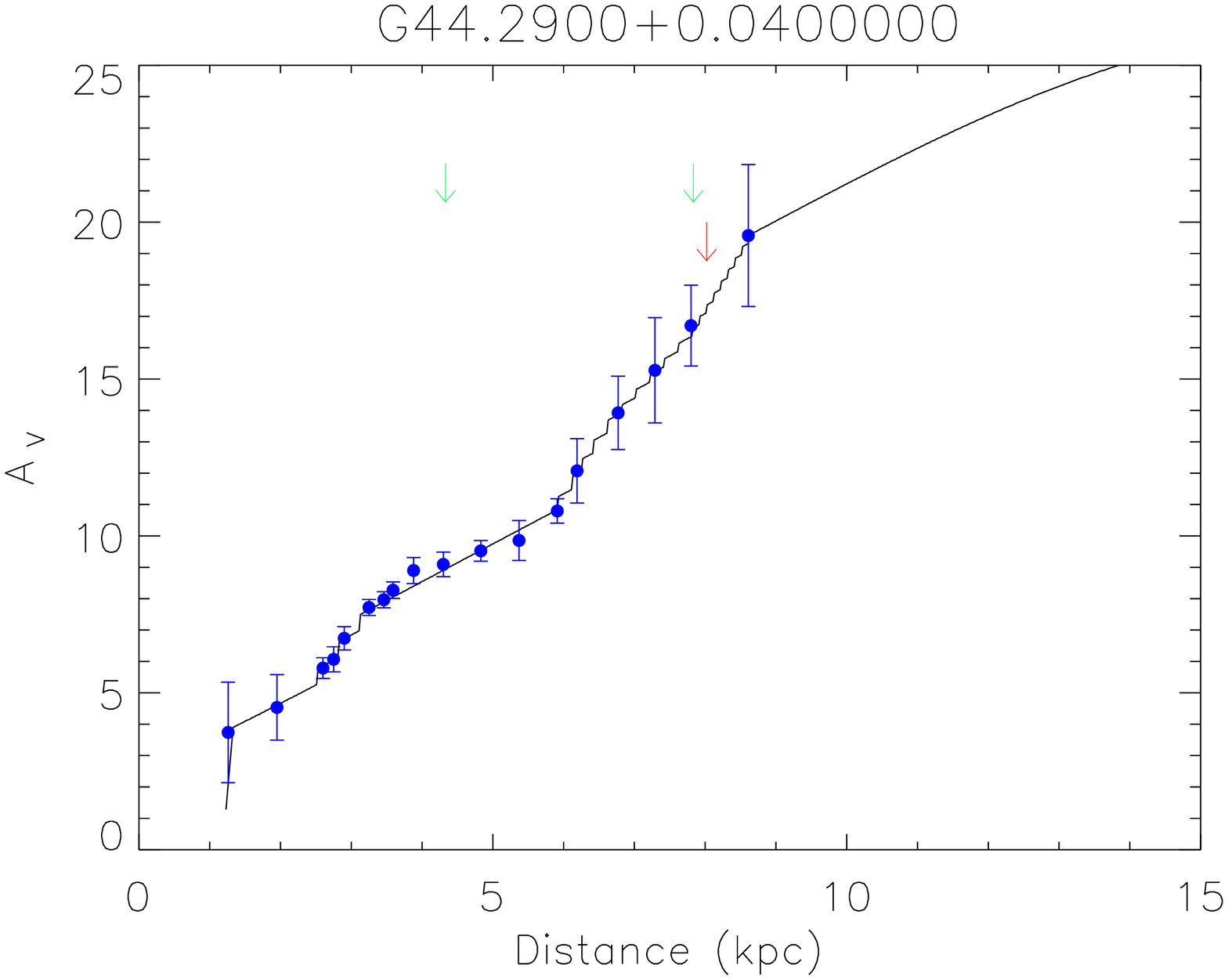}}\\
      \resizebox{80mm}{!}{\includegraphics[angle=90]{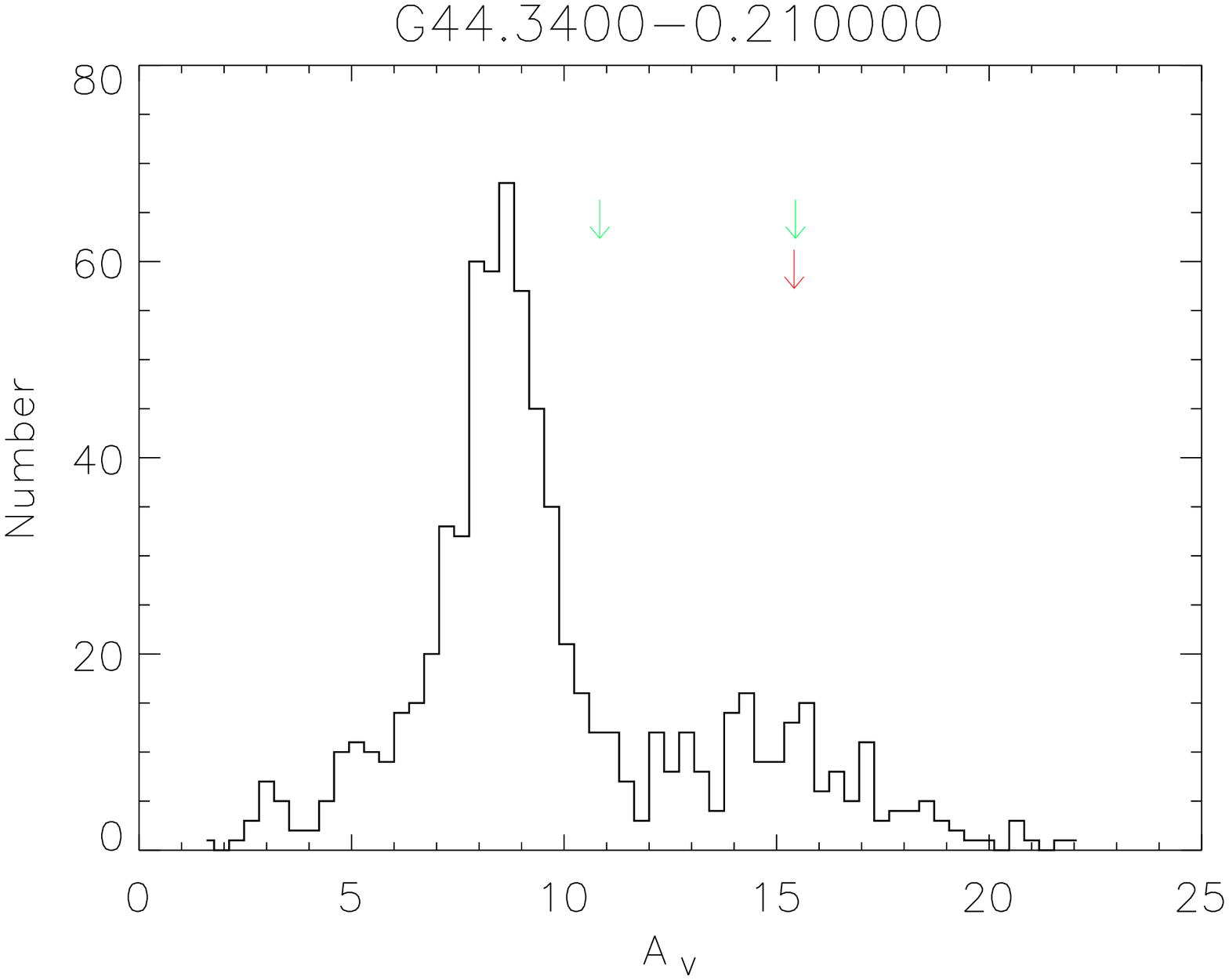}} &
      \resizebox{80mm}{!}{\includegraphics[angle=90]{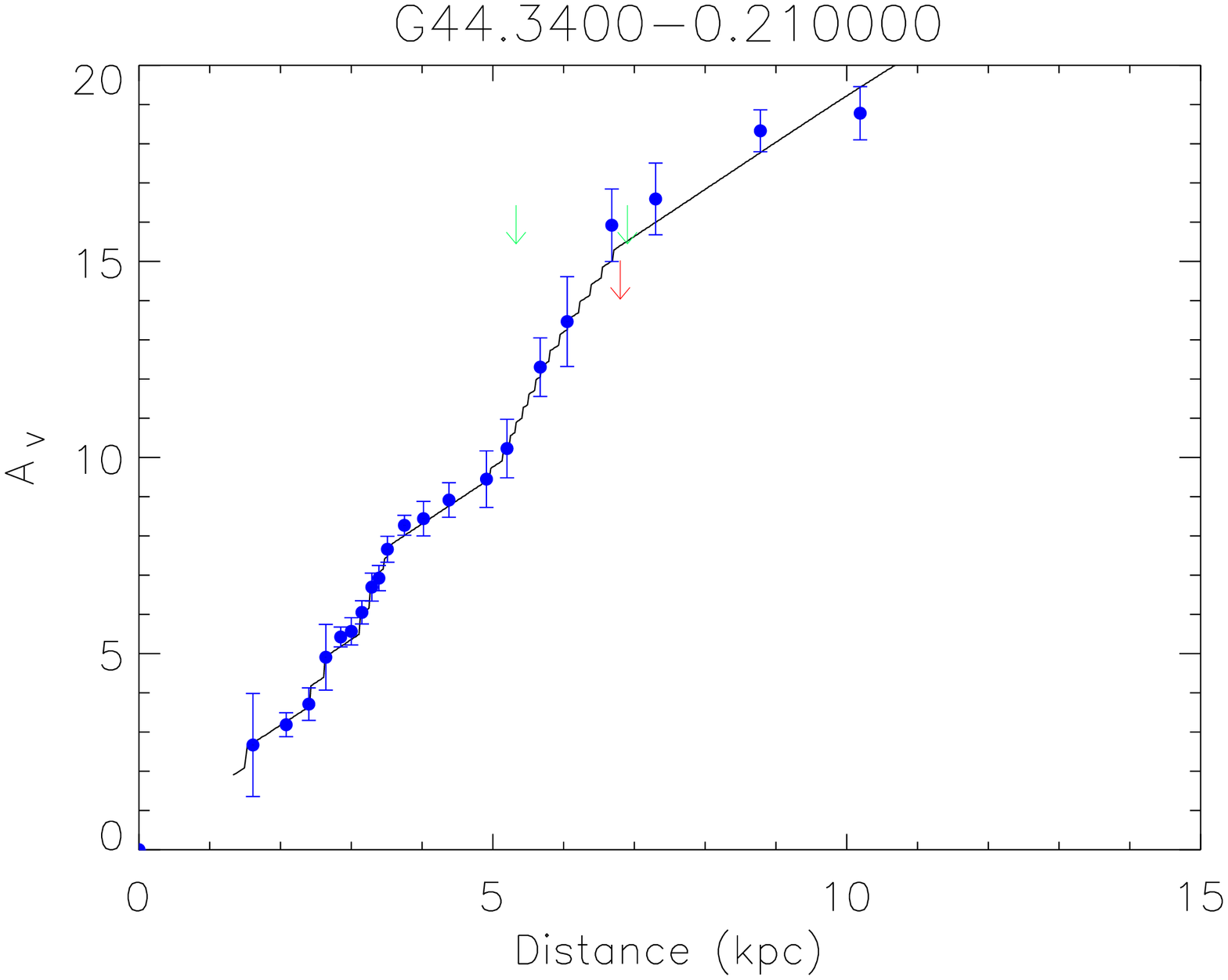}}\\
      \resizebox{80mm}{!}{\includegraphics[angle=90]{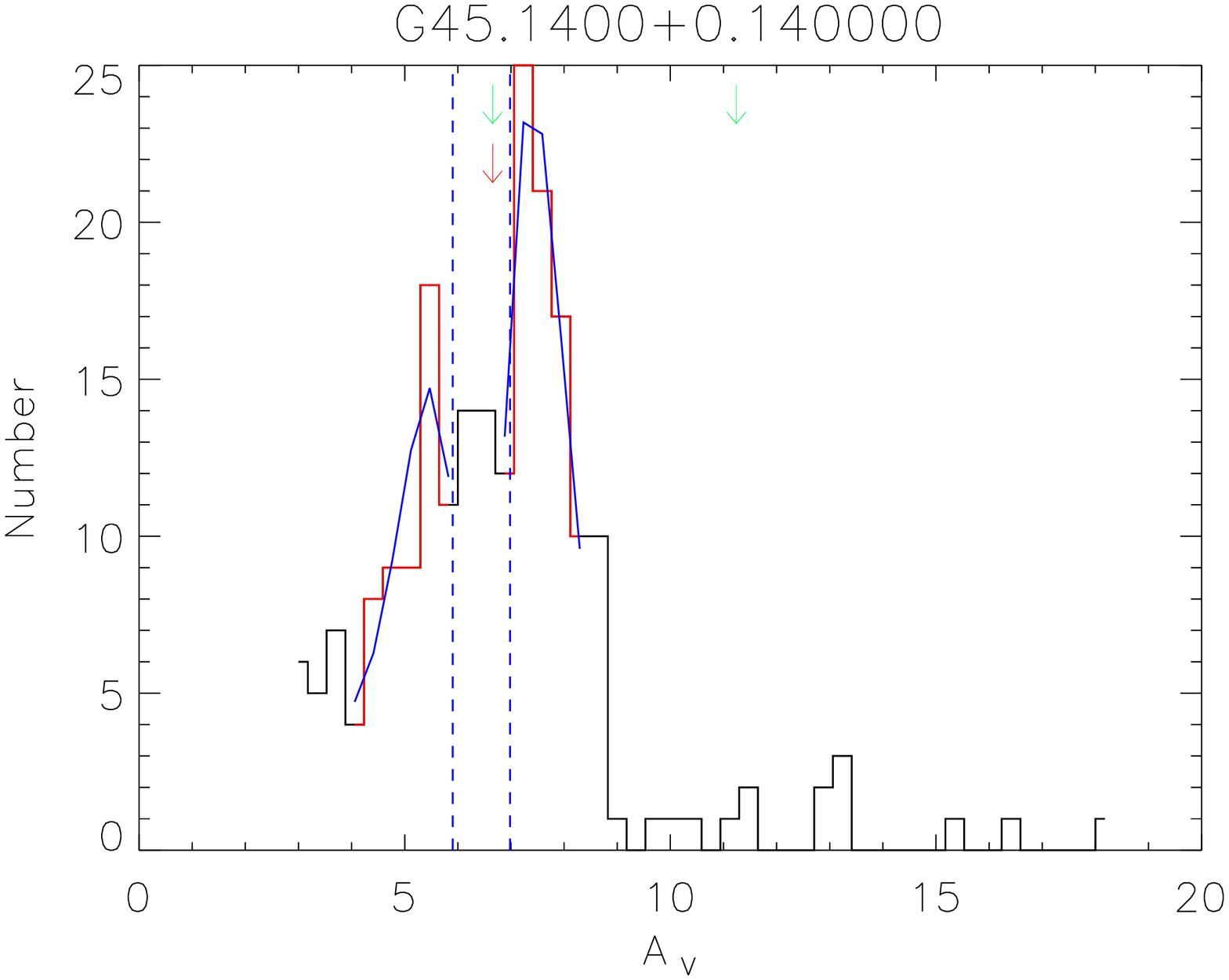}} &
      \resizebox{80mm}{!}{\includegraphics[angle=90]{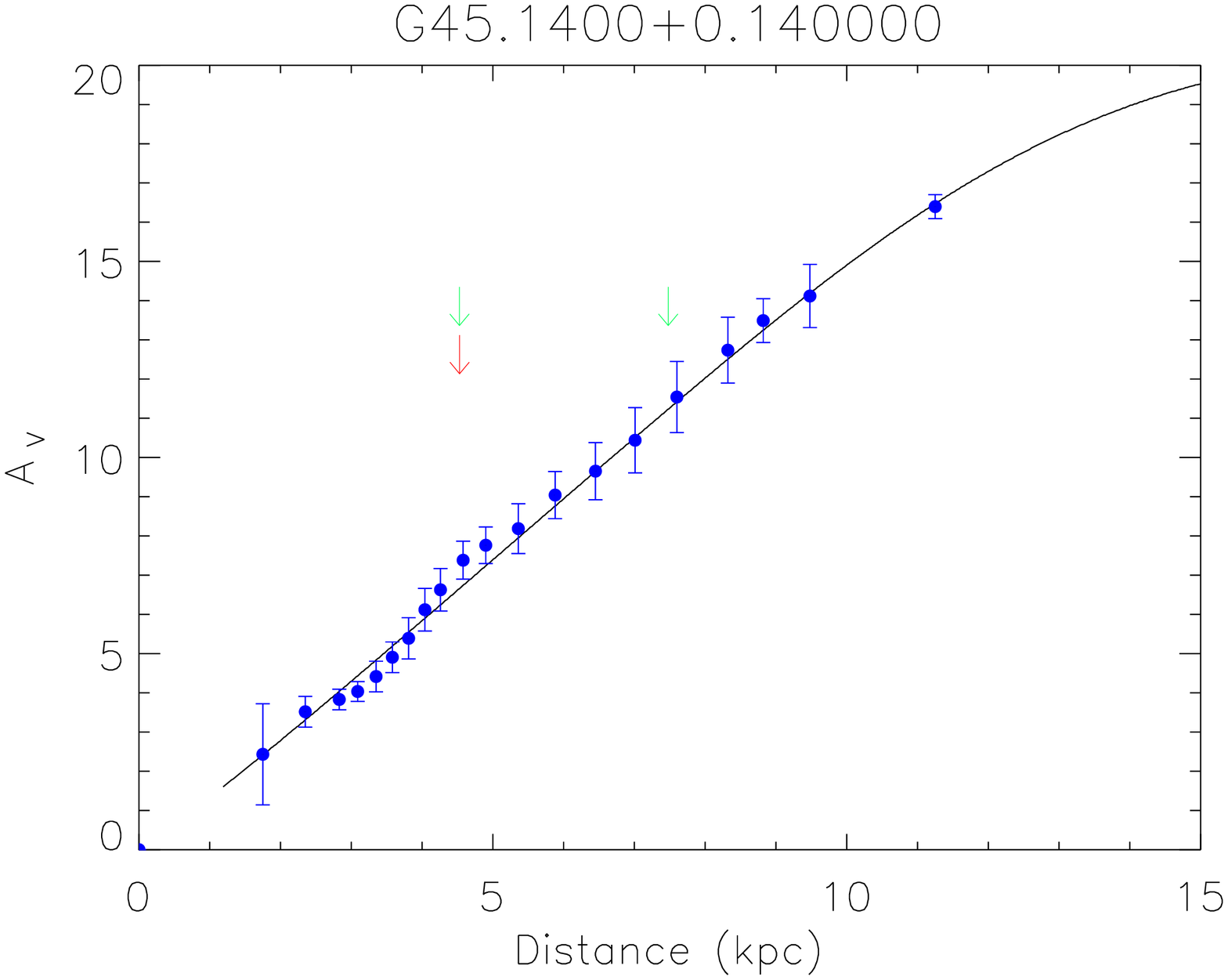}}\\
      \resizebox{80mm}{!}{\includegraphics[angle=90]{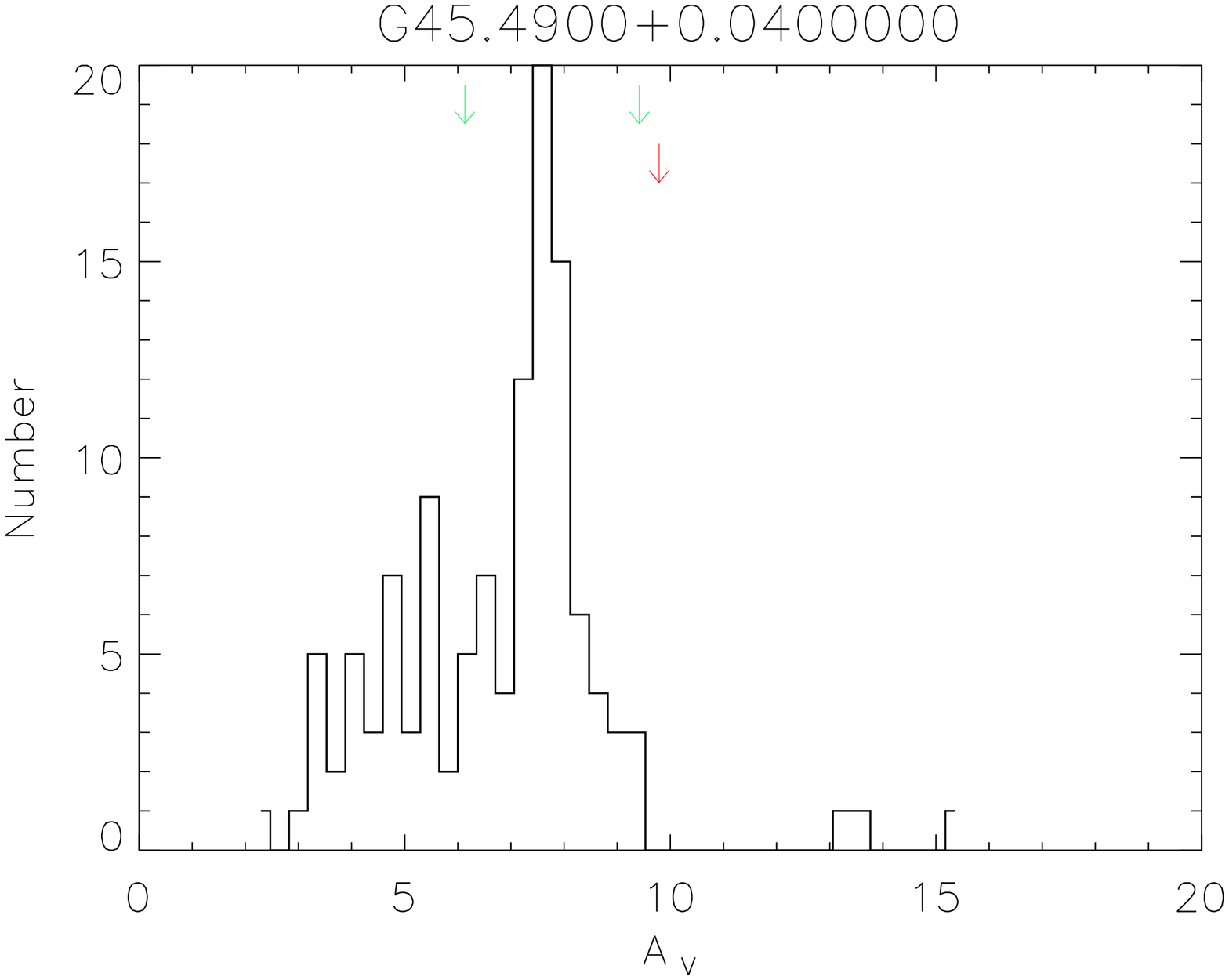}} &
      \resizebox{80mm}{!}{\includegraphics[angle=90]{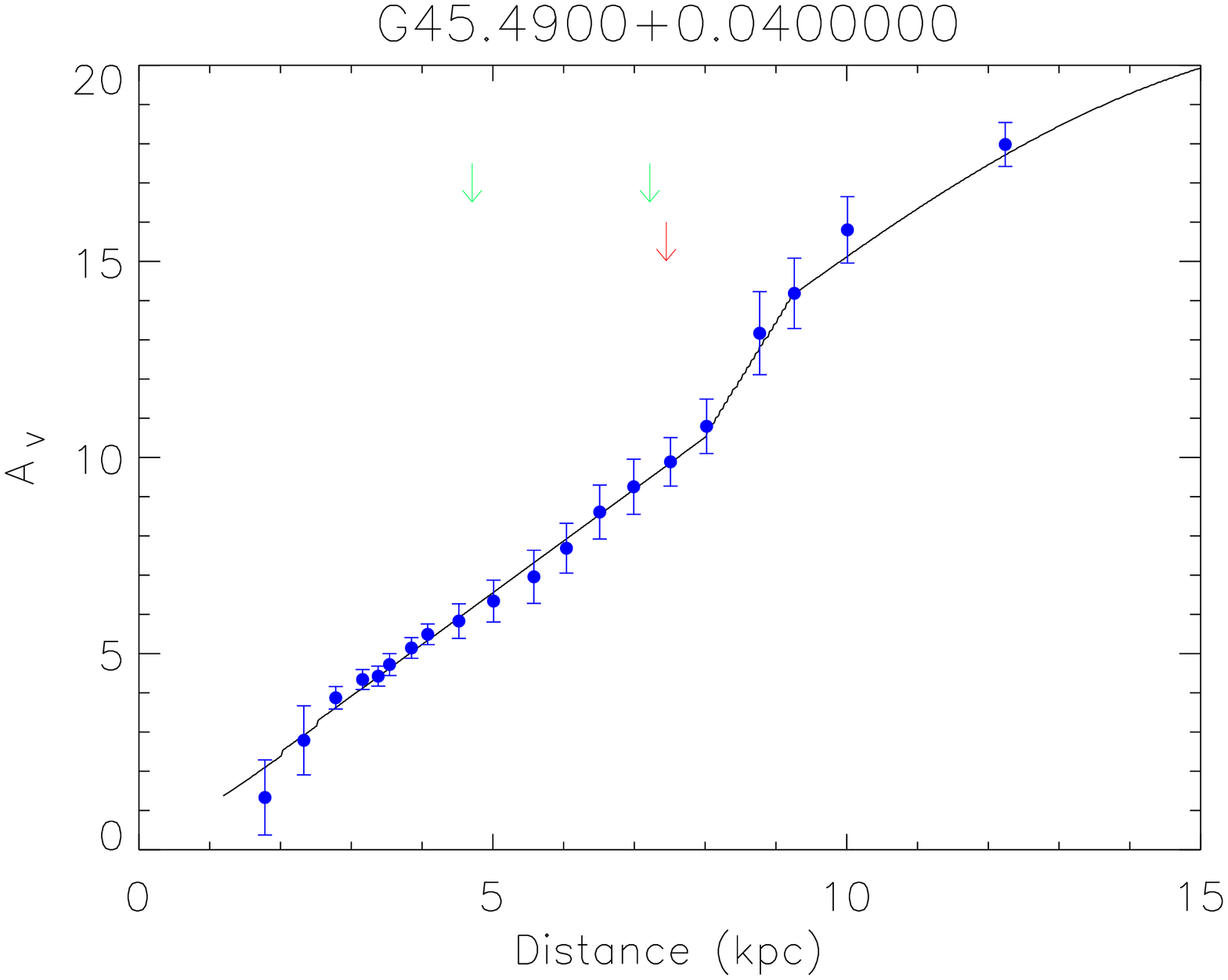}}
    \end{tabular} 
    \caption[]{\small }
    \label{fig:a6}
  \end{center}
\end{figure*}

G045.49+0.04 has been placed in group two as it is not possible to exclude the far RMS distance (Fig. \ref{fig:a6}). It has RMS distances of 4.7/7.2 kpc and a GRS distance of 7.5 kpc. There is however evidence of a molecular cloud in the A$_V$ histogram at A$_V$$\sim$6. It has an assigned distance of D = 4.6$\pm$1.4 kpc and so we are also unable to exclude the near distance. For this reason the cloud has been placed in the second subgroup of group two. There is however a second molecular cloud along the line of sight. The second cloud, G045.14+00.14, has a GRS distance of 4.5 kpc. This cloud could therefore explain the presence of the molecular cloud in the A$_V$ histogram, and so it may actually be possible to exclude the near distance.

\subsection{G030.79-0.06}

\begin{figure*}
  \begin{center}
    \begin{tabular}{cc}
      \resizebox{80mm}{!}{\includegraphics[angle=90]{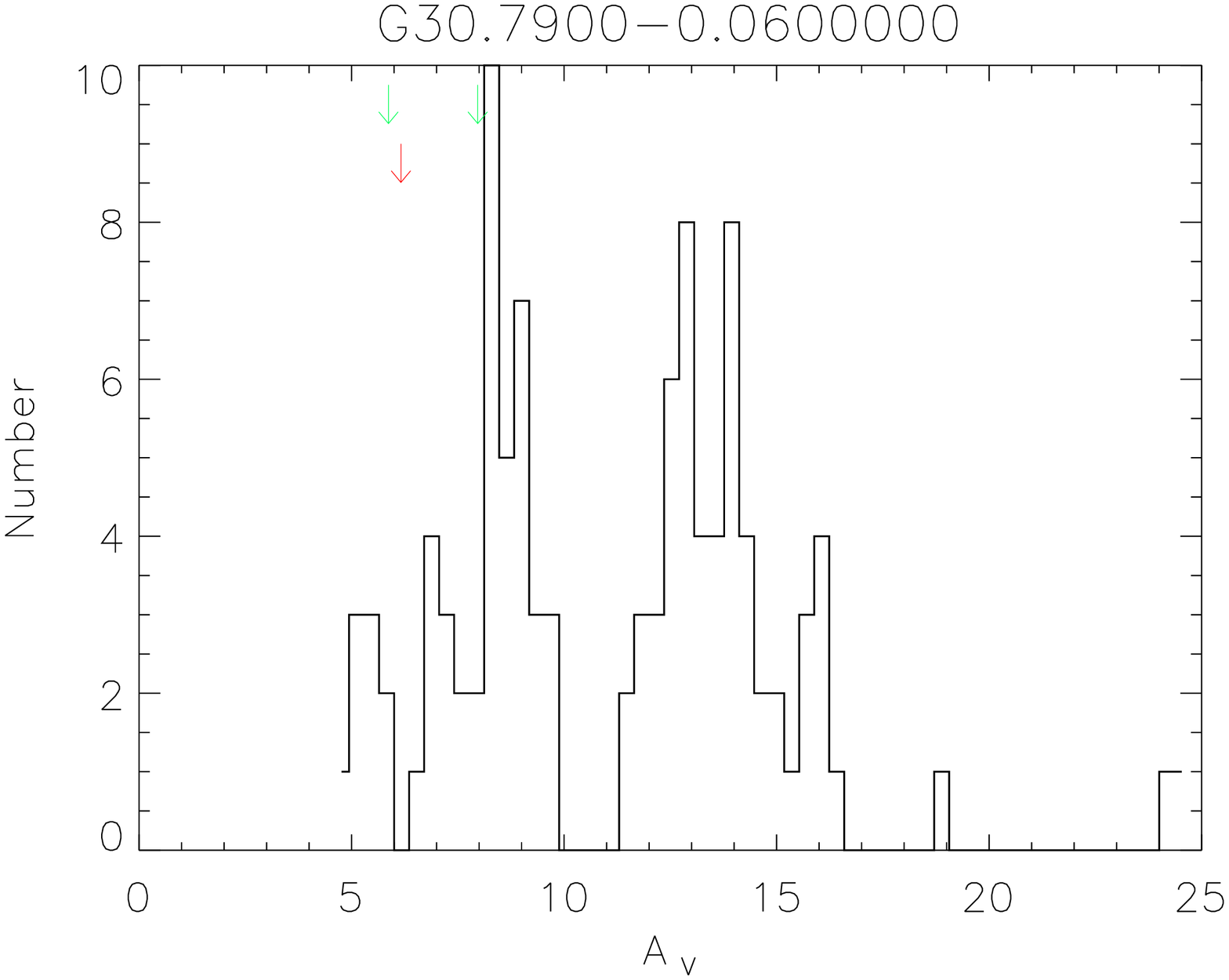}} &
      \resizebox{80mm}{!}{\includegraphics[angle=90]{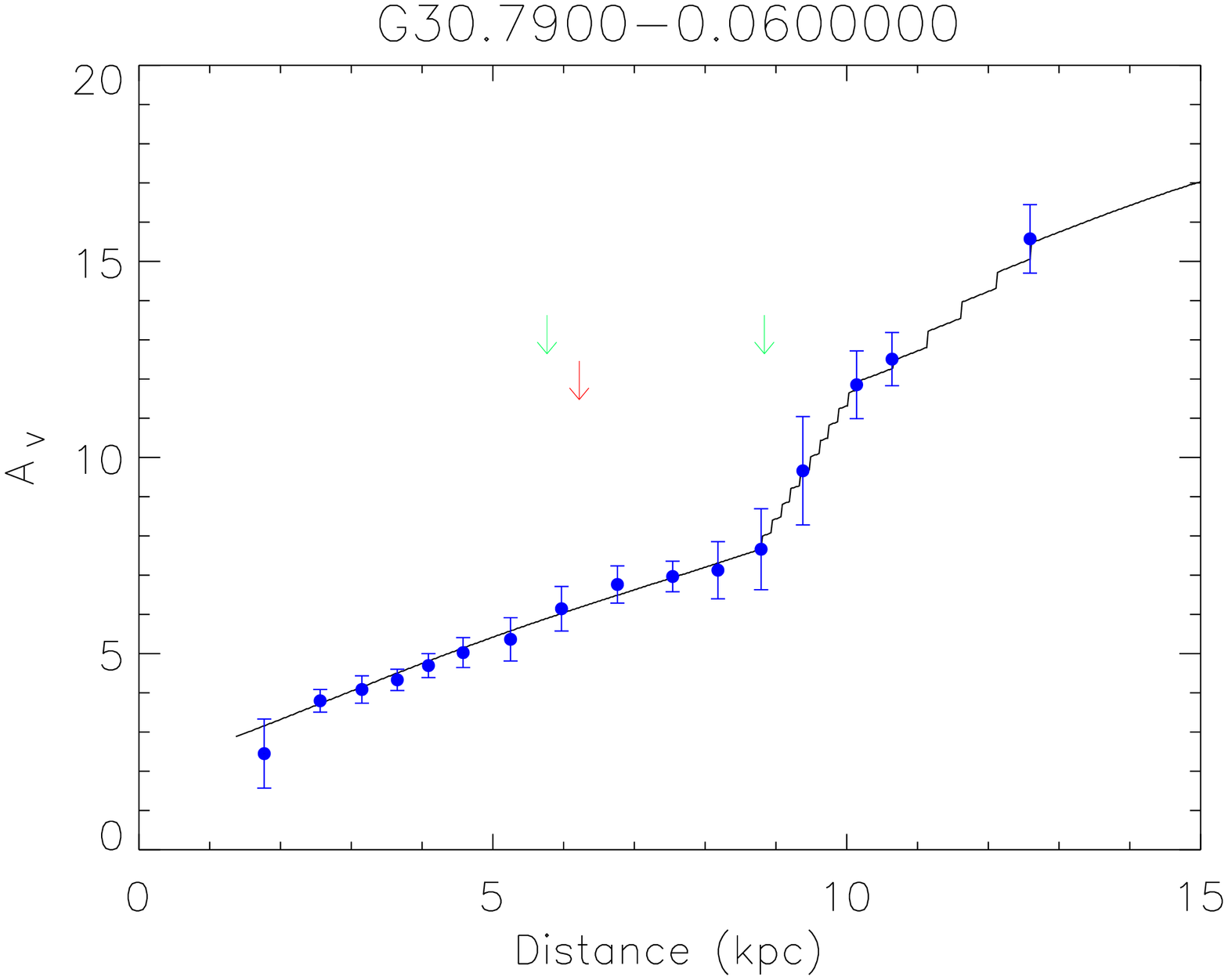}}\\
      \resizebox{80mm}{!}{\includegraphics[angle=90]{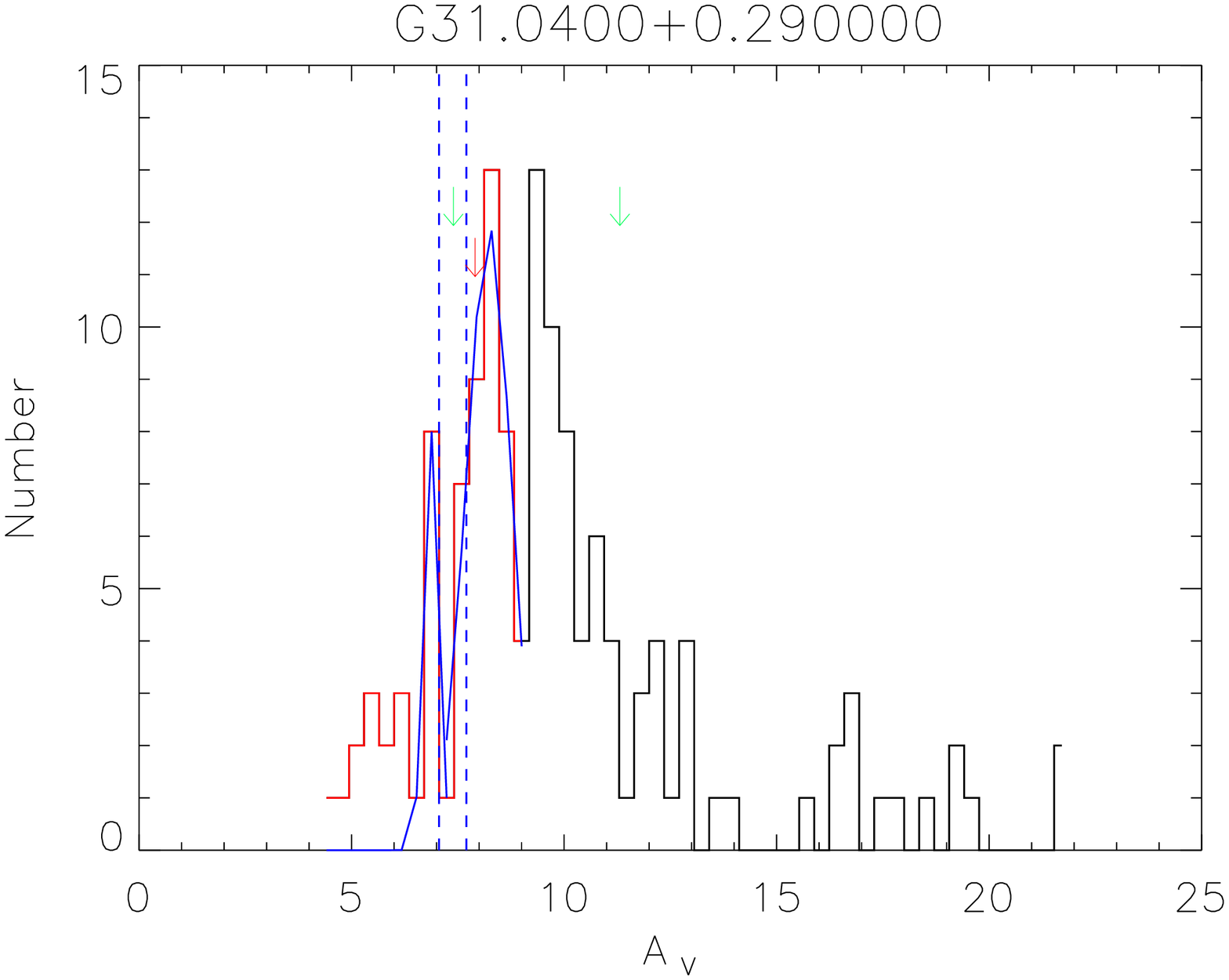}} &
      \resizebox{80mm}{!}{\includegraphics[angle=90]{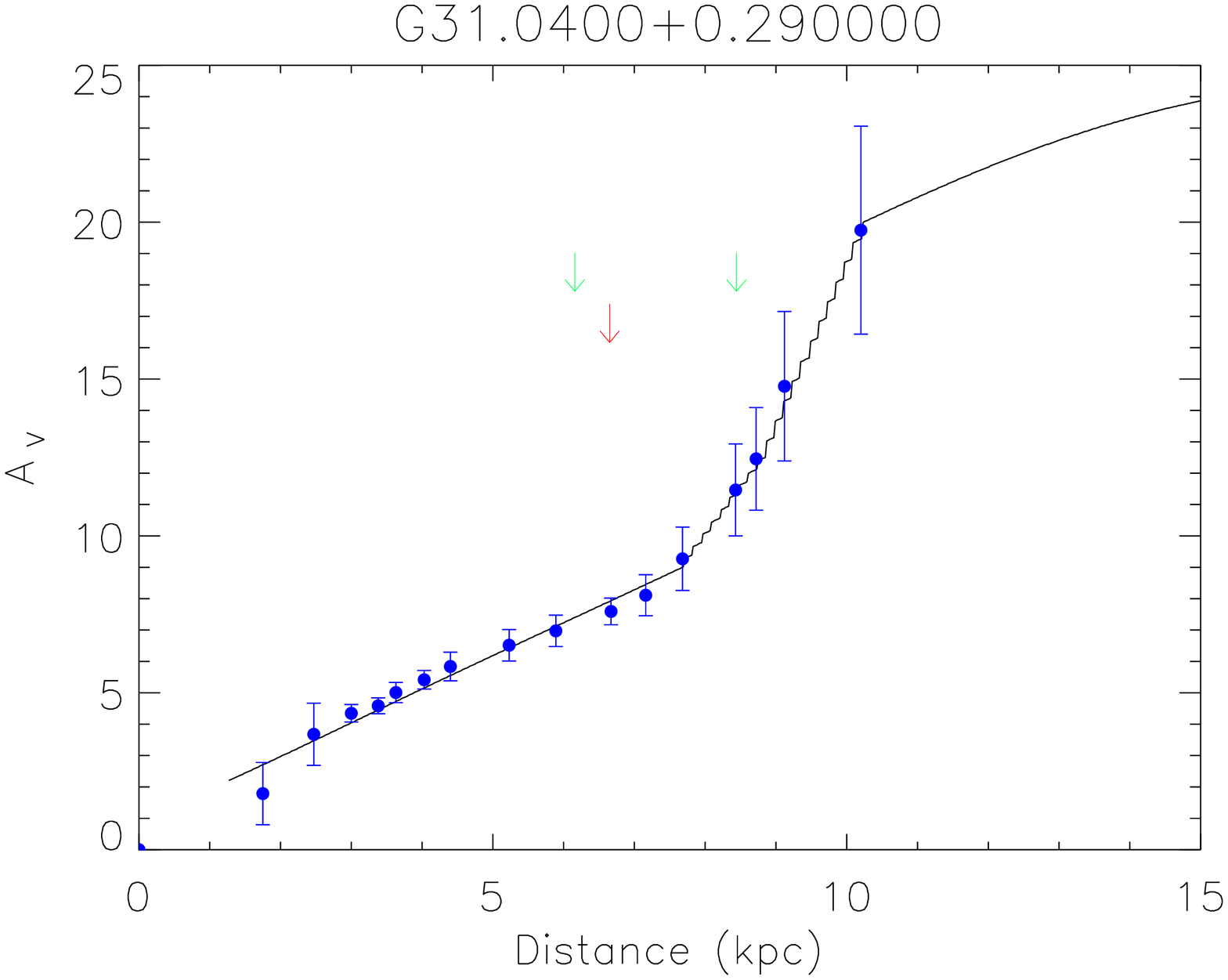}}\\
      \resizebox{80mm}{!}{\includegraphics[angle=90]{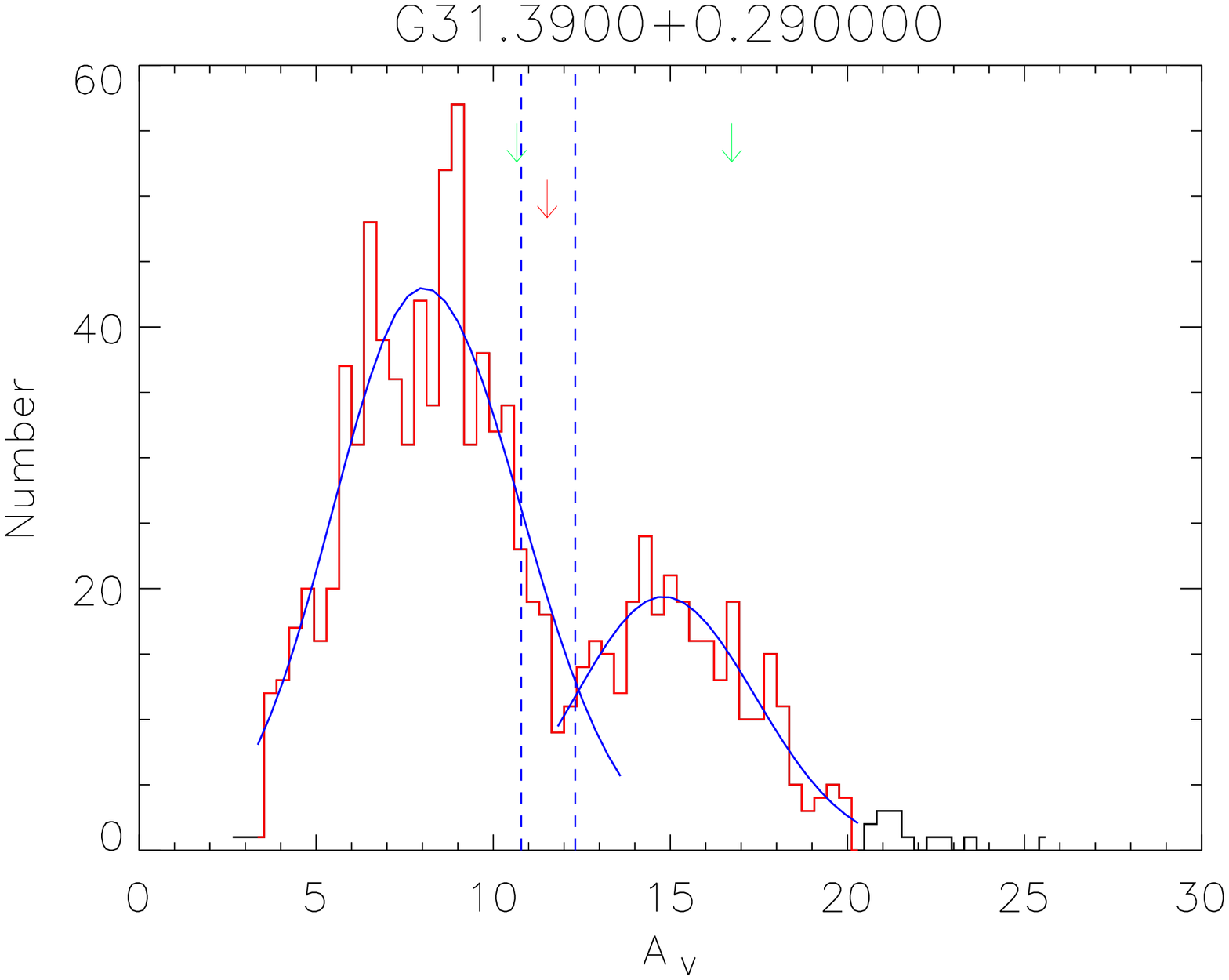}} &
      \resizebox{80mm}{!}{\includegraphics[angle=90]{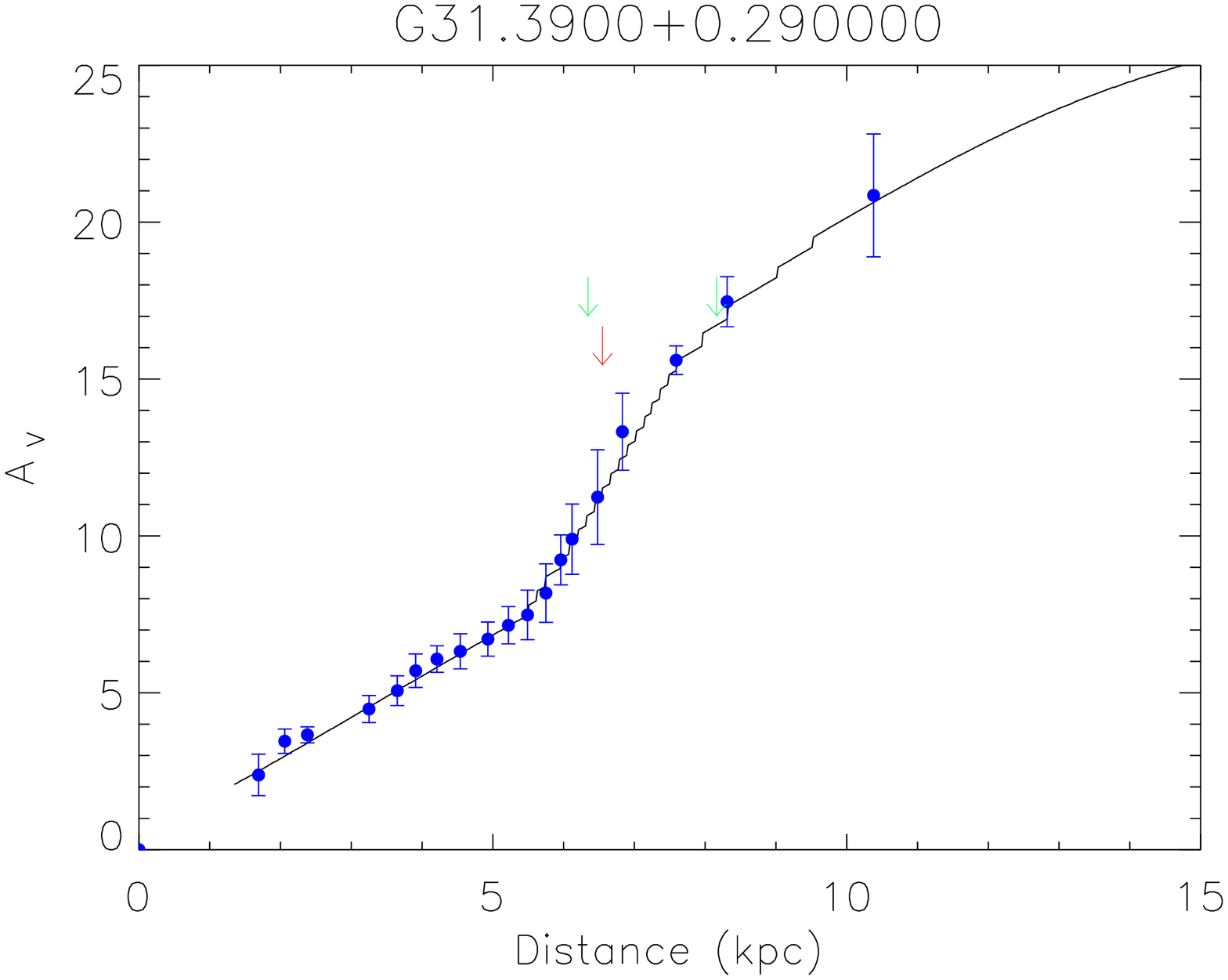}}\\
      \resizebox{80mm}{!}{\includegraphics[angle=90]{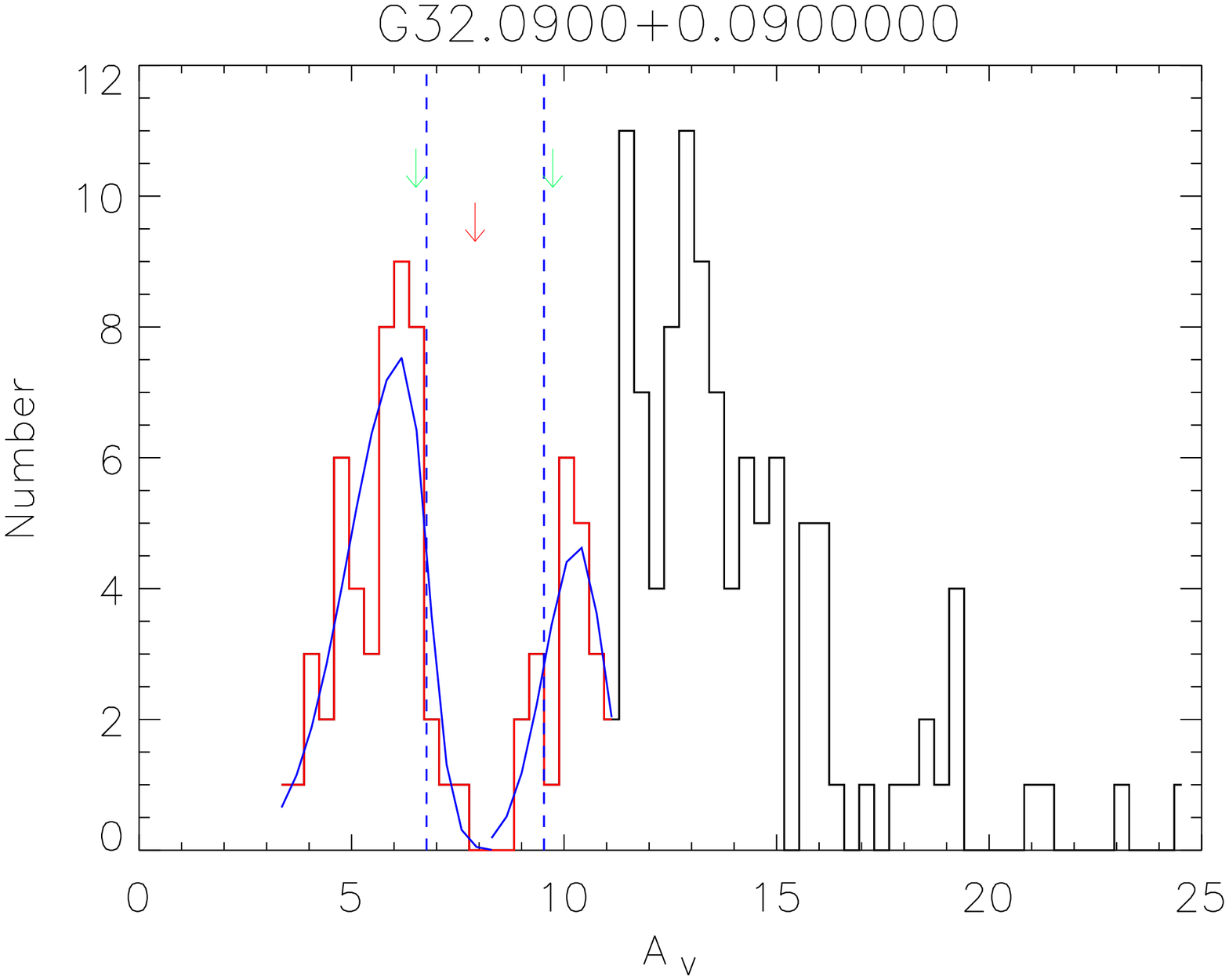}} &
      \resizebox{80mm}{!}{\includegraphics[angle=90]{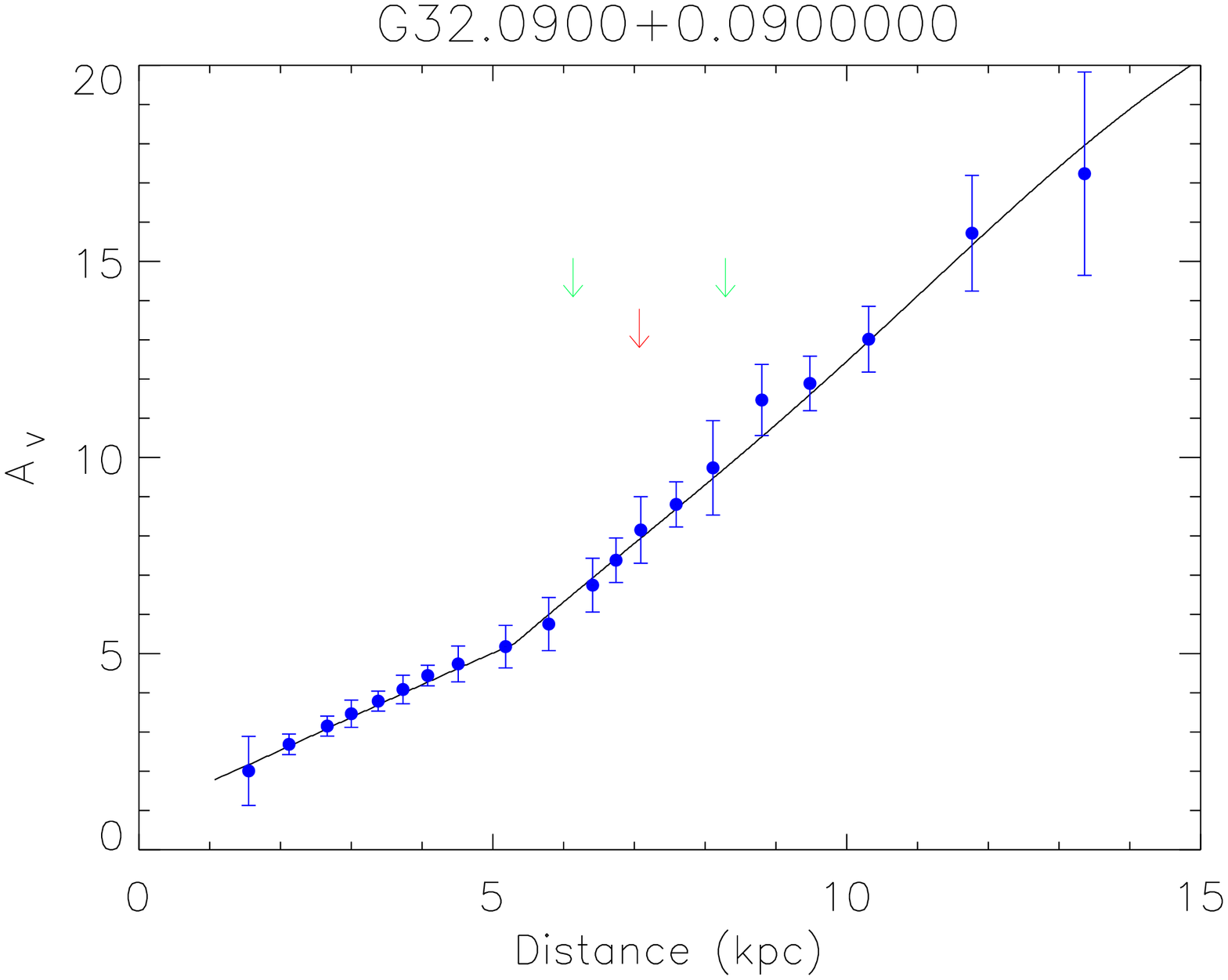}}
    \end{tabular} 
    \caption[]{\small }
\label{fig:a2}
  \end{center}
\end{figure*}

The molecular cloud G030.79-0.06 has RMS distances of 5.8/8.8 kpc and a GRS distance of 6.2 kpc. It has failed the goodness of fit test and so has been placed in group 3. There is however a very strong presence of a molecular cloud in the A$_V$ histogram, not associated with either kinematic distance (Fig. \ref{fig:a2}). The associated distance to this dip is D = 10.0$\pm$1.2 kpc. There is a second RMS source, G030.9281-00.0308, with an associated molecular cloud, G030.94-0.06, along the same line of sight. The RMS distances are 4.3/10.3 kpc and the GRS distance corresponds to the near distance at 4.4 kpc. This would appear to be the only confrontation with the results of \citet{roman09}. There is however another very CO bright molecular cloud, G031.24-0.01, with a GRS distance of 11.55 kpc that overlaps the same region of sky. \\
$\indent$As the GRS distance to the original cloud studied, G030.79-0.06, is 6.2 kpc, and the location of the dip, should it have been detected, would have been at the very near edge of the histogram (Fig. \ref{fig:a2}), it is apparent that for this line of sight we are unable to detect molecular clouds at distances much smaller than the GRS distance D = 6.2 kpc. The method failed because the line of sight extinction is too large along this particular direction. The molecular cloud G030.94-0.06 has a GRS distance of 4.4 kpc, and so it is likely that we are unable to detect the presence of this cloud in the A$_V$ histogram. As there does not appear to be a third A$_V$ gap in the histogram, it is likely that this prominent A$_V$ gap, at A$_V$$\sim$11 really is associated with the molecular cloud G031.24-0.01, at a GRS distance of 11.55 kpc. The M06 data set used is centred at G030.75+00.00 which is $\sim$30$\arcmin$ off-centre from this third molecular cloud G031.24-0.01. Therefore it is possible that the M06 data do not represent the line of sight extinction centred on this molecular cloud. For this reason the extinction measurement may not have been accurately converted to a distance measurement.

\bsp

\label{lastpage}

\end{document}